\RequirePackage{fix-cm}
\documentclass[smallextended]{svjour3}       
\smartqed

\usepackage{amsfonts} 
\usepackage{amsmath}
\usepackage{amssymb}
\usepackage{cancel} 
\usepackage{bm}
\usepackage[usenames, dvipsnames]{color} 
\usepackage{dcolumn}
\usepackage{epic} 
\usepackage{epsfig}
\usepackage{feynmp}
\usepackage{graphicx}
\usepackage{grffile}
\usepackage[breaklinks,colorlinks = true,linkcolor = red,urlcolor=blue,citecolor=red]{hyperref}
\usepackage{mathrsfs}
\usepackage{soul}
\usepackage{subfigure}
\usepackage{wrapfig} 
\usepackage{xy} 
\usepackage{xcolor}
\usepackage{amsmath,amssymb,amsfonts,amsthm}
\usepackage[ascii]{inputenc}
\usepackage{makecell}
\usepackage{hyperref}
\usepackage{cite}
\allowdisplaybreaks

\providecommand{\llangle}{\langle\!\langle}
\providecommand{\rrangle}{\rangle\!\rangle}

\numberwithin{equation}{section}

\providecommand{\url}[1]{#1}
\csname url@samestyle\endcsname

\providecommand{\BIBentrySTDinterwordspacing}{\spaceskip=0pt\relax}
\providecommand{\BIBentryALTinterwordstretchfactor}{4}
\providecommand{\BIBentryALTinterwordspacing}{\spaceskip=\fontdimen2\font plus
\BIBentryALTinterwordstretchfactor\fontdimen3\font minus
  \fontdimen4\font\relax}
\providecommand{\BIBforeignlanguage}[2]{{%
\expandafter\ifx\csname l@#1\endcsname\relax
\typeout{** WARNING: IEEEtran.bst: No hyphenation pattern has been}%
\typeout{** loaded for the language `#1'. Using the pattern for}%
\typeout{** the default language instead.}%
\else
\language=\csname l@#1\endcsname
\fi
#2}}
\providecommand{\BIBdecl}{\relax}
\BIBdecl

\begin{document}

\title{Nonlinear Fluctuating Hydrodynamics for the Classical XXZ Spin Chain}


\author{Avijit Das, Kedar Damle, Abhishek Dhar, David A. Huse, Manas Kulkarni, Christian B. Mendl and Herbert Spohn}
\authorrunning{Avijit Das et al.}

\institute{Avijit Das \at
              International Centre for Theoretical Sciences, Tata Institute of Fundamental Research,  Bengaluru 560089, India.
              \email{avijit.das@icts.res.in}           
           \and
           Kedar Damle \at
           Tata Institute of Fundamental Research,  Mumbai 400005, India.
           \and
           Abhishek Dhar \at
           International Centre for Theoretical Sciences, Tata Institute of Fundamental Research,  Bengaluru 560089, India.
           \and
           David A. Huse \at
           Department of Physics, Princeton University, Princeton, NJ 08544, USA.
           \and
           Manas Kulkarni \at
           International Centre for Theoretical Sciences, Tata Institute of Fundamental Research,  Bengaluru 560089, India.
           \and
           Christian B. Mendl \at
           Technische Universit\"{a}t Dresden, Institute of Scientific Computing, Dresden, Germany.
           \and
           Herbert Spohn \at
           Zentrum Mathematik and Physik Department, Technische Universit\"{a}t M\"{u}nchen, Garching, Germany.        
}

\maketitle
\begin{abstract}
Using the framework of nonlinear fluctuating hydrodynamics (NFH), we examine equilibrium spatio-temporal correlations in classical ferromagnetic spin chains with nearest neighbor interactions. In particular, we consider the classical XXZ-Heisenberg spin chain (also known as Lattice Landau Lifshitz  or LLL model) evolving deterministically and chaotically via Hamiltonian dynamics, for which energy and $z$-magnetization are the only locally conserved fields. For the easy-plane case, this system has a low-temperature regime
in which the difference between neighboring spin's angular orientations in the XY plane is an \textit{almost conserved} field.  According to the predictions of NFH, the dynamic correlations in this regime exhibit a heat peak and propagating sound peaks, all with anomalous broadening. We present a detailed molecular dynamics test of these predictions and find a reasonably accurate verification. We find that, in a suitable intermediate temperature regime, the system shows two sound peaks with Kardar-Parisi-Zhang (KPZ) scaling and a heat peak where the expected anomalous broadening is less clear. In high temperature regimes of both easy plane and easy axis case of LLL, our numerics show clear diffusive spin and energy peaks and absence of any sound modes, as one would expect. We also simulate an integrable version of the XXZ-model, for which the ballistic component instead moves with a broad range of speeds rather than being concentrated in narrower peaks around the sound speed.
 
\keywords{Hydrodynamics . Dynamical Correlations . Heisenberg Spin Chain}
\end{abstract}

\tableofcontents

\section{Introduction}
\label{sec1} 
For generic classical and quantum spin chains the only conservation law is the energy, perhaps in addition one  spin component (or all three). 
Momentum conservation is destroyed by the underlying lattice and in thermal equilibrium the average currents vanish. One therefore expects that a local perturbation of the thermal state will spread diffusively, a behavior which is actually observed in a large variety of systems, as prototypical examples we refer to\cite{aoki2000bulk,sirker2011conservation}. There are obvious exceptions such as integrable spin chains, for which a small perturbation induces a ballistic response. Also, at least classically, at very low temperatures the harmonic approximation generally becomes valid, which then implies ballistic transport over a suitable time scale.

The goal of our paper is to explain that, beyond the standard folklore, there can be a parameter regime, in which the dynamic correlations consist of a ``heat'' peak at the origin and in addition two sound peaks symmetrically moving to the right and left.  These peaks broaden sub-ballistically but faster than diffusion. The theoretical argument is based on nonlinear fluctuating hydrodynamics (NFH), which refers to long wavelength behavior and should thus be equally valid for both classical and quantum chains. 
For quantum chains, because of numerical limitations, it is difficult to pin down the phenomenon, even less so the precise scaling form. In this contribution we thus restrict our study to classical spin chains of XXZ type. We note that for off-lattice models, where there is  momentum conservation (e.g anharmonic oscillator chains), the presence of sound peaks with anomalous scaling has been observed in several recent studies and is well understood within the framework of NFH \cite{MendlSpohn2013,das2014numerical}. The results presented here for the XXZ model bear close resemblance to those seen for the rotor model, where early results in \cite{PhysRevLett.84.2144,PhysRevLett.84.2381,Yang2003} indicated a diffusive to super-diffusive transport, and has recently been understood in the framework of NFH \cite{das2014role,spohn2014fluctuating}. It should also be noted that there are no phase transitions (from diffusive to anomalous transport regimes), rather the anomalous scaling is observed over very long transient time-scales.

For our argument we require that the anisotropy parameter $\Delta=|J_z/J_{x}|=|J_z/J_{y}|$, where  $J_z$ is the nearest neighbor coupling between $z$-components of spin and $J_{x}=J_{y}$ between the $x$ or $y$-components respectively, satisfies $\Delta < 1$. Then, at high temperatures, energy and the $z$-component of the spin diffuse. However at low temperatures the spin motion is confined to a plane orthogonal to the $z$-axis (the easy-plane) and phase differences between neighboring spins are small.  To achieve differences of order $\pi$ ( in other words ``phase slips'' or, equivalently, ``umklapp'') is an activated process and is thus strongly thermally suppressed in a low temperature regime.  In this regime, the phase differences are an almost conserved field, so there is a broad range of time scales where this conservation law dictates the hydrodynamics. Under such conditions, nonlinear fluctuating hydrodynamics can be applied. The theory predicts the dynamical correlations contain a central non-propagating heat peak and left- and right-moving sound peaks. The three peaks broaden as nontrivial powers of time according to characteristic explicitly known scaling functions \cite{SpohnAHC2014,Spohn2016book}.

To illustrate the difference between sound peaks and ballistic broadening, we also simulate the integrable spin chain of Fadeev and Takhtajan \cite{faddeev2007hamiltonian}. 
The infinite number of conservation laws
then leads to a structured scaling function which scales self-similarly as $\sim t^{-1}f(x/t)$. \\
There has been previous work on non-integrable classical models (and their KPZ connection) such as the Fermi-Pasta-Ulam chain \cite{MendlSpohn2013,das2014numerical}, the discrete nonlinear Schr\"{o}dinger equation \cite{PhysRevA.88.021603,PhysRevA.92.043612,MendlSpohnNLS2015}, coupled rotors \cite{das2014role,spohn2014fluctuating,Spohn2016book}, and one-dimensional hard-point systems \cite{PhysRevE.90.012147}.
However, to our knowledge the present paper is the first exploration of NFH in classical spin chains.

\section{High-temperature, non-integrable and integrable Lattice Landau-Lifshitz equations}
\label{sec2}\noindent
\textbf{LLL equations} --- We consider spins of unit length on the one-dimensional lattice, $\vec{S}_j = (S_j^x,S_j^y,S_j^z)$ with $| \vec{S}_j| = 1$, $j \in \mathbb{Z}$.
The standard LLL interaction is quadratic in the spins with Hamiltonian
\begin{equation}\label{2.1}
H = - \sum_{j \in \mathbb{Z}} \big( S_j^xS_{j+1}^x + S_j^yS_{j+1}^y + \Delta S_j^zS_{j+1}^z\big),
\end{equation}
$\Delta$ the asymmetry parameter, $\Delta \geq 0$. The LLL equations of motion then read
\begin{equation}\label{2.2}
\tfrac{d}{dt}\vec{S}_j = \{\vec{S}_j,H\}= \vec{S}_j \times \vec{B}_j,\quad \vec{B}_j = -\nabla_{\vec{S}_j} H,
\end{equation}
where the Poisson bracket between two functions, $g_1,g_2$, of the spin variables is defined by $\{g_1,g_2 \} = \sum_j\epsilon_{\alpha\beta\gamma} \big(\partial g_1/\partial {S_j^\alpha}
\big) \big(\partial g_2/\partial {S_j^\beta}\big) S_j^\gamma$ with the usual summation convention. Clearly $| \vec{S}_j(t)| = 1$ for all times. The Hamiltonian character of the dynamics can be seen also by introducing the position-like angular variable $\phi_j \in S^1$ and the conjugate canonical momentum-like variable $s_j \in [-1,1]$
defined through
\begin{equation}\label{2.3}
S_j^x = f(s_j) \cos \phi_j,\quad S_j^y = f(s_j) \sin \phi_j,\quad S_j^z = s_j ,
\end{equation}
where $f(x) =(1-x^2)^{1/2}$. Indeed, one checks that $\{s_i,\phi_j\} = \delta_{ij}$, $\{\phi_i,\phi_j\} = 0$, $\{s_i,s_j\} = 0$. In these variables the hamiltonian \eqref{2.1}  reads
\begin{equation}\label{2.4}
H = - \sum_{j \in \mathbb{Z}} \Big( f(s_j) f(s_{j+1}) \cos( \phi_{j+1}-\phi_j) + \Delta s_js_{j+1} \Big).
\end{equation}
Thus at low energies the phases tend to align, while $\Delta$ sets the interactions between the $z$ components. The isotropic model corresponds to $\Delta =1$, easy-plane to $\Delta <1$, and easy-axis to $\Delta>1$. In the new variables the equations of motion become
\begin{align}\label{2.5}
\begin{split}
{}&\tfrac{d}{dt}\phi_j = -f'(s_{j})f(s_{j+1})\cos(\phi_{j+1} - \phi_{j}) - f(s_{j-1})f'(s_{j})\cos(\phi_{j} - \phi_{j-1}) \\
&\qquad\qquad\quad-\Delta(s_{j-1} + s_{j+1}),
\end{split}\nonumber\\
{}&\tfrac{d}{dt}s_j = f(s_{j})f(s_{j+1})\sin(\phi_{j+1} - \phi_{j}) - f(s_{j-1})f(s_{j})\sin(\phi_{j} - \phi_{j-1}).
\end{align}
For $|s_j(t)| <1$ the angle $\phi_{j}(t)$ is defined modulo $2\pi$.
For $s_j(t) = \pm 1$ the angle $\phi_{j}(t)$ is ill-defined.
However, in our context, trajectories where a spin precisely hits either the north or south pole of the unit sphere ($|s_j(t)|=1$) have measure zero and therefore can be ignored.

The $s$-$s$ interaction could have additional contributions. One example is the ionic potential $\sum_j s_j^2$ \cite{Zagorodny2004}. Many of our  results are valid in greater generality, but we explore the simplest case \eqref{2.5}.

The LLL dynamics has two locally conserved fields, namely the $z$-component of the spin, $s_j$, and the energy
\begin{equation}\label{energy}
e_j = -f(s_j)f(s_{j+1}) \cos(r_j) - \Delta s_js_{j+1},
\end{equation}
where for convenience we have introduced the phase difference $r_j = \phi_{j+1} - \phi_{j}$. From the equations of motion \eqref{2.5} one 
deduces the form of the spin current, $\mathcal{J}_{j}^\mathsf{s}$, and energy current, $\mathcal{J}_{j}^\mathsf{e}$, as
\begin{align}\label{es_currents}
\mathcal{J}_{j}^\mathsf{s} &=  - f(s_{j-1})f(s_j) \sin(r_{j-1}),\\
\mathcal{J}_{j}^\mathsf{e} &= f(s_{j-1})f(s_j)f'(s_j)f(s_{j+1}) \sin(r_{j-1} +r_{j}) \nonumber\\
&\qquad\quad +\Delta f(s_j)f(s_{j+1})s_{j-1}\sin(r_j) + \Delta f(s_j)f(s_{j-1})s_{j+1}\sin(r_{j-1}).
\end{align}
{\bf High-temperature diffusive regime} --- 
We first assume that there are no further conservation laws. The equilibrium Gibbs measures are then given by the two-parameter family
\begin{equation}\label{2.8}
Z_N(\beta,h)^{-1} \exp\Big[ - \beta \Big( H - h \sum_j s_j\Big)\Big] \prod_{j} dr_j ds_j, \quad \beta \geq 0, \,\,h\in \mathbb{R} ,
\end{equation}
for a chain of length $N$ spins with periodic boundary conditions and partition function $Z_N(\beta, h)$.
Infinite volume equilibrium averages will be denoted by $\langle \cdot \rangle_{\beta,h}$. 
We note that the Hamiltonian is even and the currents are odd in $r_j$. Hence  
\begin{equation}\label{2.9}
\langle \mathcal{J}_{j}^\mathsf{s} \rangle_{\beta,h} =0,\quad  \langle \mathcal{J}_{j}^\mathsf{e} \rangle_{\beta,h}=0.
\end{equation}
But then also their derivatives with respect to $\beta,h$ vanish and, using the definition of the Drude weight given in \cite{MendlSpohnCurrent}, one concludes that
both Drude weights are zero. Thus the conventional expectation is to have  a diffusive spreading of the equilibrium time-correlations for spin and energy.
Since they have opposite signature under time reversal, the cross diffusion coefficient should vanish.

\begin{figure}[b]
  \includegraphics[width=0.49\textwidth]{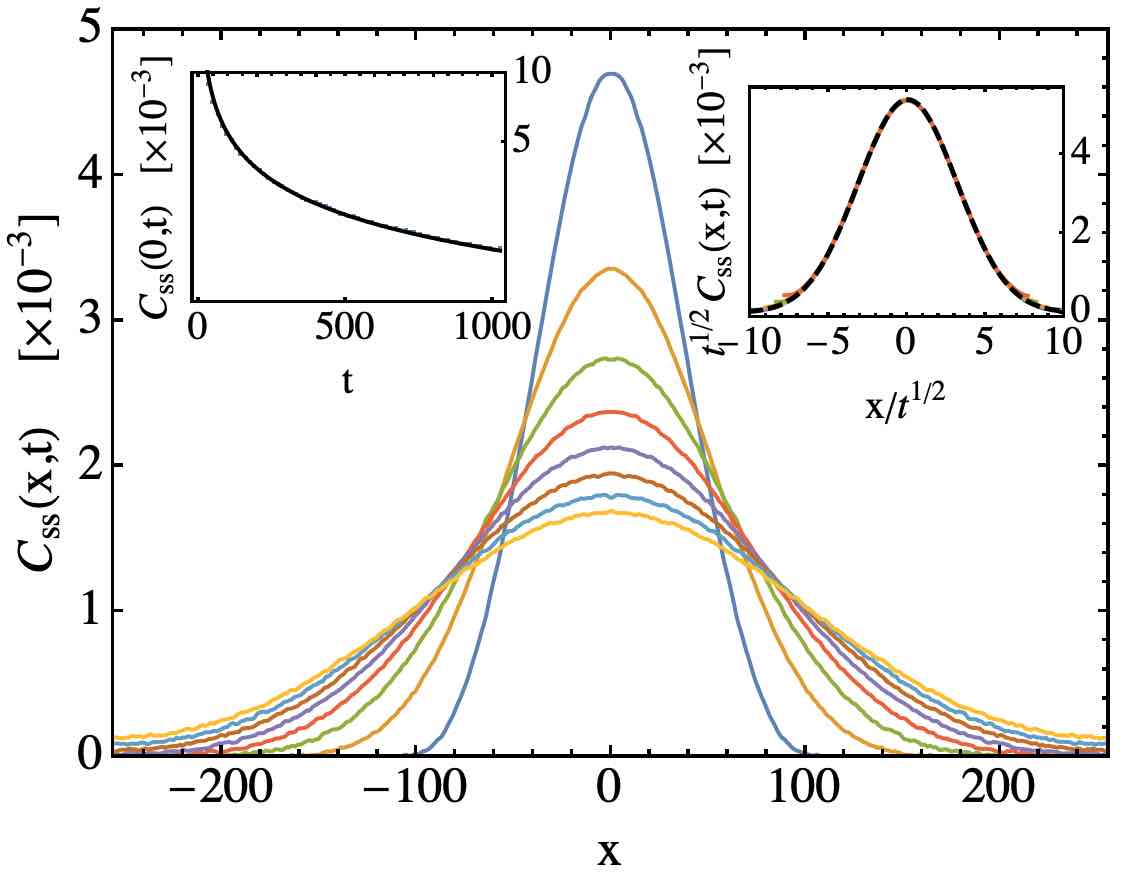}
  \includegraphics[width=0.49\textwidth]{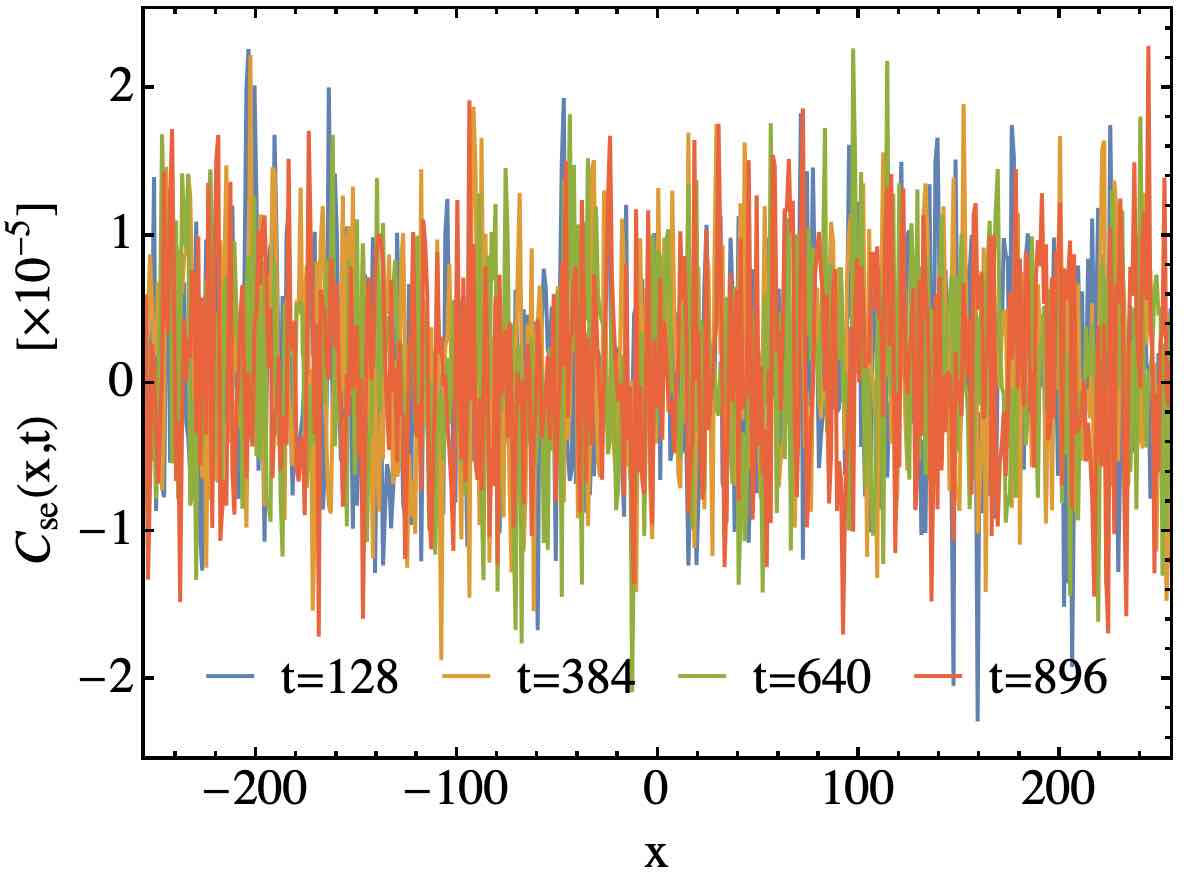}
  
  \includegraphics[width=0.49\textwidth]{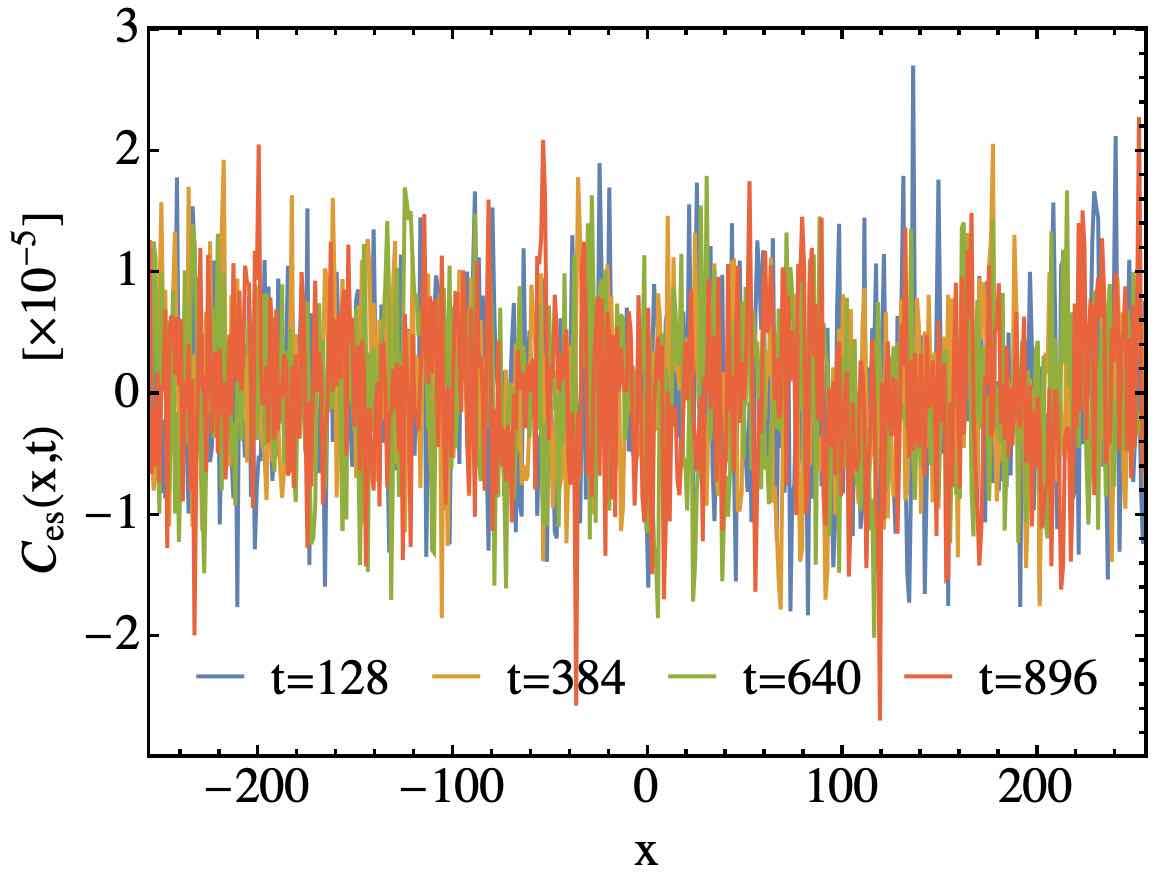}
  \includegraphics[width=0.49\textwidth]{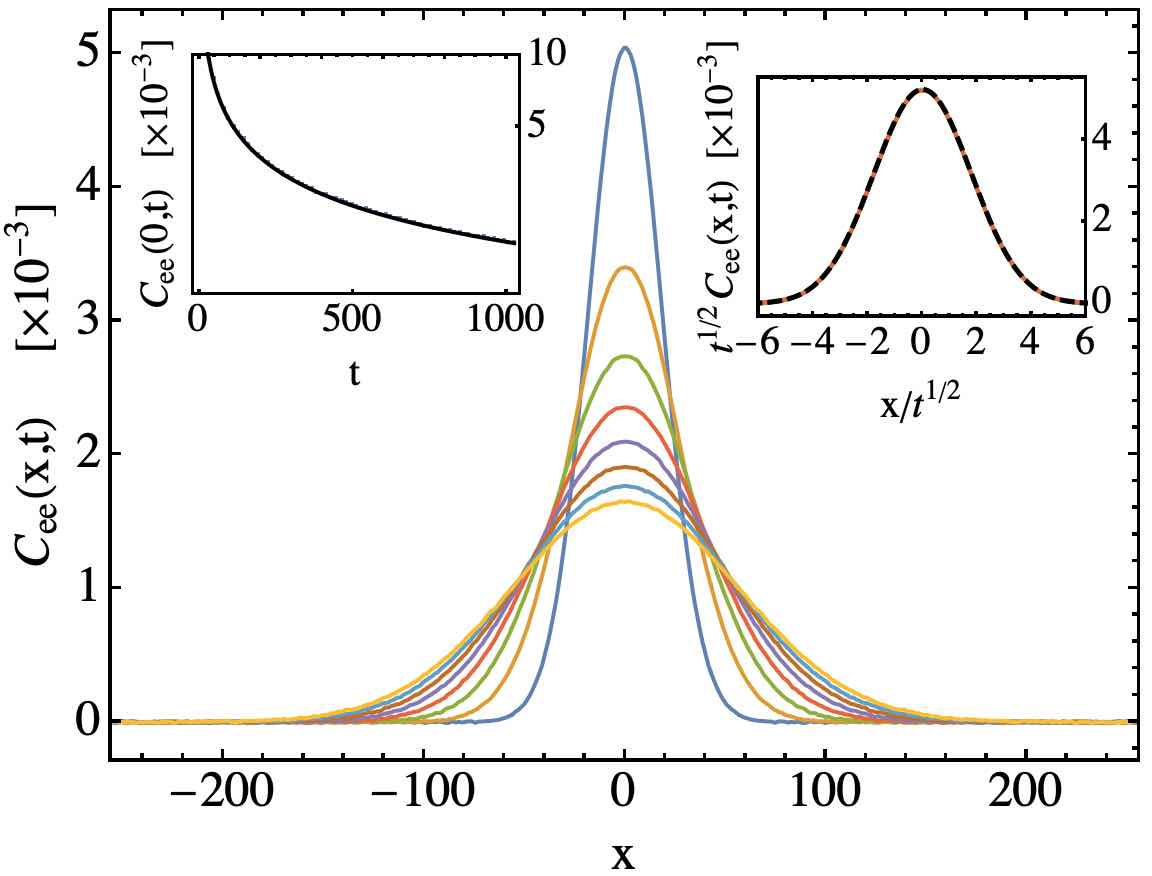}
\caption{Plots of $C_{ab}(x,t)$ for XXZ easy plane regime at high temperature. Parameters are $\Delta=0.5, \beta=1$, $N=512$ and $h=0$. $C_{ss}(x,t)$ and $C_{ee}(x,t)$ are plotted at time $128, 256,...,1024$. Insets show the slow temporal decay of $C_{ss}(0,t)$ and $C_{ee}(0,t)$, and the diffusive scaling of $C_{ss}(x,t)$ and $C_{ee}(x,t)$ with the Gaussian fits at times $640, 768, 896$ and $1024$. The diffusion constants are 4.95 and 1.72 for spin and energy respectively, obtained from the corresponding Gaussian fits and the decay of $C_{ss}(0,t),C_{ee}(0,t)$. As expected, energy and spin are uncorrelated.}
\label{highTh0}
\end{figure} 

To confirm, we performed molecular dynamics simulations at  inverse temperature $\beta=1$. In Fig.~(\ref{highTh0}) we show  numerical results for the spin and energy correlations defined by
\begin{eqnarray}\label{corrse}
C_\mathrm{ss}(j,t) &&=\langle s_j(t) s_{0}(0) \rangle_{\beta,h}^\mathrm{c},\nonumber\\
C_\mathrm{ee}(j,t) &&= \langle e_j(t) e_0(t) \rangle_{\beta,h}^\mathrm{c}
\end{eqnarray} 
where  $\langle \dots \rangle_{\beta,h}^\mathrm{c}$ denotes the connected correlation defined as $\langle Q_j(t) Q_0(0) \rangle^{\mathrm{c}}_{\rm eq} := \langle ( Q_j(t) - \langle Q_0 \rangle_{\rm{eq}})( Q_0(0) - \langle Q_0 \rangle_{\rm{eq}}) \rangle_{\rm{eq}}$.

These simulations are for system size $N=512$ and were run up to  time $t=1024$ [simulation details are given later in Sec.~\ref{sec4}]. We see that spin and energy autocorrelations indeed show diffusive behavior, while there are no cross correlations.

\begin{figure}[b]
  \includegraphics[width=0.49\textwidth]{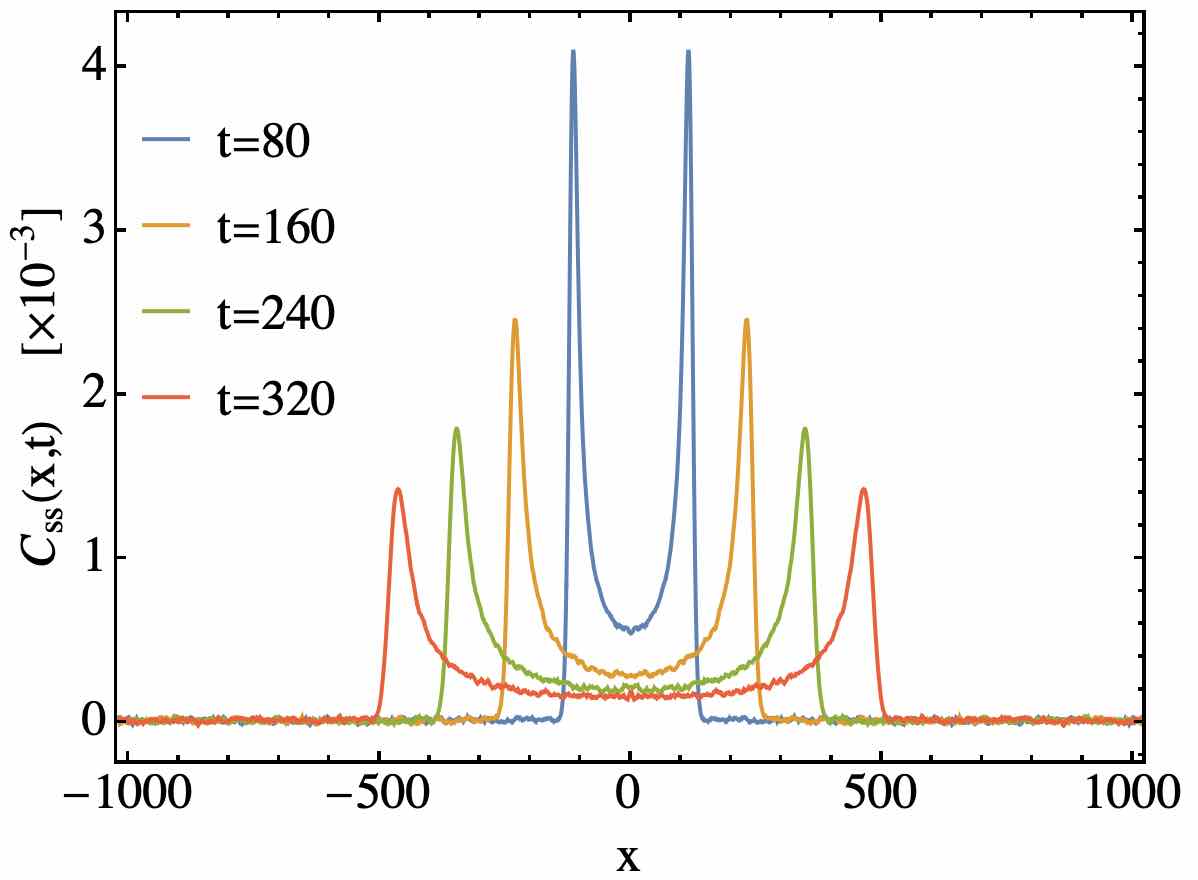}
  \includegraphics[width=0.49\textwidth]{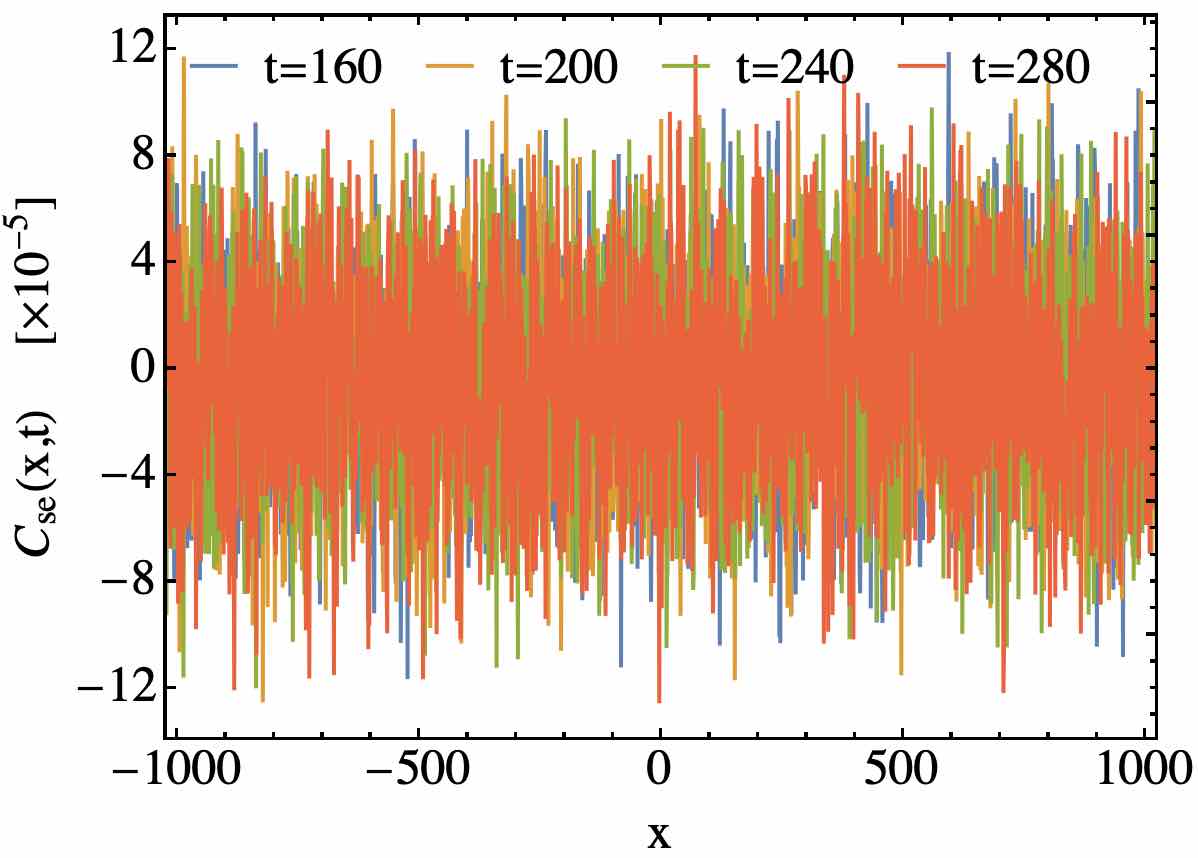}
  
  \includegraphics[width=0.49\textwidth]{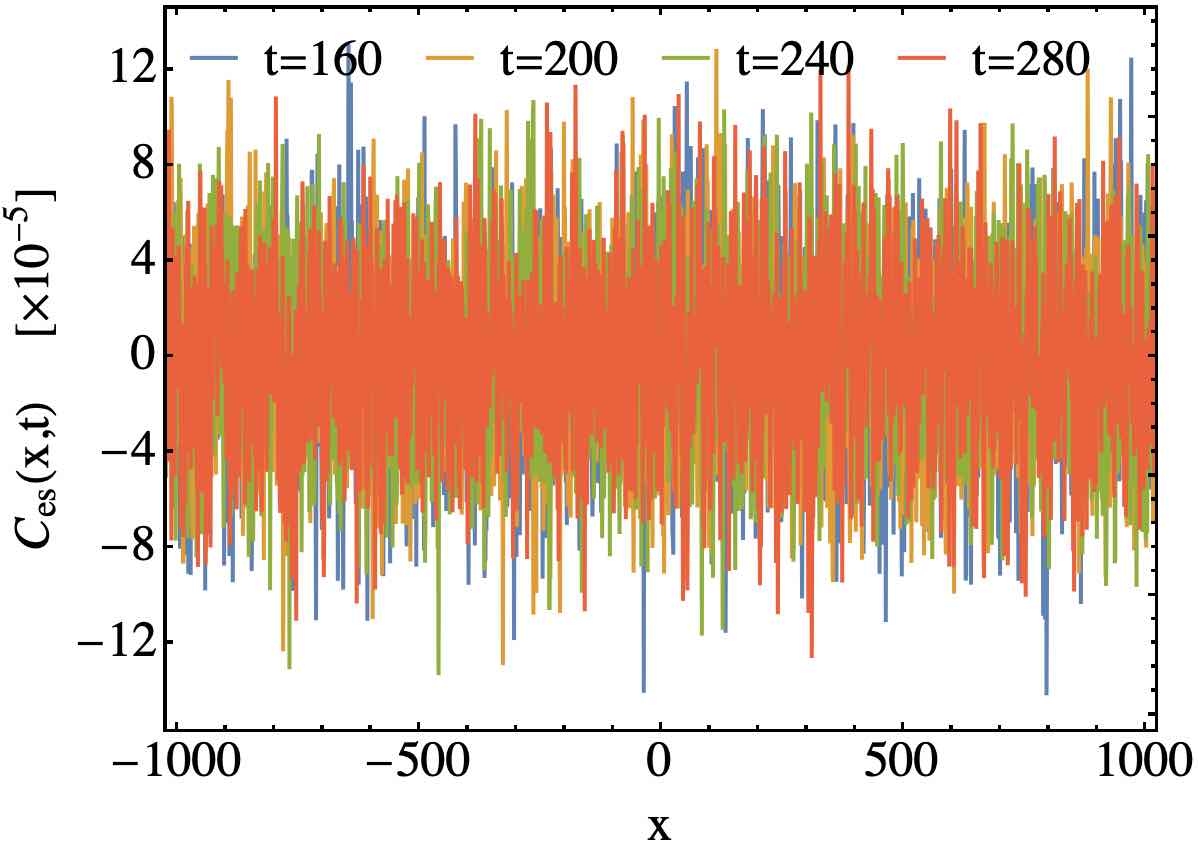}
  \includegraphics[width=0.49\textwidth]{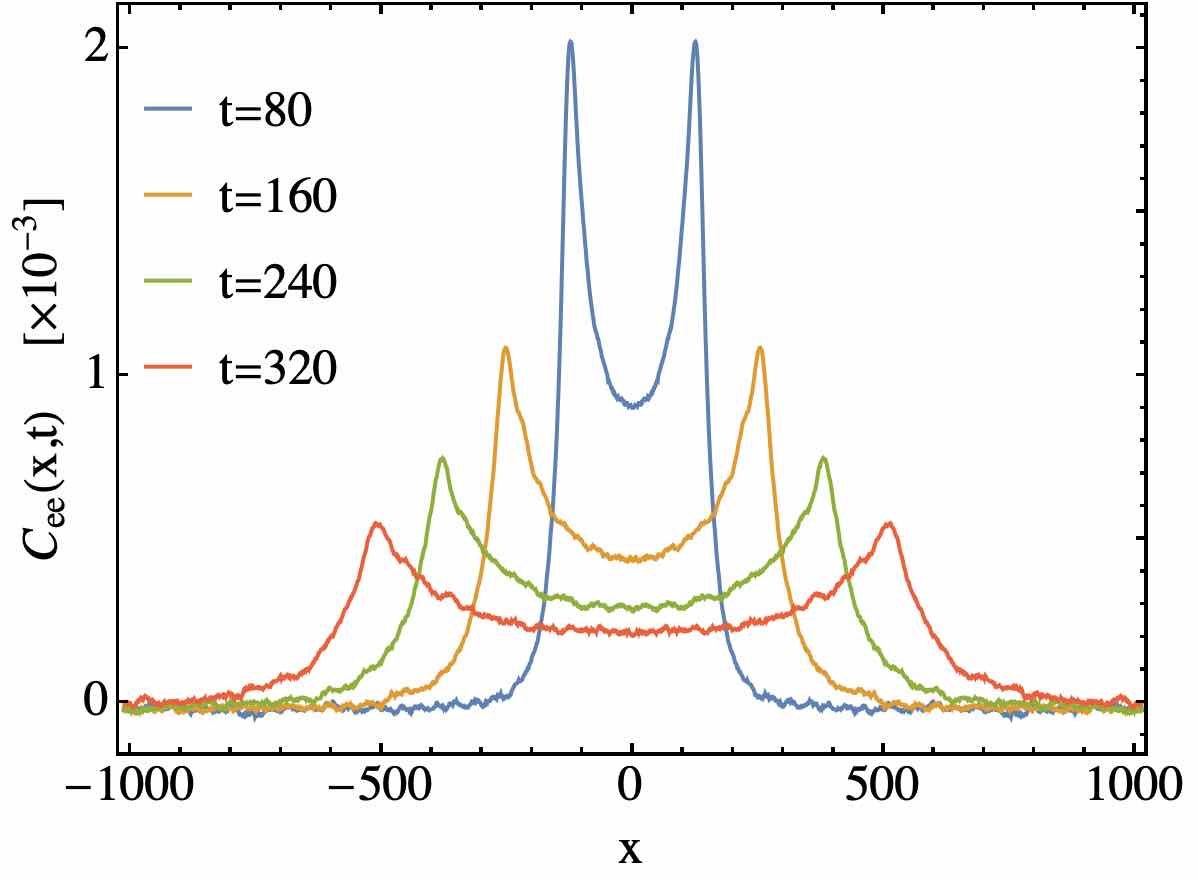}
\caption{Plots of $C_{ab}(x,t)$ for the integrable LLL model in easy plane regime at high temperature at different times. Parameters are $\rho=1.0, \beta=1.0$ and $N=2048$. The energy and spin are uncorrelated. Energy correlation reaches the boundary of the system faster than the spin correlation. From the maxima of the peaks, the estimated speed is 1.4448 for the spin mode, whereas it is 1.5888 for the energy mode. Scaling plots are shown in Fig.~(\ref{FT_scaling}).}
\label{FT}
\end{figure} 

The novelty of our contribution is to establish that at lower temperatures the dynamical properties change dramatically through the appearance of ballistic sound propagation. But before embarking on that discussion we use the opportunity to illustrate that equilibrium time-correlations for an integrable
spin-chain are dominated by a broad ballistic spreading, in contrast to the sub-ballistic broadening of the two sound peaks in the case of a nonintegrable chain. The form of dynamical correlations in classical integrable models have been discussed in several earlier work including \cite{zhao06,PhysRevE.94.062130} for the Toda chain and in \cite{prosen2013macroscopic} for a particular integrable spin chain, which we will now discuss.
\medskip\\ 
{\bf Integrable LLL model} --- 
Faddeev and Takhtajan \cite{faddeev2007hamiltonian} discovered an integrable version of the LLL model, which still has nearest neighbor coupling but is no longer quadratic. Their Hamiltonian is given by 
\begin{equation}\label{2.10}
H = - \sum_{j \in \mathbb{Z}} h(\vec{S}_j,\vec{S}_{j+1})
 \end{equation}
 with local energy 
\begin{eqnarray}\label{2.11}
&&h(\vec{S},\vec{S'}) = \log\big|\cos(\rho S^z)\cos(\rho S'^z) + (\cot(\rho))^2\sin(\rho S^z)\sin(\rho S'^z)\nonumber\\
&&\hspace{80pt}+ (\sin(\rho))^{-2} G(S^z) G(S'^z) (S^xS'^x + S^yS'^y)\big|, \nonumber\\
&&G(x) = \big(1-x^2\big)^{-\tfrac{1}{2}}\big(\cos(2\rho x)-\cos(2\rho)\big)^{\frac{1}{2}}
\end{eqnarray}
with $\rho \geq 0$. The hamiltonian \eqref{2.10} seems to be the only known integrable classical spin chain. Easy plane corresponds to $\rho>0$, while in the limit $\rho \to 0$
one recovers the isotropic interaction 
 \begin{equation}\label{2.12}
h(\vec{S},\vec{S'}) = \log\big(1+ \vec{S}\cdot\vec{S'}\big).
 \end{equation}
The  infinitely extended Faddeev-Takhtajan spin chain has a countable number of locally conserved fields, which are constructed by successive differentiations of the 
$R$-matrix, see \cite{faddeev2007hamiltonian}.

\begin{figure}[t]
\begin{minipage}[c]{\textwidth}
 \includegraphics[width=0.49\textwidth]{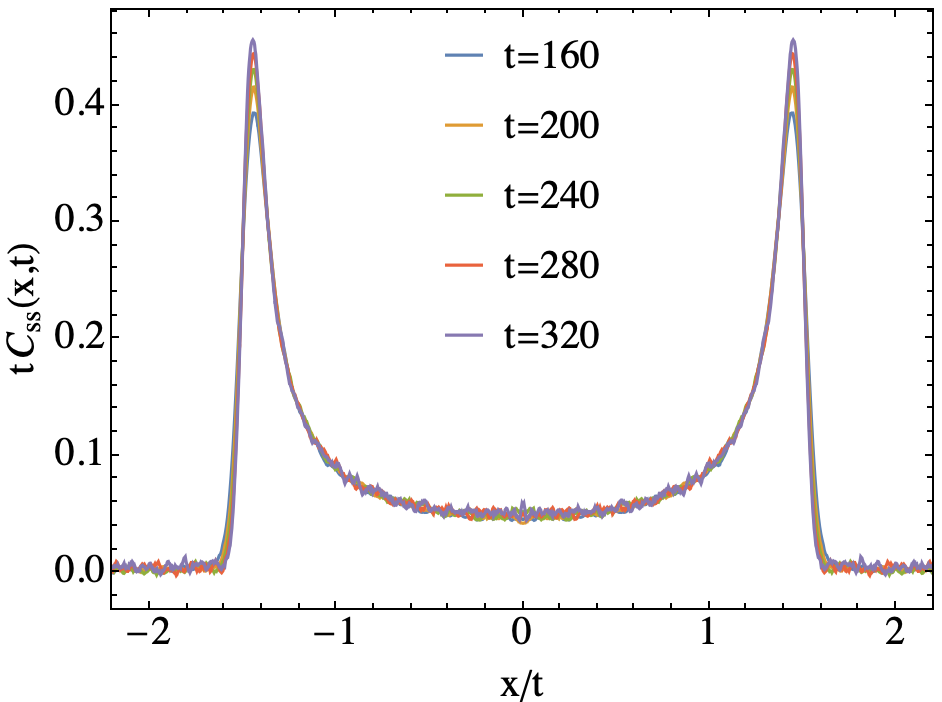} 
  \includegraphics[width=0.49\textwidth]{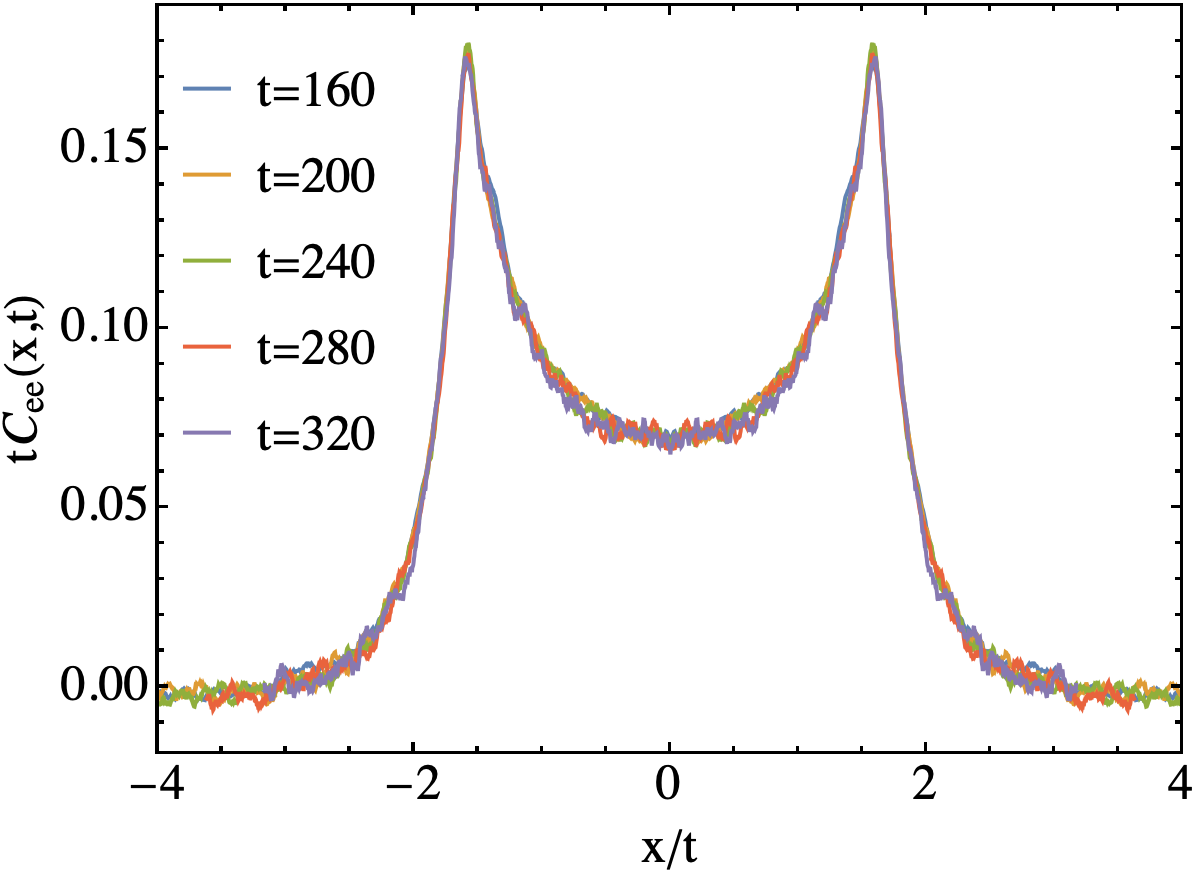}
\end{minipage}   

\caption{Scaling plots of $C_{ss}(x,t)$ and $C_{ee}(x,t)$ for the integrable LLL model in easy plane regime at high temperature. Parameters are $\rho=1.0, \beta=1.0$ and $N=2048$. The figures show the ballistic scaling of $C_{ss}(x,t)$ and $C_{ee}(x,t)$ at different times.}
\label{FT_scaling}
\end{figure} 
Here we focus on the $z$-component of the magnetization, $s_j$, and the local energy, $e_j$, as the first items in the list of conservation laws. We display
their time-correlations at $\rho =1$ with inverse temperature $\beta=1$ and magnetic field $h=0$, see  Fig.~(\ref{FT}). In the scaling plot  Fig.~(\ref{FT_scaling}), we see that the energy and spin correlations show good ballistic scaling already at short times.  In \cite{prosen2013macroscopic} simulations of the spin current correlations are reported at  parameters $\rho = 1,\beta = 0.25, h =0$. 

Without losing integrability, the Hamiltonian \eqref{2.11} can be analytically continued to purely imaginary $\rho$, which amounts to replace the trigonometric functions by their hyperbolic cousin \cite{faddeev2007hamiltonian}. Physically this corresponds to easy-axis regime. We refer to the discussions in  \cite{prosen2013macroscopic,AvijitFT}, also reporting on parameters with diffusive spreading.

\begin{figure}[h]
  \includegraphics[width=0.49\textwidth]{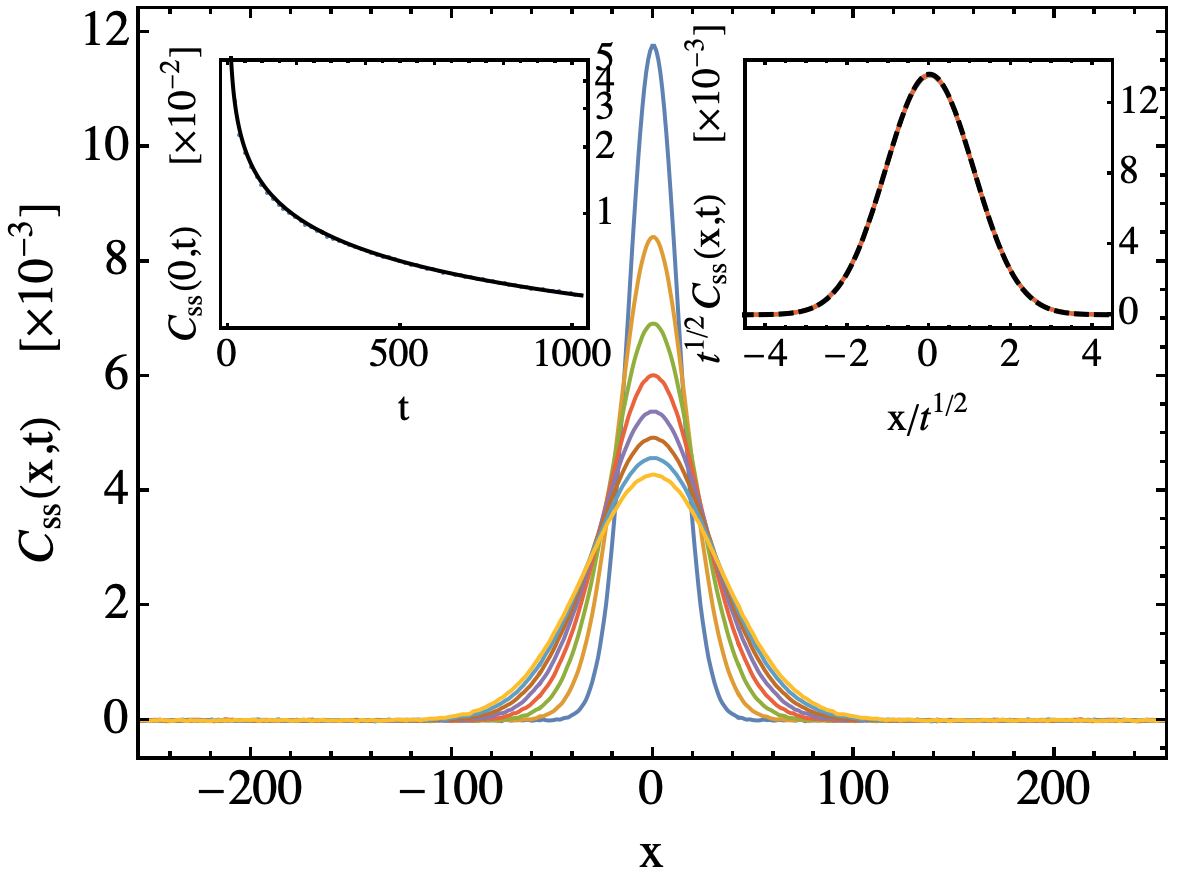}
  \includegraphics[width=0.49\textwidth]{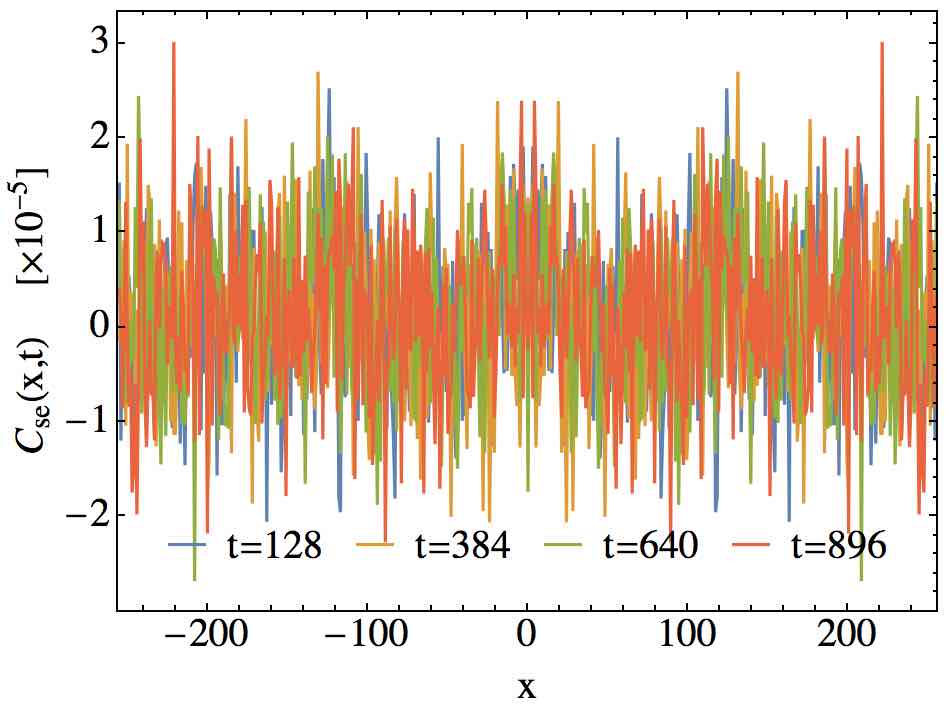}
  
  \includegraphics[width=0.49\textwidth]{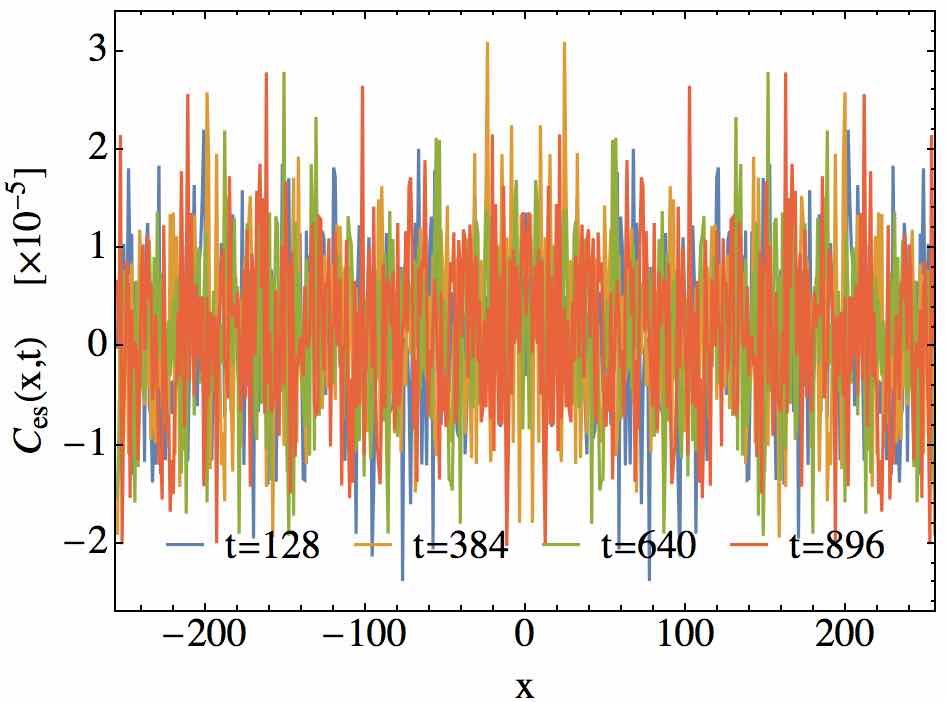}
  \includegraphics[width=0.49\textwidth]{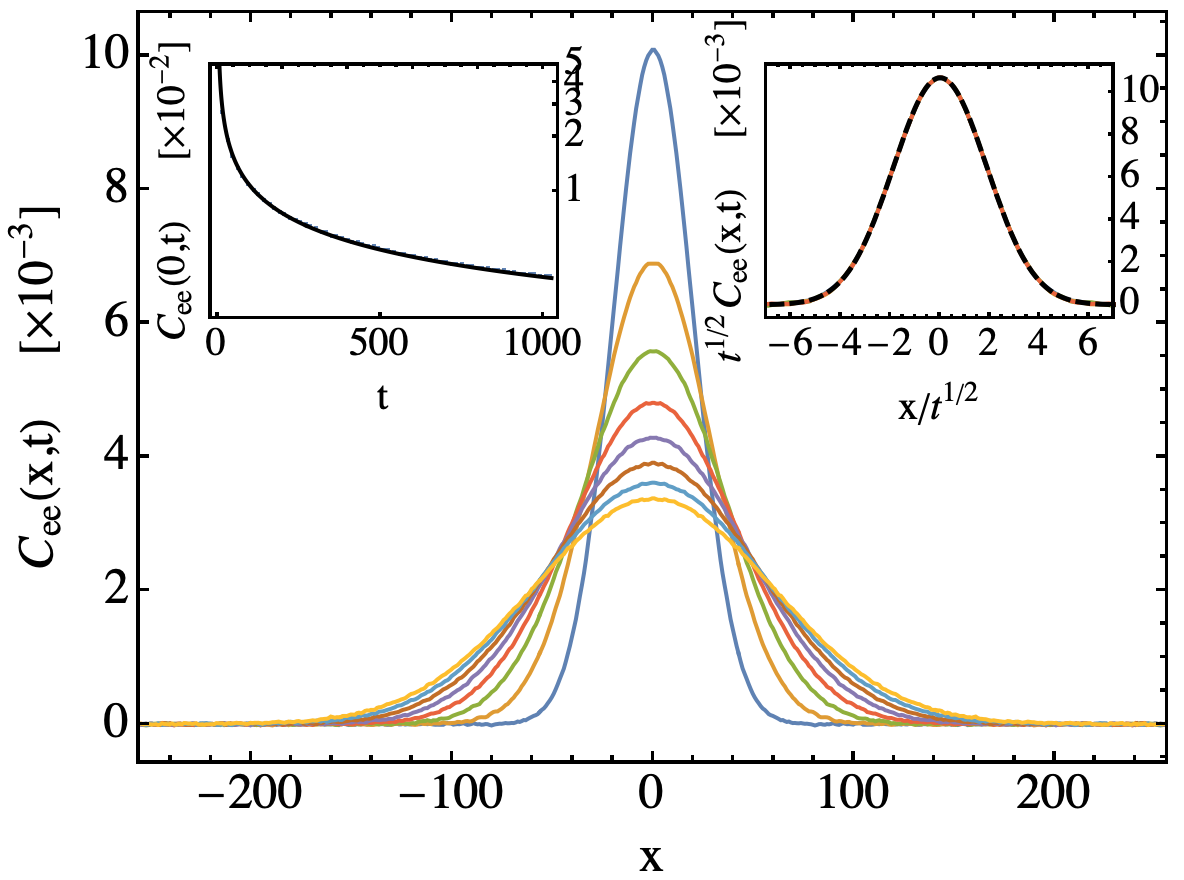}
\caption{Easy axis data $\Delta = 1.5$, $\beta = 0.1$, $N = 512$. We observe perfect diffusive behaviour (see inset). }
\label{axis_highT}
\end{figure} 

\section{Easy plane at low-temperatures}\label{sec3}
\noindent
\textbf{Low-temperature effective hamiltonian} --- We now return to the nonintegrable XXZ spin chain.  As the temperature is lowered, if we impose easy-axis anisotropy, $\Delta > 1$, the spreading of spin and energy correlations is still diffusive, as confirmed by further molecular dynamics simulations.
In Fig.~(\ref{axis_highT}) we show the easy axis data. 
Apparently diffusion still holds at $\Delta  =1$, although with slow convergence \cite{AvijitChaos}.  However, for easy-plane, $\Delta < 1$,
phase differences become locally almost conserved, which drastically changes the dynamical behavior, as will now be explained in detail.   
The equations of motion are given by \eqref{2.5}. 

\begin{figure}[b]
  \includegraphics[width=0.49\textwidth]{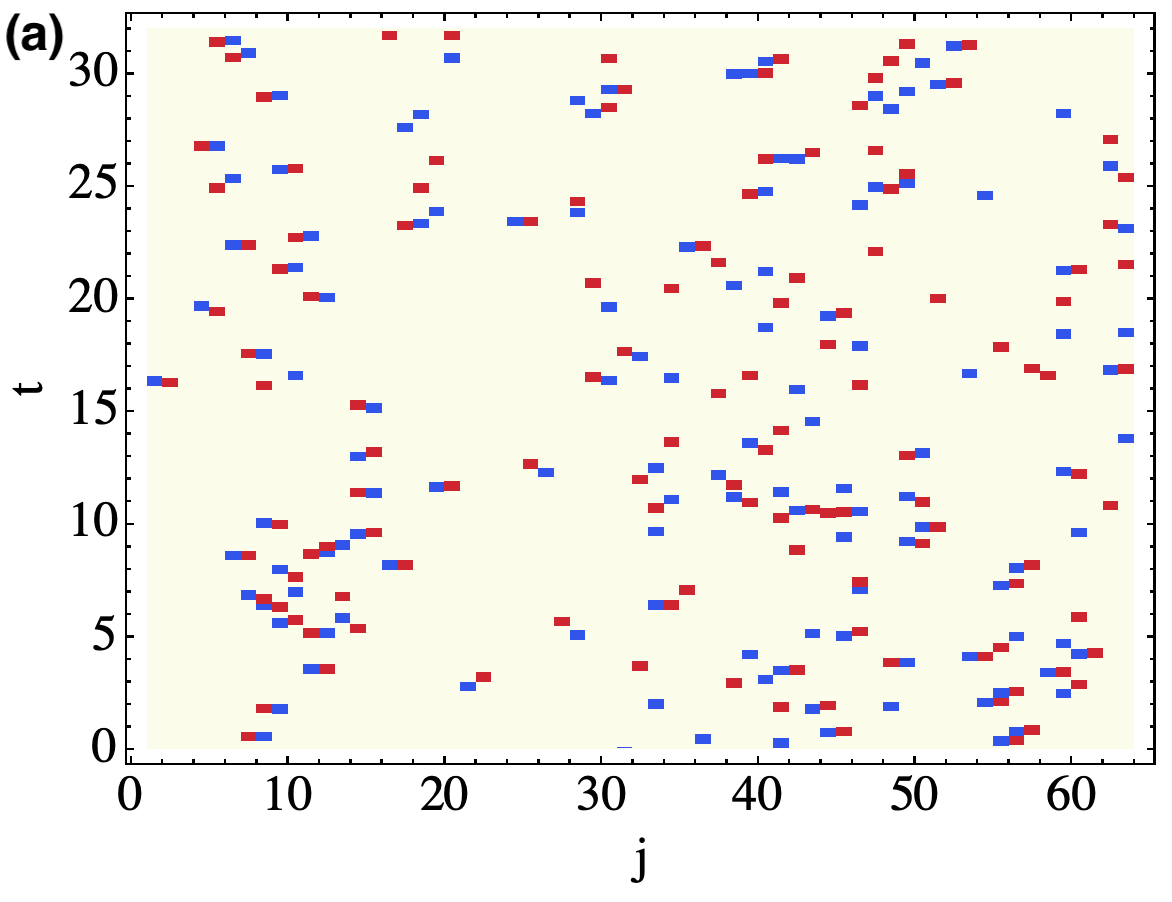}
  \includegraphics[width=0.49\textwidth]{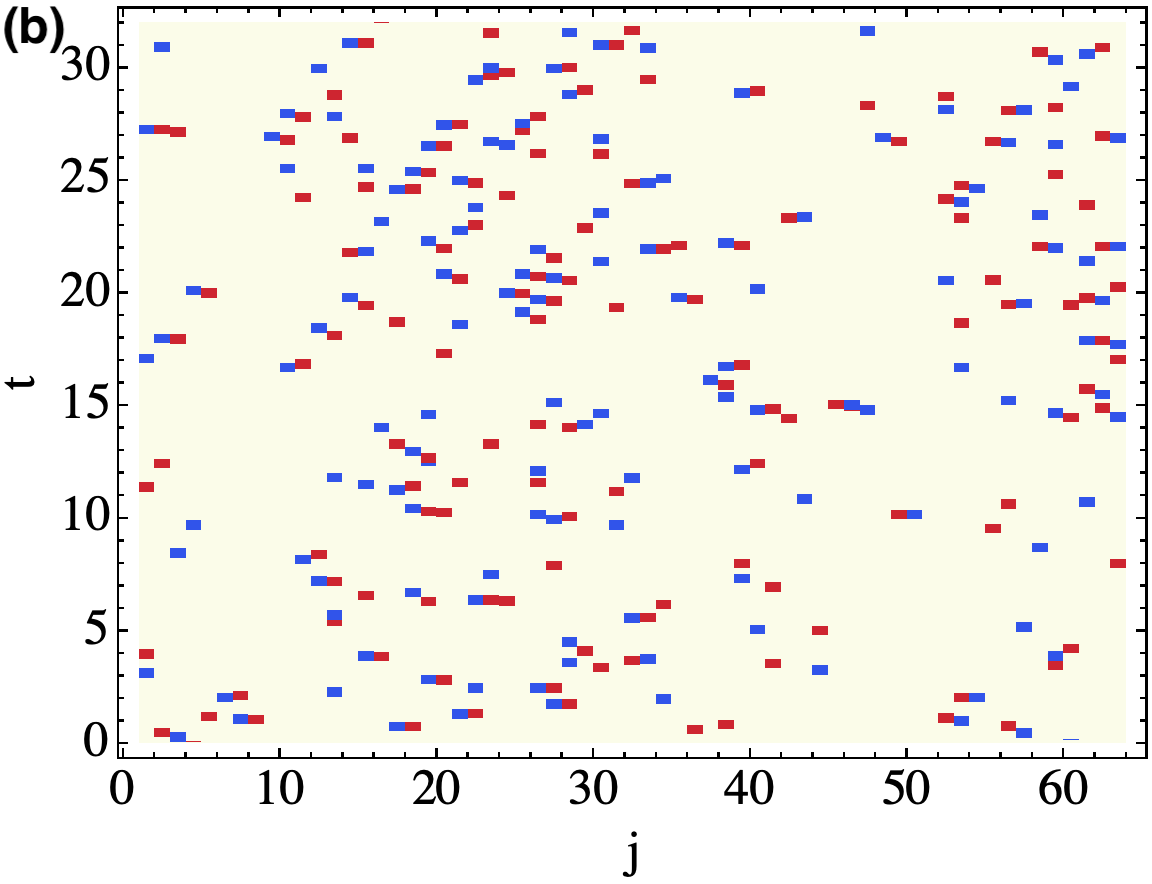}
  \includegraphics[width=0.49\textwidth]{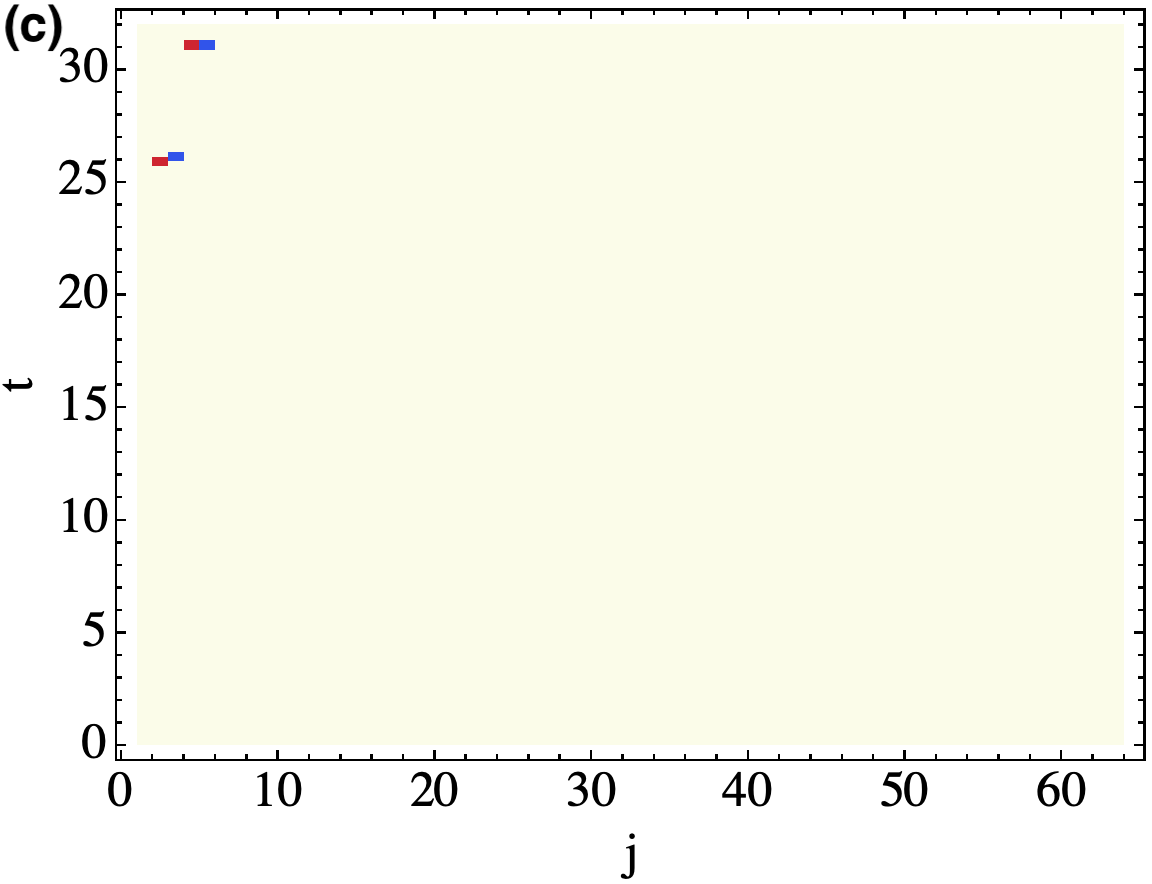}
    \includegraphics[width=0.49\textwidth]{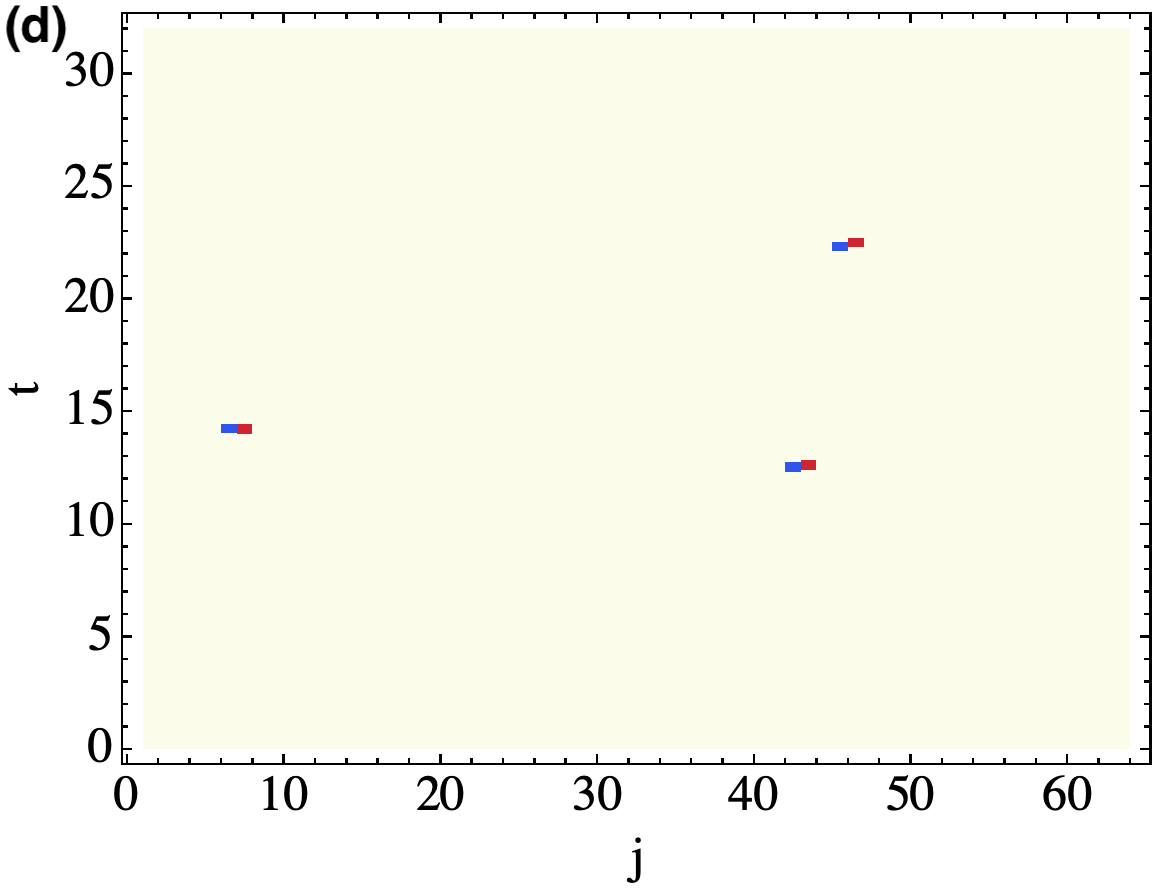}
  
	\caption{(a) Phase slip events at $\beta = 1, h = 0$, (b) $\beta = 1, h = 0.1$, (c) $\beta = 5, h = 0$ and (d) $\beta = 5, h = 0.1$ in easy plane regime with $\Delta=0.5$. It shows that the phase slip events become rare with decreasing temperature and decreasing external magnetic field. Red (blue) boxes indicate sites  where phase slips occur, i.e, $\delta n_j$ is +1 (-1) (Eq. \eqref{3.3}).}
\label{phase_slips}
\end{figure}

Phase differences are defined through
\begin{equation}\label{3.1}
r_j = \phi_{j+1}-\phi_j ~, \quad\mathrm{mod}\,~2\pi,
\end{equation} 
where we choose coordinates such that  $r_j \in [-\pi,\pi]$. $r_j = 0$ corresponds to $\phi_j = \phi_{j+1}$ which is the minimum of the cosine-potential.
The dynamics of the $r_j(t)$'s has the following generic structure. One starts from their extended version with $\tilde{r}_j \in \mathbb{R}$ governed by 
\begin{equation}\label{3.2}
\tfrac{d}{dt} \tilde{r}_j(t) = g_j(t) - g_{j+1}(t)
\end{equation} 
for a given collection of smooth functions $\{g_j(t), j \in \mathbb{Z}\}$. The restriction to the unit circle $S^1$ is achieved by setting
\begin{equation}\label{3.3}
\tilde{r}_j(t) = r_j(t) + 2\pi n_j(t)
\end{equation} 
with $n_j(t)$ the integer winding number for bond $j$; we may choose $n_j(t=0) = 0$, while $r_j(t)$ is a smooth function on $S^1$. 
By construction the $\tilde{r}_j$'s 
have the form of a local conservation 
law, which in integrated version reads
\begin{equation}\label{3.4}
\sum_{j} \big(\tilde{r}_j(t) - \tilde{r}_j(0)\big) = \sum_j \int_0^t ds\big(g_j(s) - g_{j+1}(s)\big) = 0 ,
\end{equation}
for a system with periodic boundary conditions.  On the other hand, 
\begin{equation}\label{3.5}
 \sum_{j}\big(r_j(t) - r_j(0)\big) =  
- 2\pi\sum_{j} n_j (t) ~. 
\end{equation}
Hence $r_j(t)$ is locally conserved only if the total winding number remains constant: $\sum_j n_j(t) = 0$.  The dynamical events $r_j(t) = \pm\pi$ are the ``phase slip'' processes where the winding number changes.
Thus the field of phase differences $r_j$ is locally conserved only until the first phase slip event. However, as we show in Fig.~(\ref{phase_slips}), at low temperatures most phase slip events come in closely-spaced pairs with no change in the total winding number, so the coarse-grained dynamics actually respects this conservation law until one has unpaired phase slip events.

As illustrated in Fig.~(\ref{phase_slips}a) and (\ref{phase_slips}b), for high temperatures there are lots of phase slips. However for easy-plane and low enough temperatures the $z$-component is approximately constant 
and the phase difference is trapped by the cosine-potential. Then phase slips are very much suppressed, see Fig.~(\ref{phase_slips}c) and (\ref{phase_slips}d). Therefore, there is an emergence of an approximately conserved quantity. 
In other words, phase slips are  thermally activated process which, in a low-temperature regime, 
can safely be ignored on the time scales reached by our simulation and, in fact, much longer. Phase differences are thus \textit{approximately conserved} in this regime.

Let us first attempt a rough estimate for the presence of a third conservation law. We assume constant $r_j = r$ and $s_j = s$. Adding also an external field, $h$, the energy
of this configuration equals
\begin{equation}\label{3.6}
e_\mathrm{g}(r,s) = -(1 - s^2)\cos r -\Delta s^2   - hs. 
\end{equation}
Its minimum is located at $r=0,s = h/[2(1-\Delta)]$
and we require $|h| < 2(1-\Delta)$ to ensure that the minimum lies inside $\{|s| < 1\}$. Let us compute the energy required for a phase slip event caused by motion of a single spin. For the case where $s$ remains fixed but $r$ changes to $\pi$, the  energy barrier $E_{\rm slip}$ can be easily computed and leads to
\begin{equation}\label{3.7}
E_{\rm slip}= 4 \left(1- \tfrac{1}{4}(1 -\Delta)^{-2}h^2\right)  >1 
\end{equation}
On the other hand for the case where the angle $r$ remains fixed and the spin moves to the north pole (given by $s=1$),  the energy barrier  is 
\begin{equation}\label{3.7}
E_{\rm slip} = 2\left( 1-\tfrac{h}{2} - \tfrac{1}{2} (1-\Delta)^{-1} {\Delta h} \right)~.
\end{equation}

\begin{figure}[t]
\begin{minipage}[c]{0.51\textwidth}
  \includegraphics[width=\textwidth]{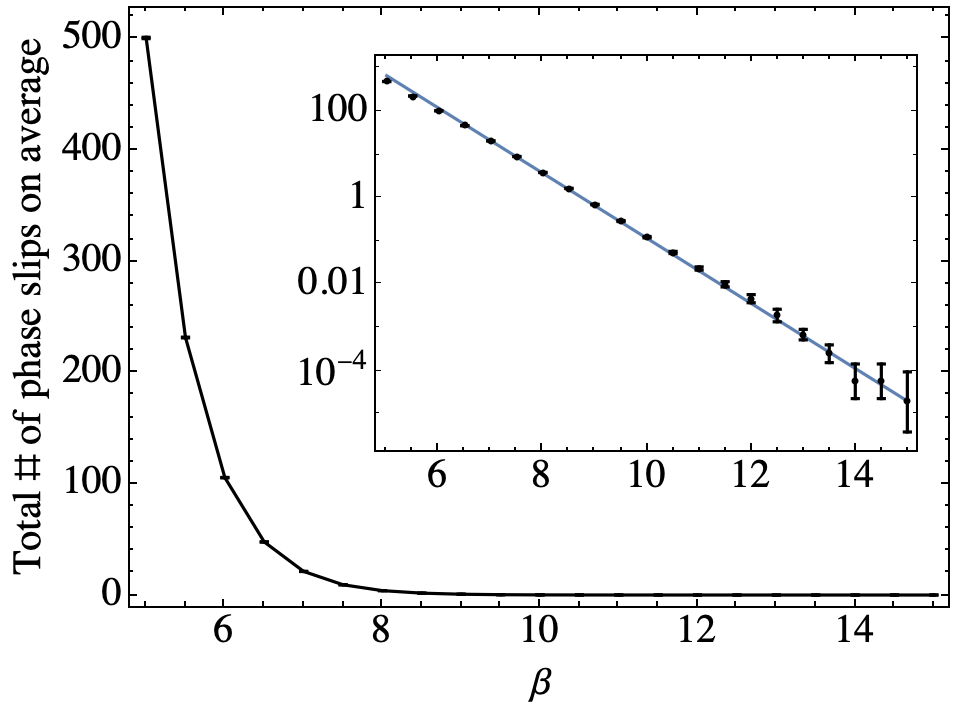}
  \end{minipage}   
\begin{minipage}[c]{0.47\textwidth}
\caption{The figure illustrates that the average number of phase slip events decays exponentially with $\beta$. The parameters are -- $h = 0, \Delta = 0.5, N=1024, t=512$, $10^5$ initial states. Inset shows the same plot in log scale. In log scale the slope is -1.73 which is consistent with our theoretical estimation within  error bars.}
\label{phase_slips_vsT}
\end{minipage}
\end{figure} 

We  expect phase slips to occur at a rate $\sim e^{-\beta \Delta E_{\rm slip}}$ where  $\beta$ is the inverse temperature, and so $\beta E_{\rm slip} >1$ could be a rough criterion for small number of phase-slips and a new approximate conservation law. In Fig.~(\ref{phase_slips}) we show space-time plots showing phase-slip events seen in simulations at (a) high temperature $(\beta = 1)$, zero magnetic field, (b) high temperature $(\beta = 1)$, finite magnetic field $(h=0.3)$, (c) low temperature $(\beta = 5)$, zero magnetic field and (d) low temperature $(\beta = 5)$, finite magnetic field $(h=0.3)$ with parameters $\Delta=1/2, N= 64$.  In Fig.~(\ref{phase_slips_vsT}), we show the dependence of average total number of observed phase slips on the inverse temperature $\beta$ up to $ t=512$ seen in the low temperature simulations (with parameters $\Delta=1/2$, $h = 0$).  We see that $E_{\rm slip} $ takes the values $4$ and $2$ from the two energy estimates  mentioned    above. From the simulations  we find the expected activated dependence form, however with a barrier  $\Delta E \approx 1.73$.
To understand such behavior we have to investigate in more detail the actual process of phase slips. Let us consider the spin chain with zero external magnetic field and in the easy-plane regime $(h=0, \Delta <1)$.  In the ground state all spins are aligned and lying on the $xy$ plane.  We are interested in finding the minimum-energy spin configuration in which there is a phase slip. What we find at the lowest temperatures where such phase slip events do happen in our simulations is that the phase slip event is centered on a single spin $j$ that moves out of the $xy$ plane and is very close to either $\theta_j=0$ or $\theta_j=\pi$.  The nearby spins $i$ move to near the configuration of their angles $\theta_i$ that minimizes the energy given  the special orientation of the one ``central'' spin $j$, with all spins except the central one oriented near the same $\phi_i$.  When we do this minimization for $\Delta=0.5$ and $h=0$, the resulting minimum energy of the phase slip event is $E_{\rm slip} \cong 1.73$ \cite{ongoing}.  As expected, the measured density of phase slip events shows a thermally activated dependence on temperature as $\sim \exp(-\beta E_{\rm slip})$, as shown in Fig.~(\ref{phase_slips_vsT}). These phase slip events are produced at this low density by the chaotic equilibrium dynamics of the LLL chain.

To work with an almost conserved field is somewhat vague and it is more convenient
to modify the dynamics such that $r_j(t)$ is strictly locally conserved. Of course, this is a valid approximation only in a regime with a very low density of phase slips. 
We rewrite our Hamiltonian in a slightly more general form as 
\begin{equation}\label{3.8}
H_\mathrm{lt} =  \sum_{j \in \mathbb{Z}} \big(f(s_j) f(s_{j+1})U(r_j) - \Delta s_j s_{j+1}\big) . 
\end{equation}
If one would set $U(x) = -\cos(x)$, included to have motion on $S^1$, then $H_\mathrm{lt} = H$.
To suppress phase slips entirely we modify $U(x)$ by adding  infinitely high potential barriers at $x = \pm \pi$.
Then up to the first phase slip event, the dynamics generated by $H$ agrees  with the dynamics generated by $H_\mathrm{lt}$, but they differ later on.
Actually, for what we want to show the precise shape of the potential barriers plays no role, as long as phase slips are forbidden.\medskip\\

\textbf{Nonlinear fluctuating hydrodynamics (NFH) in the low temperature regime} ---
As before, the hamiltonian equations of motions for \eqref{3.8} are
\begin{equation}\label{3.9}
\tfrac{d}{dt}\phi_j = \partial_{ s_j}H_\mathrm{lt},\quad \tfrac{d}{dt}s_j = - \partial_{ \phi_j}H_\mathrm{lt}.
\end{equation}
The conserved fields are phase difference, spin, and energy,
\begin{equation}\label{3.10}
r_j,\quad s_j, \quad e_j = f_j f_{j+1}U_j -\Delta s_js_{j+1}, 
\end{equation}
where for later convenience we introduced the shorthands $f_j = f(s_j)$ and $U_j = U(r_j)$.  From the equations of motion we obtain for the current of the phase difference,
\begin{equation}\label{3.10}
\mathcal{J}_{j}^\mathsf{r} = -f'_j f_{j+1}U_j - f_{j-1}f'_jU_{j-1} + \Delta(s_{j-1} + s_{j+1}),
\end{equation}
for the spin current,
\begin{equation}\label{3.11}
\mathcal{J}_{j}^\mathsf{s} = - f_{j-1}f_j U'_{j-1},
\end{equation}
and for the energy current,
\begin{equation}\label{3.12}
\mathcal{J}_{j}^\mathsf{e} = - f_{j-1}f_j f'_j f_{j+1}\big(U'_{j-1}U_j + U_{j-1}U'_j \big)
+\, \Delta\big( f_{j-1} f_jU'_{j-1}s_{j+1} +f_{j} f_{j+1}U'_{j}s_{j-1} \big).
\end{equation}

The grand canonical ensemble of $H_{\mathrm{lt}}$ for a finite system with $N$ lattice sites and periodic boundary conditions is given by
\begin{equation}\label{3.13}
Z_N(\nu,h,\beta)^{-1}\, \exp \Big[-\beta \Big( H_{\mathrm{lt}} - \nu \sum_{j=1}^{N} r_j - h \sum_{j=1}^{N} s_j \Big)\Big] \prod_{j=1}^{N} \mathrm{d} r_j 
 \mathrm{d} s_j\, 
\end{equation}
with the normalizing partition function
\begin{equation}\label{3.14}
Z_N(\nu,h,\beta) = \int_{{([-1,1]\times[-\pi,\pi]})^N} \exp\Big[-\beta \Big( H_{\mathrm{lt}} - \nu \sum_{j=1}^{N} r_j - h \sum_{j=1}^{N} s_j \Big)\Big] \prod_{j=1}^{N} \mathrm{d} r_j\mathrm{d} s_j,
\end{equation}
where $\nu$ is the ``chemical potential''  for the additional conserved field $r_j$.  Infinite volume averages with respect to \eqref{3.13} are denoted by
$\langle \cdot \rangle_{\nu,h,\beta}$. 
The canonical free energy is defined as
\begin{equation}\label{3.15}
F(\nu,h,\beta) = - \beta^{-1} \lim_{N \to\infty} \tfrac{1}{N} \log Z_N(\nu,h,\beta).
\end{equation}
The infinite volume equilibrium averages of $r_j$, $s_j$, $e_j$ are
\begin{equation}\label{3.16}
\begin{split}
\mathsf{r} &= \langle r_j \rangle_{\nu,h,\beta} = - \partial_{\nu} F(\nu,h,\beta), \qquad \mathsf{s} = \langle s_j \rangle_{\nu,h,\beta} = - \partial_{h} F(\nu,h,\beta), \\[1ex]
\mathsf{e} &=  \langle e_j \rangle_{\nu,h,\beta} = \partial_{\beta} (\beta\,F(\nu,h,\beta)) +  \nu\mathsf{r} + h\mathsf{s},
\end{split}
\end{equation}
independent of $j$ because of translation invariance. By convexity of $F$, these relations define the inverse mapping $(\mathsf{r},\mathsf{s},\mathsf{e}) \mapsto (\nu(\mathsf{r},\mathsf{s},\mathsf{e}), h(\mathsf{r},\mathsf{s},\mathsf{e}), \beta(\mathsf{r},\mathsf{s},\mathsf{e}))$. 

Under the constraints \eqref{3.7}  
the LLL equilibrium time-correlations of $s_j,e_j$ should be well approximated by the same time-correlations as computed from the dynamics governed by
  $ H_{\mathrm{lt}}$.
But $ H_{\mathrm{lt}}$ is just one particular anharmonic chain and, as explained in \cite{MendlSpohnNLS2015}, the time-correlations can be predicted from nonlinear  fluctuating hydrodynamics. 
We do not repeat here  the details, but merely point out that in normal mode representation one arrives at a three-component fluctuating field, $\vec{\phi}(x,t)=
\big(\phi_{-1}(x,t),\phi_{0}(x,t), \phi_{1}(x,t)\big)$. The Euler currents have to be expanded to second order, which in approximation then  leads to the coupled Langevin equations 
\begin{equation}\label{3.17}
\partial_t \vec{\phi}(x,t) + \partial_x \left[\text{diag}(-c,0,c)\vec{\phi} + \langle \vec{\phi}, \vec{G}\vec{\phi} \rangle - D\partial_x\vec{\phi} + B\vec{\xi}\, \right] =0.
\end{equation}
Here $D$ is a constant diffusion matrix and $B\vec{\xi}(x,t)$ is Gaussian white noise, both related through fluctuation-dissipation as $BB^T = 2D$.  This part of the equation is a phenomenological ansatz for the effective noise and dissipation produced by the deterministic chaos. However the sound speed, $c$, and the three symmetric coupling matrices $\vec{G} $ have to be computed from the underlying microscopic model. 
In particular $\vec{G}$ determines the dynamical universality class. Fortunately, for the LLL chain the magic identity (proven below)
\begin{equation}\label{3.18}
\big\langle \vec{\mathcal{J}}_j \big\rangle_{\nu,h,\beta} = \langle ( \mathcal{J}_{j}^\mathsf{r}, \mathcal{J}_{j}^\mathsf{s}, \mathcal{J}_{j}^\mathsf{e} ) 
\rangle_{\nu,h,\beta} = (-h, -\nu, -h\nu)
\end{equation}
is available. Using this property the precise form of $\vec{G}$ and its relation to second derivatives of the free energy have been established
in \cite{MendlSpohnNLS2015}.

We turn to the proof of \eqref{3.18}.\medskip\\
(i) For the phase difference current we obtain
\begin{eqnarray}\label{3.19}
&&\hspace{-20pt}\langle\mathcal{J}_{j}^\mathsf{r}\rangle_{\nu,h,\beta} = \big\langle
-f'_j f_{j+1}U_j - f_{j-1}f'_jU_{j-1} + \Delta(s_{j-1} + s_{j+1})\,
 \big\rangle_{\nu,h,\beta}\nonumber\\
&&\hspace{26pt}= \beta^{-1} Z_{N}^{-1} \int \big(\partial_{s_{j}} \mathrm{e}^{-\beta H_\mathrm{lt}} \big) \big( \mathrm{e}^{\beta\nu \sum_j r_j + \beta h \sum_j s_j} \big) = -h.
\end{eqnarray}

(ii) Correspondingly, for the spin current we obtain
\begin{eqnarray}\label{3.20}
&&\hspace{-20pt}\langle\mathcal{J}_{j}^\mathsf{s}\rangle_{\nu,h,\beta} = -\langle   f_{j-1}f_j U'_{j-1}\rangle_{\nu,h,\beta}\nonumber \\
&&\hspace{26pt}= \beta^{-1} Z_{N}^{-1} \int f_{j-1}f_j\big(\partial_{r_{j-1}} \mathrm{e}^{-\beta H_\mathrm{lt}} \big) \big(\mathrm{e}^{\beta\nu \sum_j r_j + \beta h \sum_j s_j} \big) \nonumber \\
&&\hspace{26pt}= -\nu,
\end{eqnarray}
where we used partial integration in the last step.\medskip\\
(iii) For the energy current there are more terms to be considered,
\begin{eqnarray}\label{3.21}
&&\hspace{-30pt}\langle\mathcal{J}_{j}^\mathsf{e}\rangle_{\nu,h,\beta} =
  \big\langle  - f_{j-1}f_j f'_j f_{j+1}\big(U'_{j-1}U_j + U_{j-1}U'_j \big)\nonumber\\
&&\hspace{66pt}+\, \Delta\big( f_{j-1} f_jU'_{j-1}s_{j+1} +f_{j} f_{j+1}U'_{j}s_{j-1} \big) \big\rangle_{\nu,h,\beta}\nonumber\\[1ex]
&&\hspace{0pt}= \beta^{-1} Z_{N}^{-1} \int \Big[    \big( f'_{j}f_{j+1}U_{j} -\Delta s_{j+1} \big)
\partial_{r_{j-1}} \mathrm{e}^{-\beta H_\mathrm{lt}}  \nonumber\\
&&\hspace{66pt}  +\big(f_{j-1}f'_{j} U_{j-1} -\Delta s_{j-1}\big) \partial_{r_j} \mathrm{e}^{-\beta H_\mathrm{lt}} \Big] \, \mathrm{e}^{\beta\nu \sum_j r_j + \beta h \sum_j s_j} \nonumber\\[1ex]
&&\hspace{0pt}= -\nu \, Z_{N}^{-1} \int \Big[ f_{j-1}f'_jU_{j-1} + f'_{j}f_{j+1}U_j\nonumber\\[1ex]
&&\hspace{86pt}-\Delta(s_{j-1} + s_{j+1}) \Big] \mathrm{e}^{-\beta (H_\mathrm{lt} - \nu \sum_j r_j - h \sum_j s_j)}
\nonumber\\[1ex]
&&\hspace{0pt}= -h\nu.
\end{eqnarray}

\begin{figure}[b]
  \includegraphics[width=0.32\textwidth]{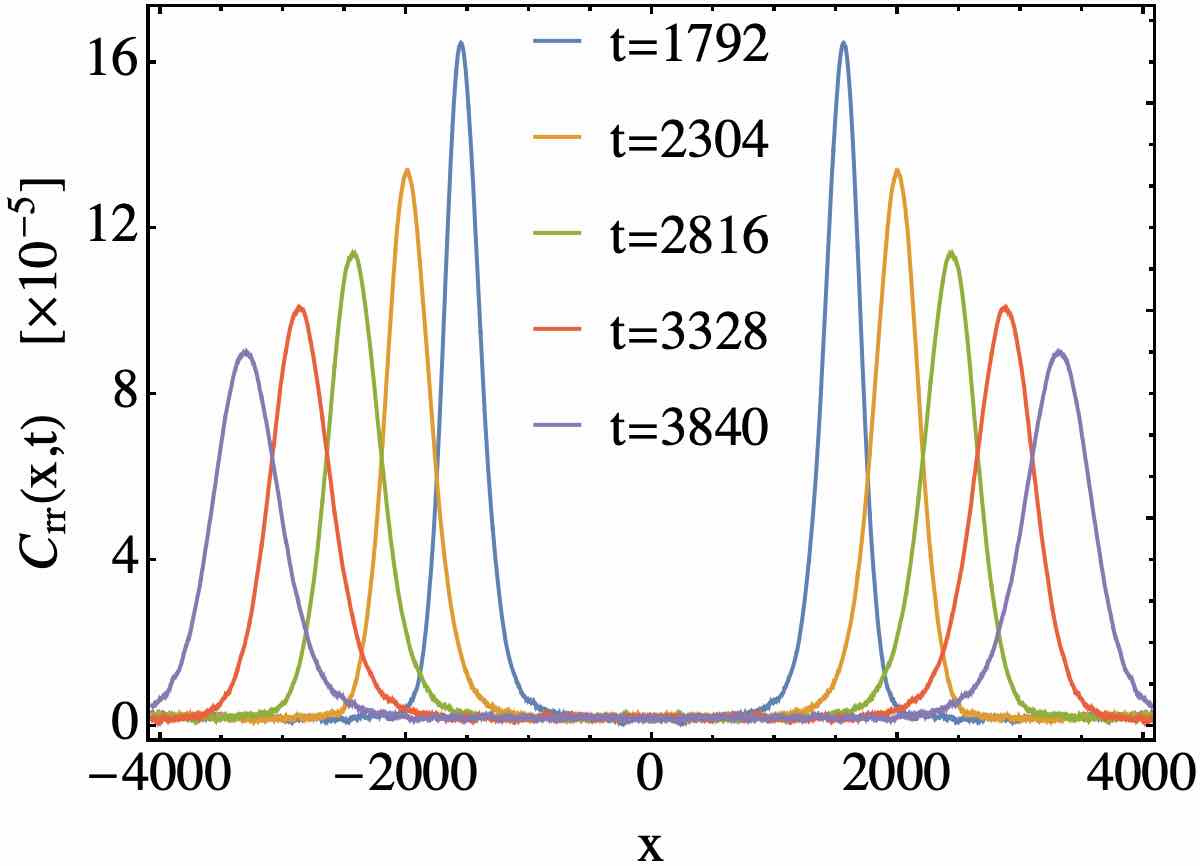}
  \includegraphics[width=0.32\textwidth]{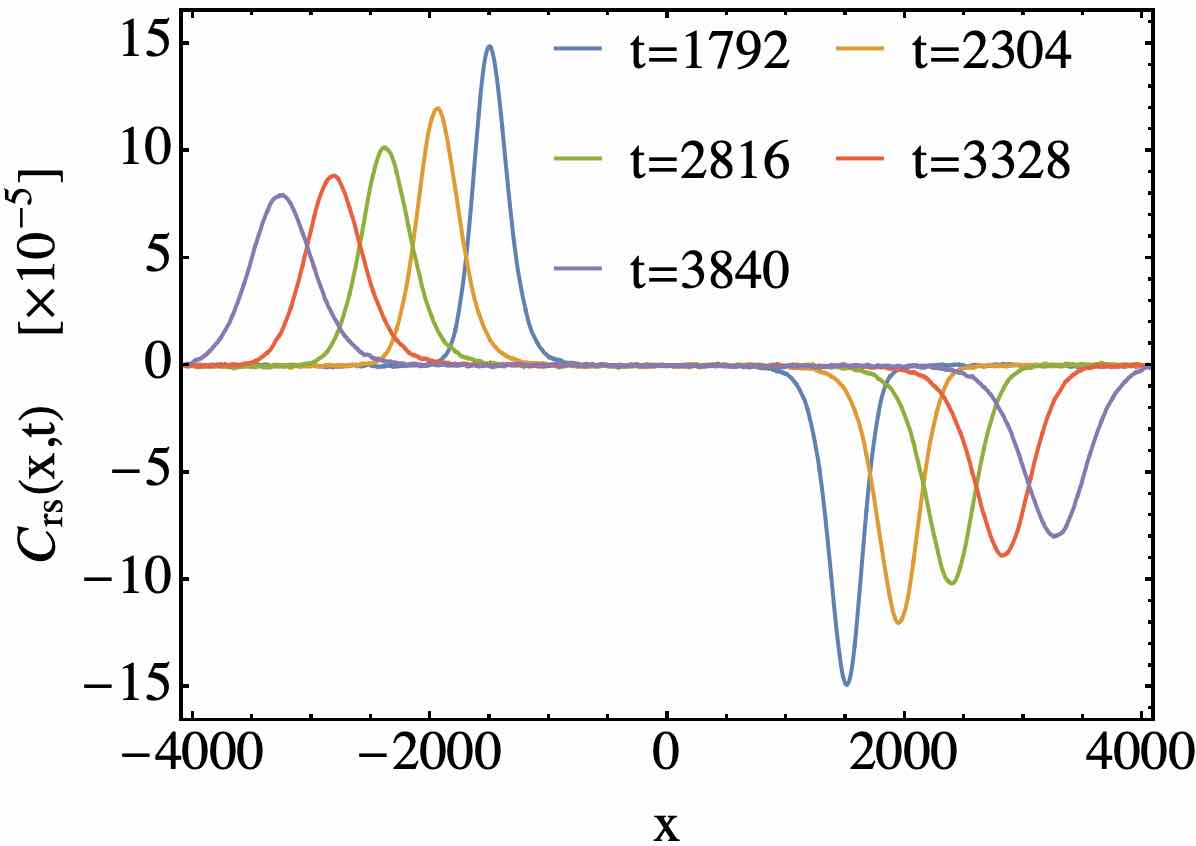}
  \includegraphics[width=0.32\textwidth]{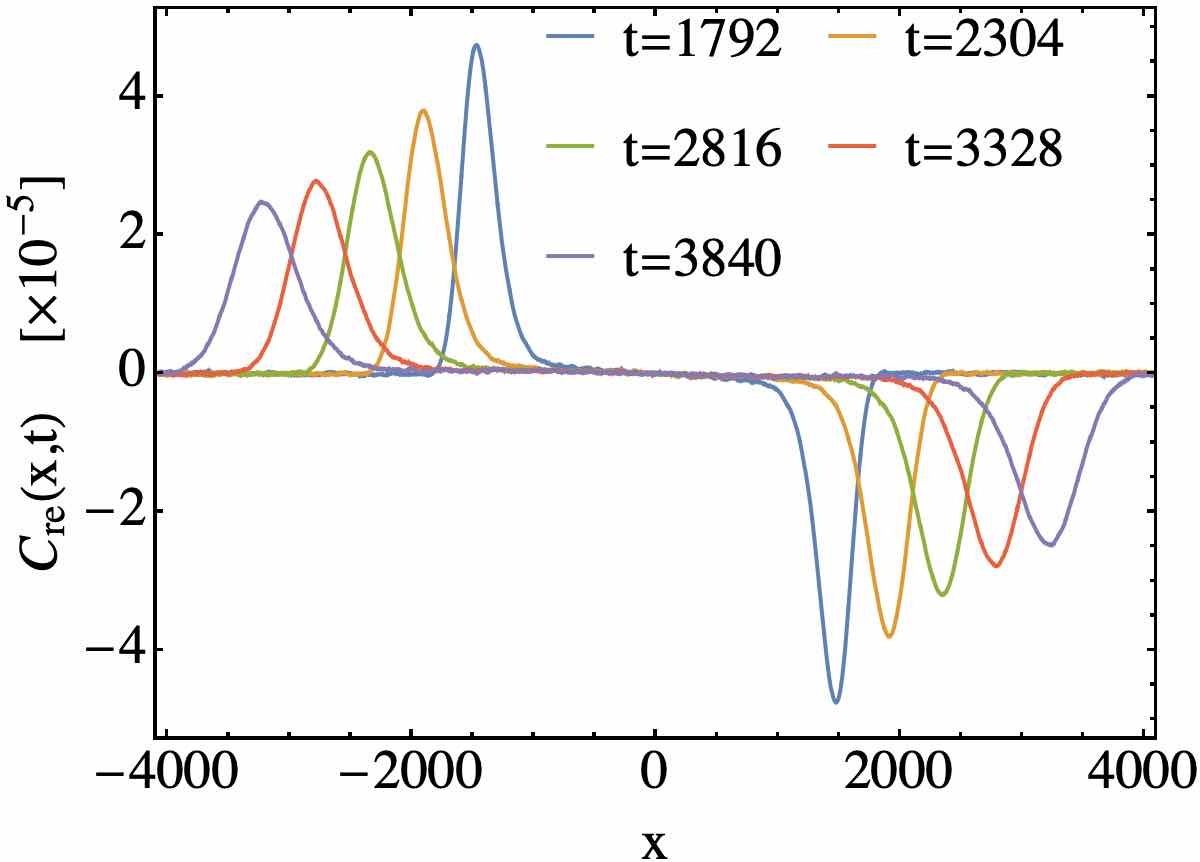}
  
  \includegraphics[width=0.32\textwidth]{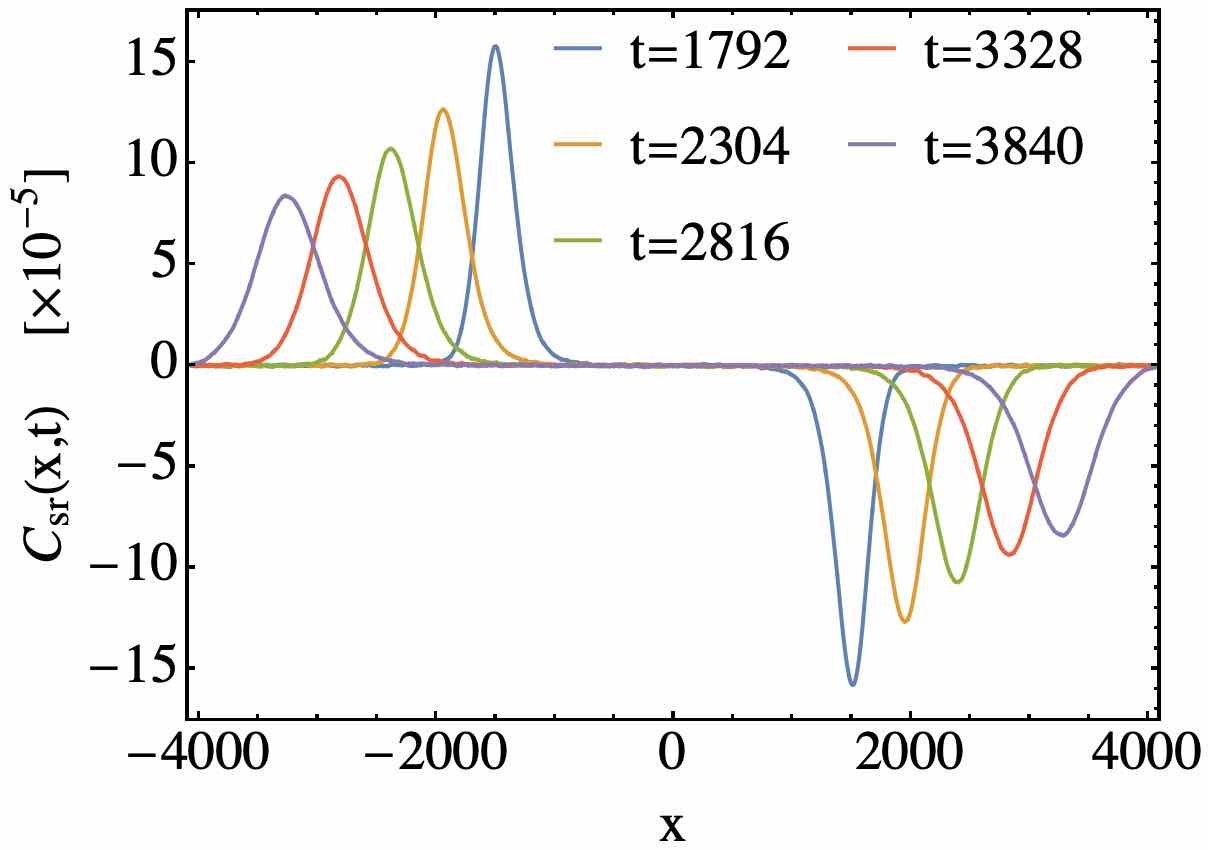}
  \includegraphics[width=0.32\textwidth]{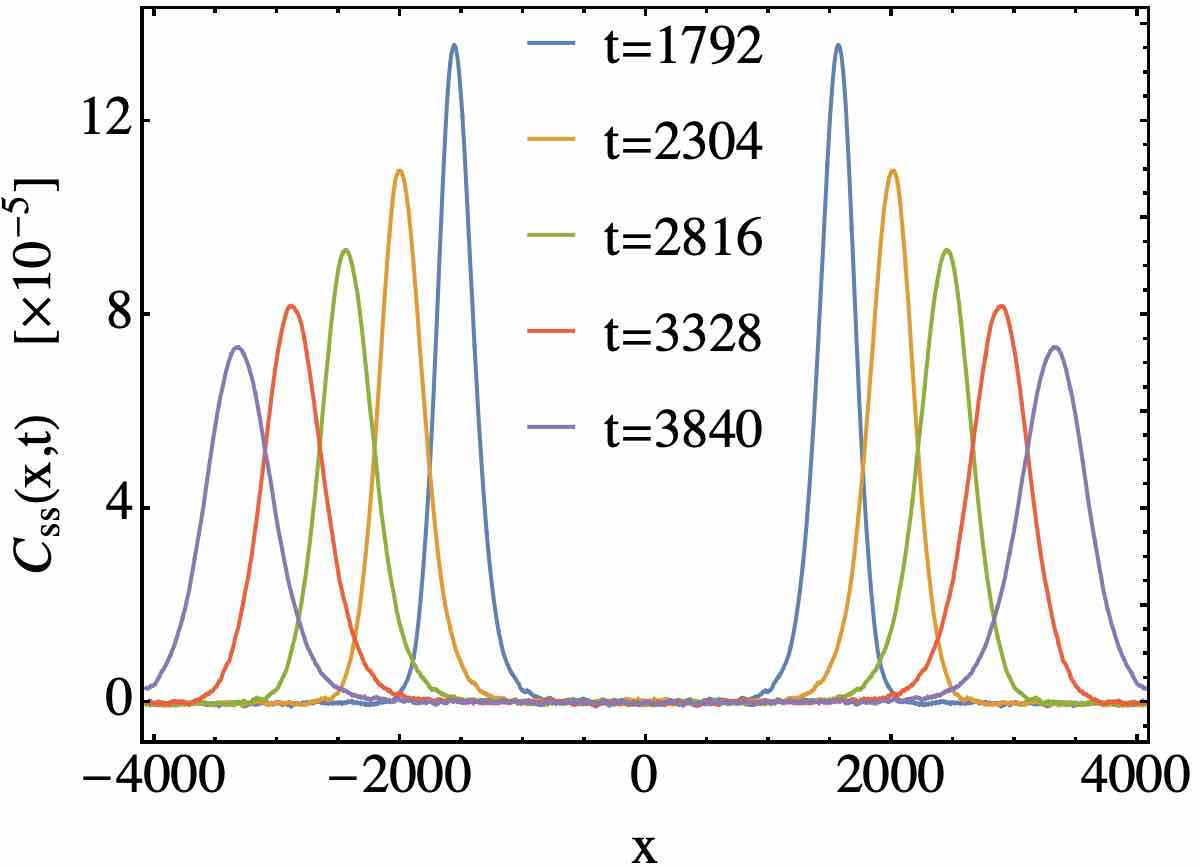}
  \includegraphics[width=0.32\textwidth]{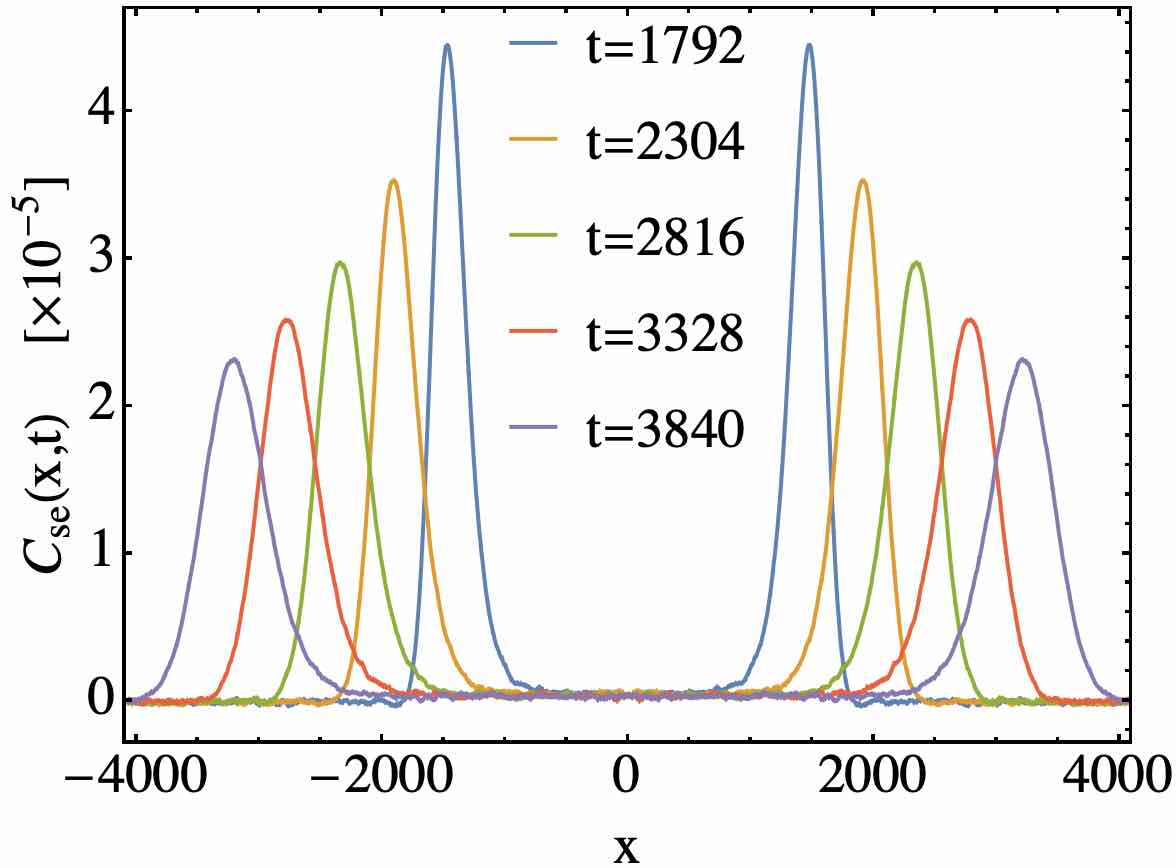}
  
  \includegraphics[width=0.32\textwidth]{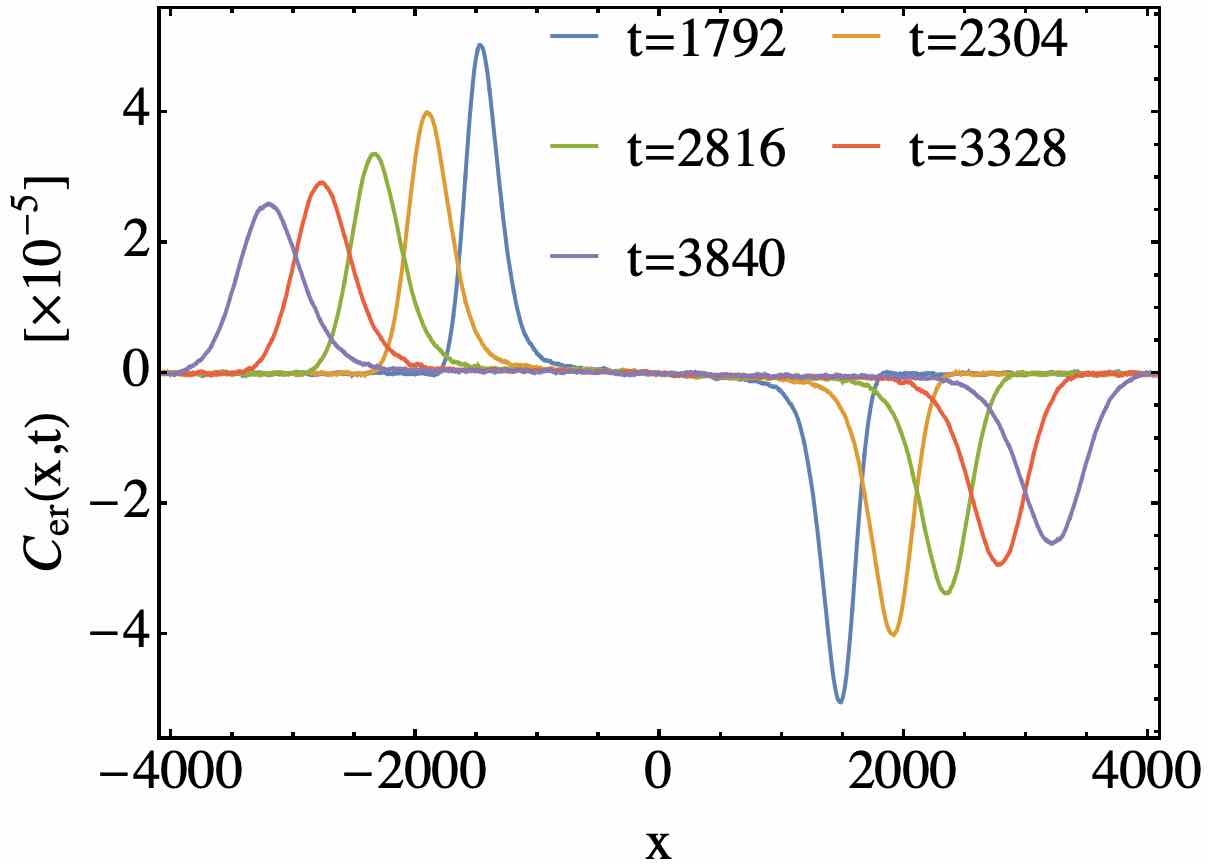}
  \includegraphics[width=0.32\textwidth]{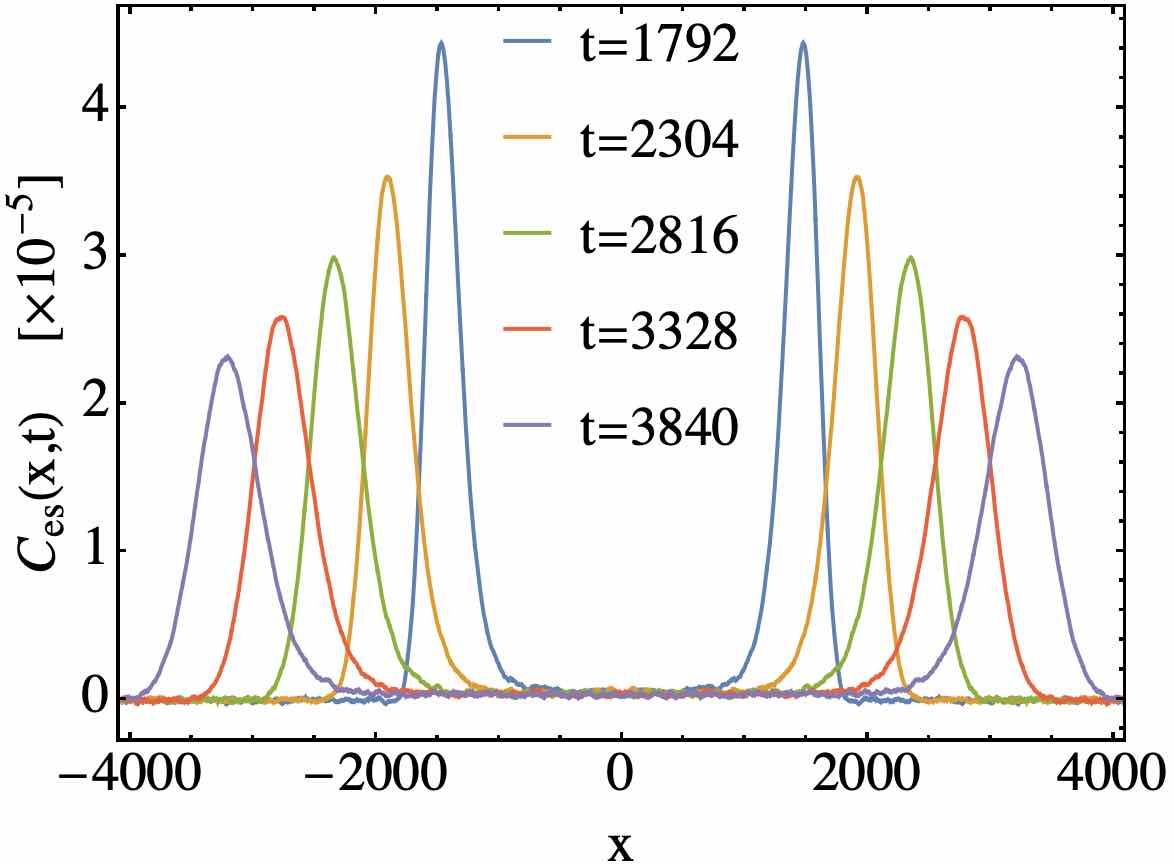}
  \includegraphics[width=0.32\textwidth]{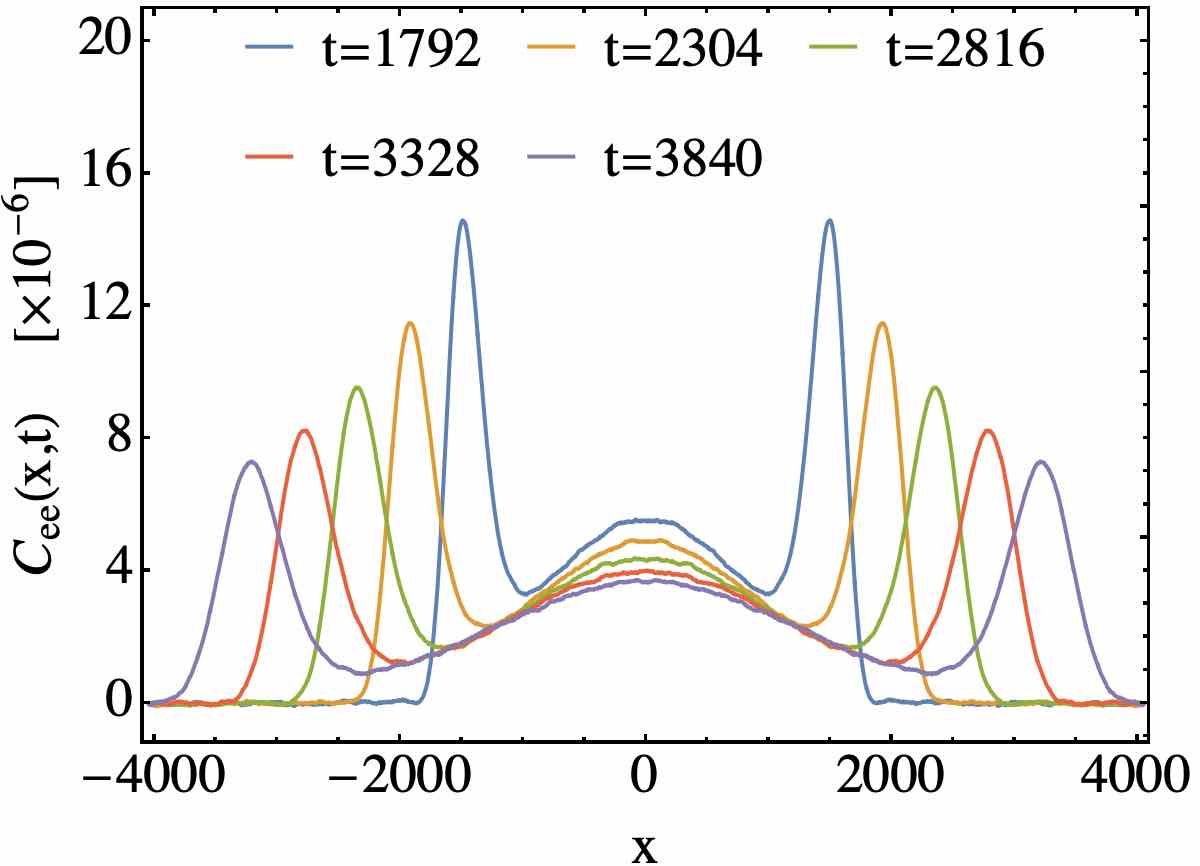}
  \caption{Parameters: $\Delta=0.5, \beta=10.0, h=0.3, N=8192$ -- RK-$4$ with $dt=0.005$: Plot of $C_{ab}(x,t)$ at different times.}
\label{figrk4b0.3}
\end{figure}

\begin{figure}[b]
  \includegraphics[width=0.32\textwidth]{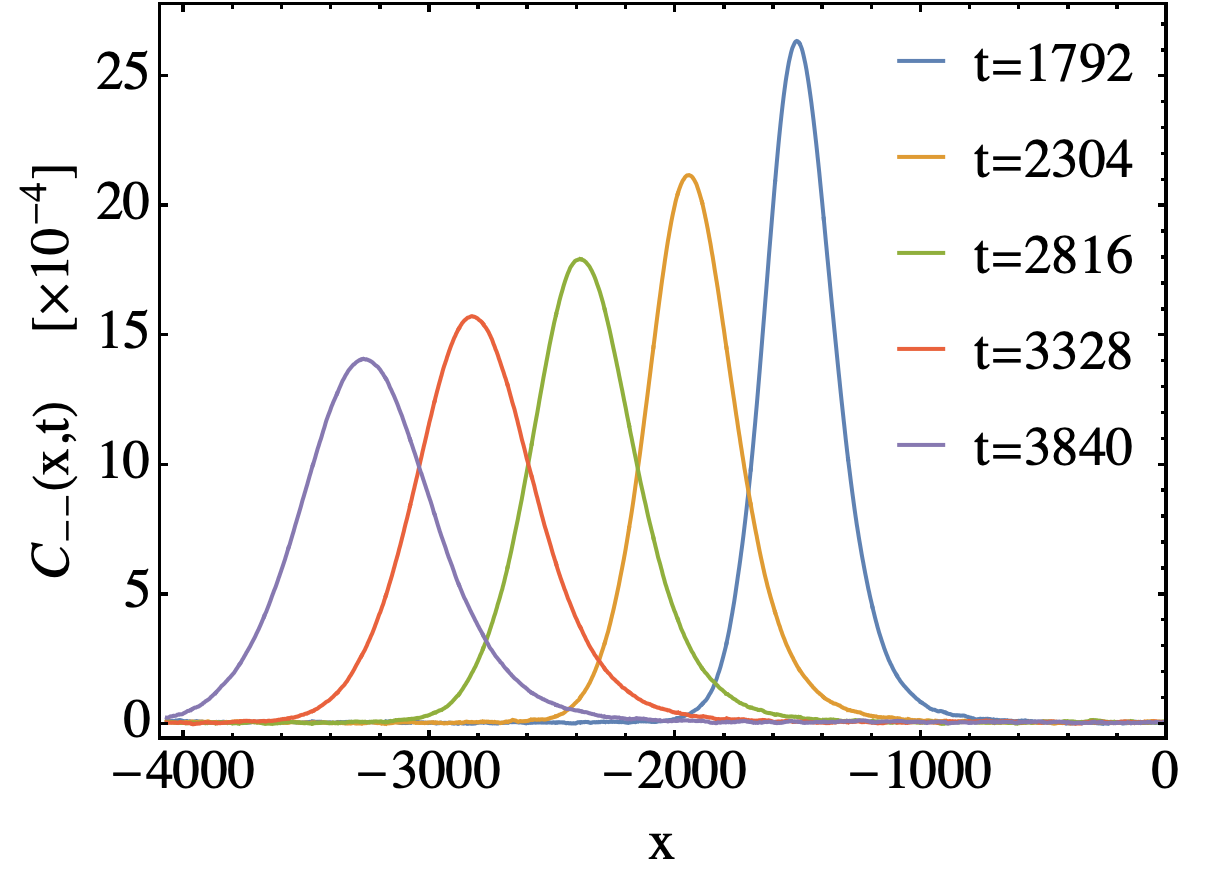}
  \includegraphics[width=0.32\textwidth]{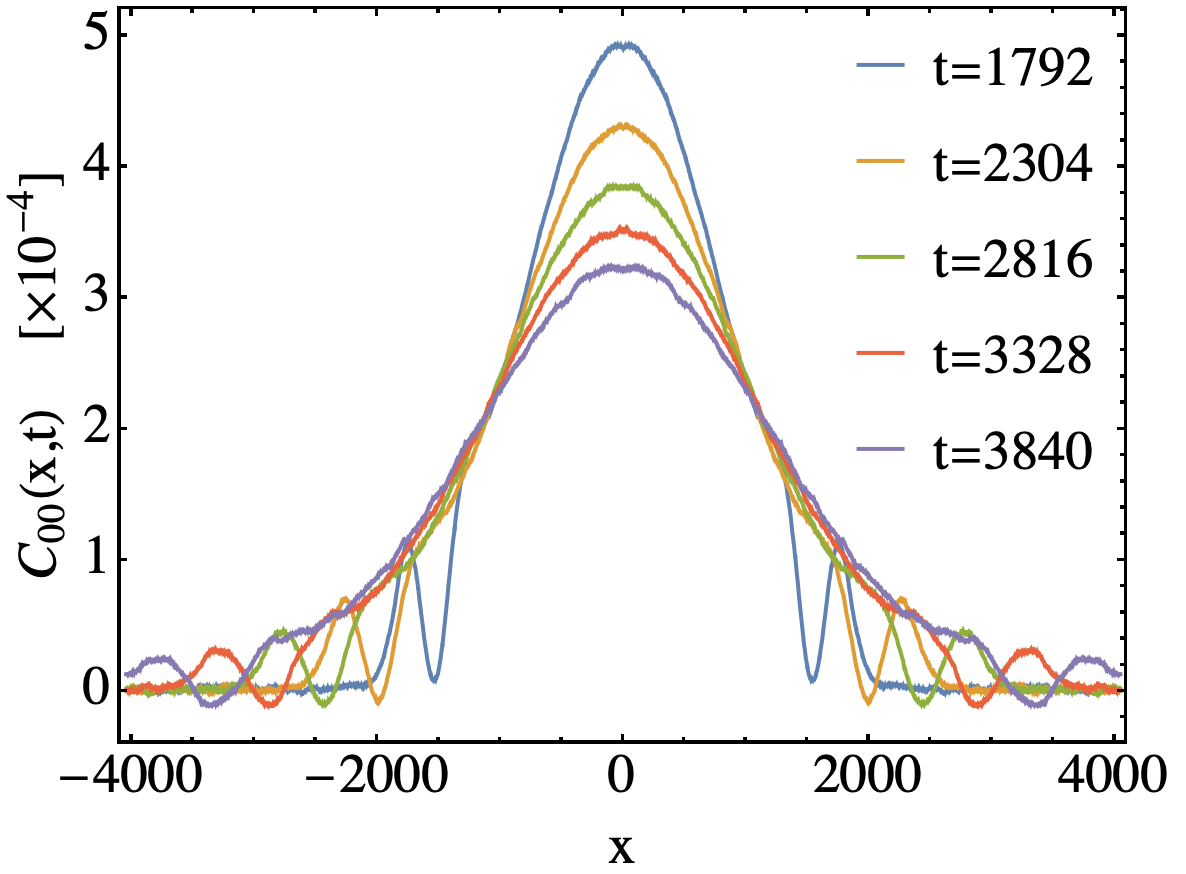}
   \includegraphics[width=0.32\textwidth]{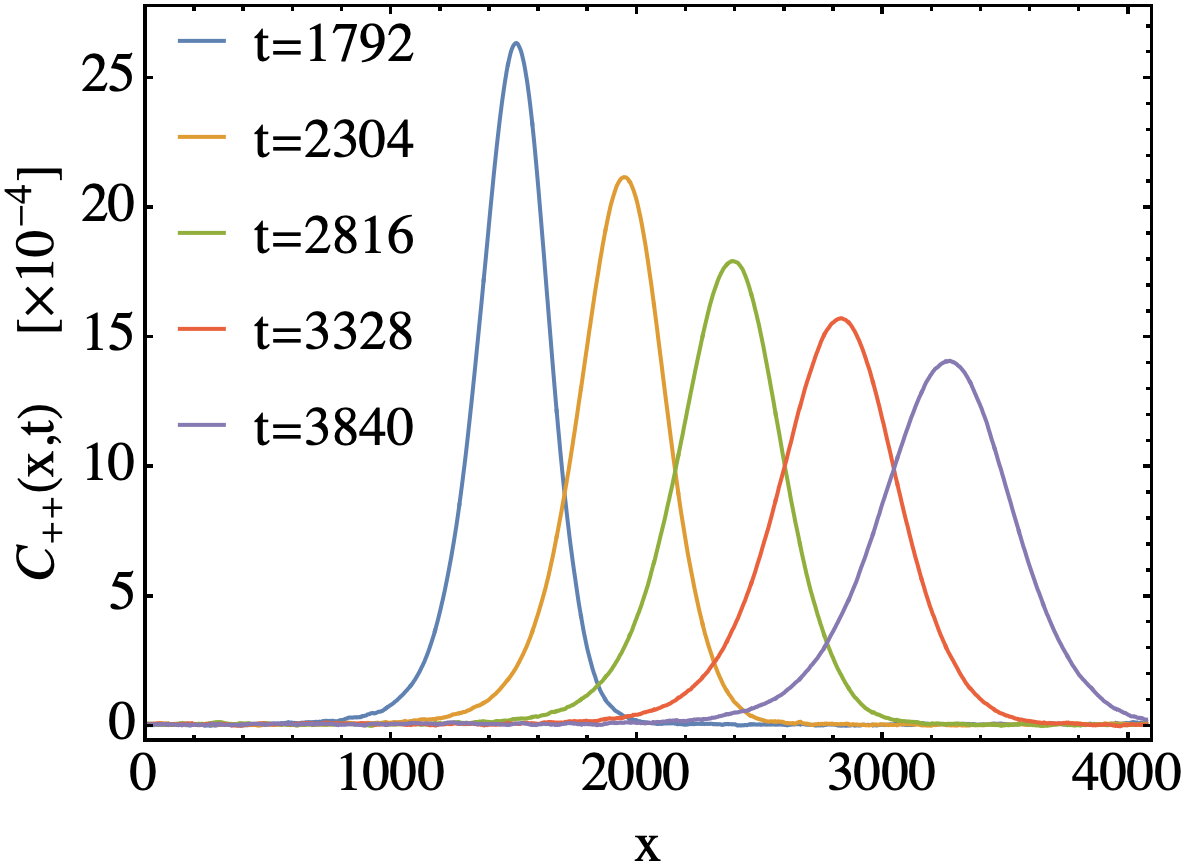}
  
  \includegraphics[width=0.32\textwidth]{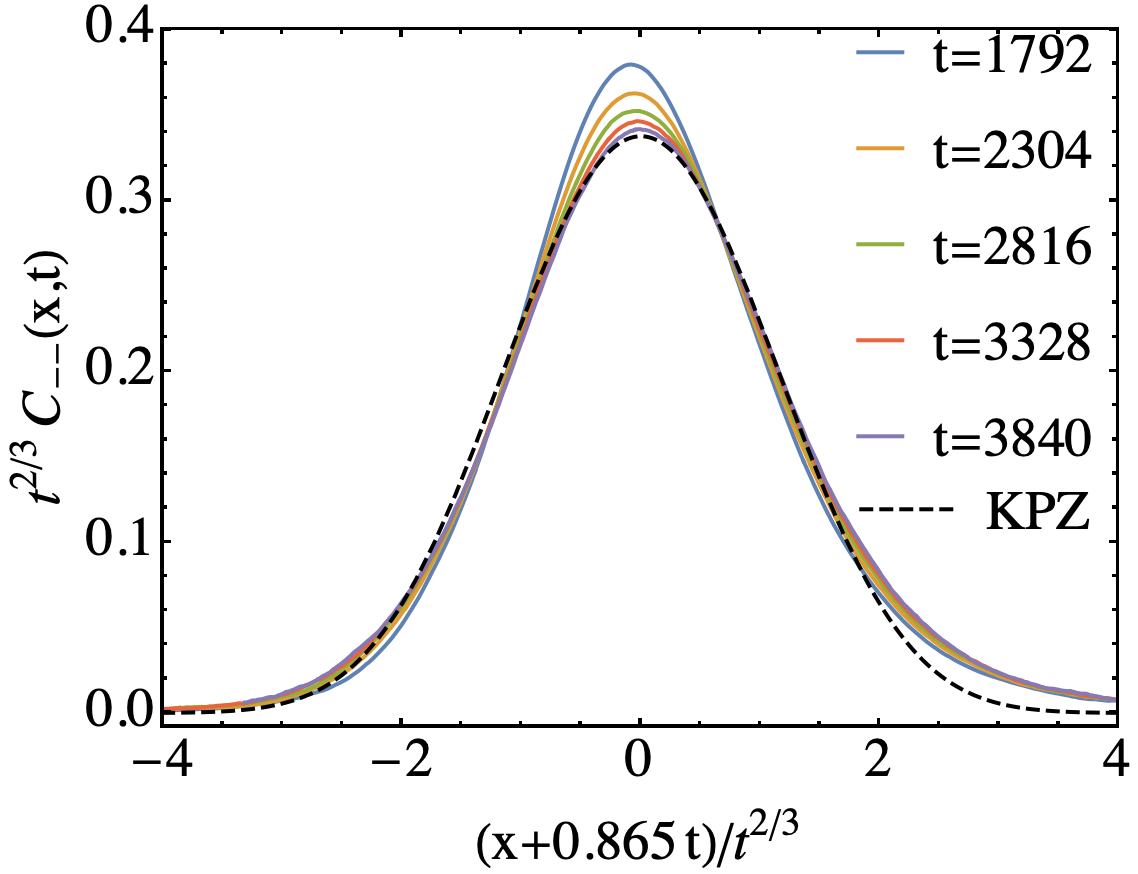}
  \includegraphics[width=0.32\textwidth]{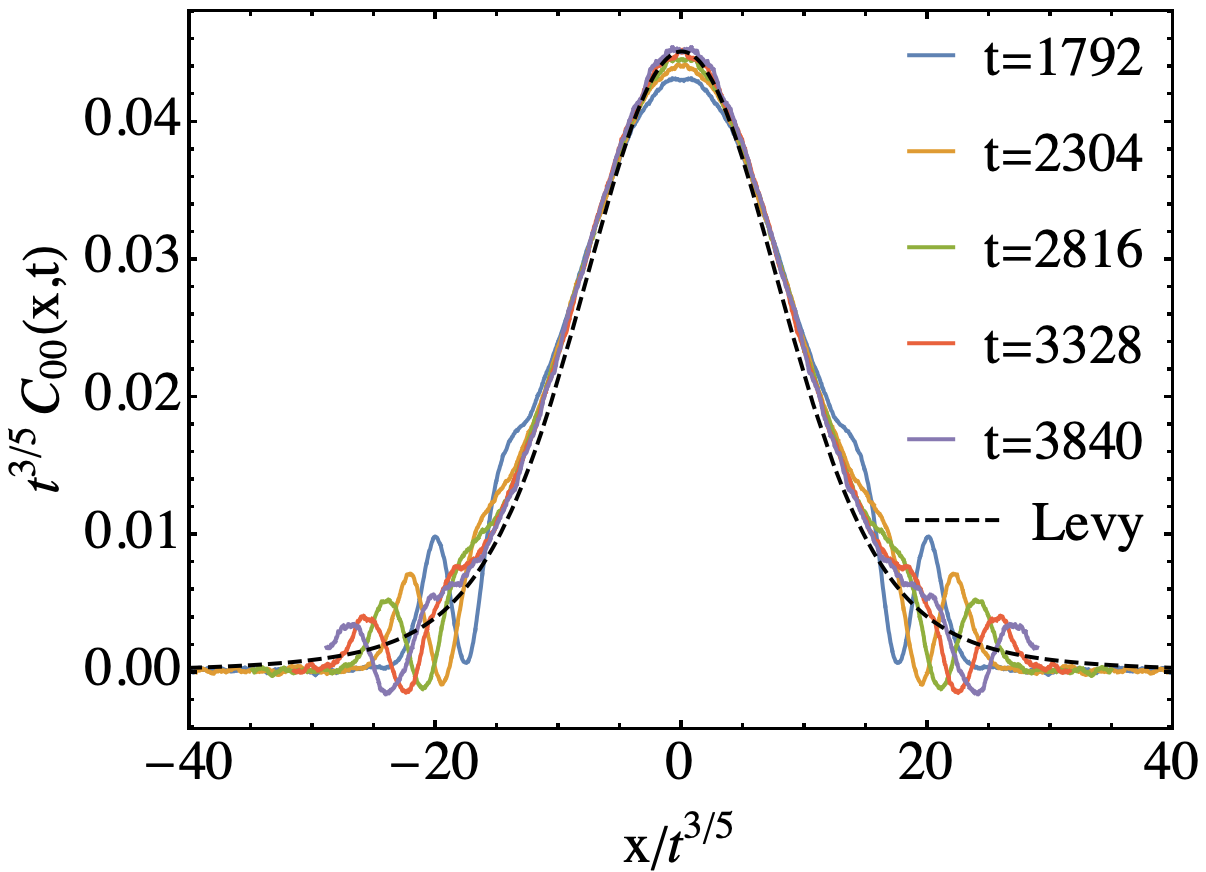}
  \includegraphics[width=0.32\textwidth]{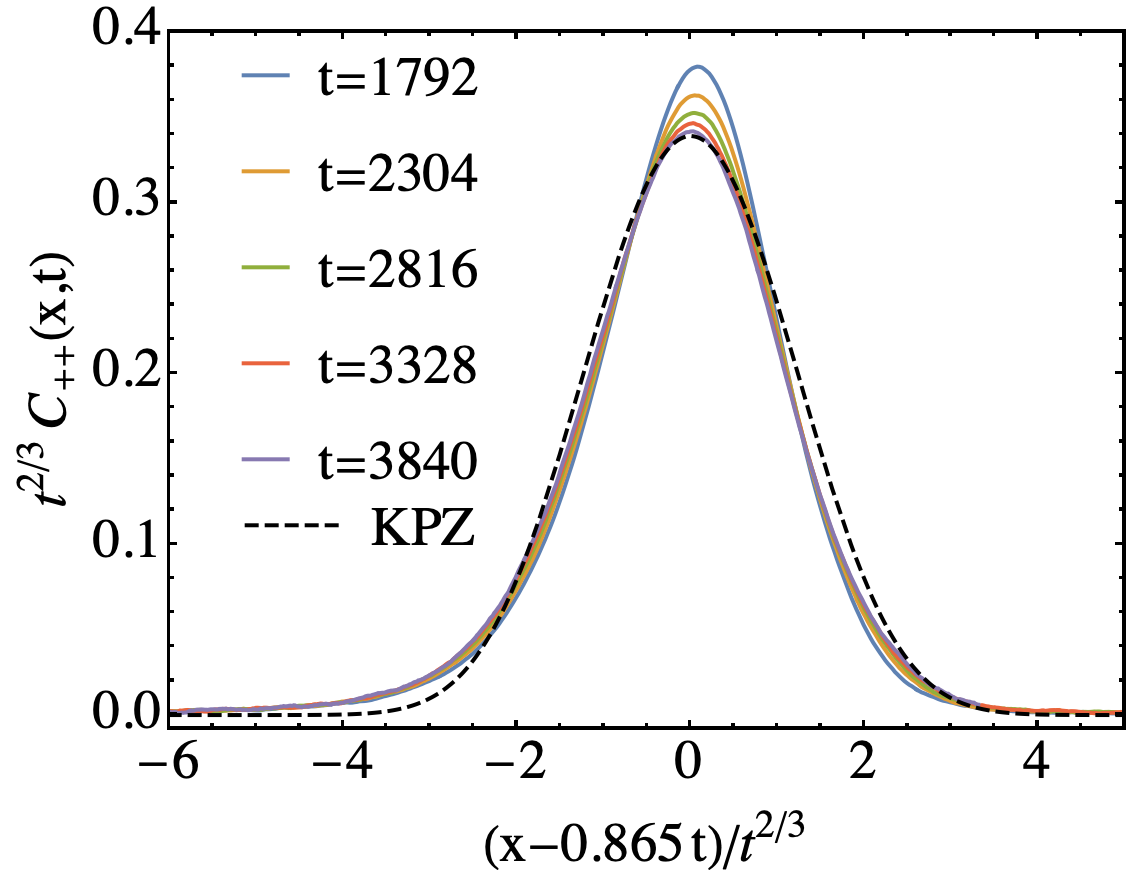}
  
  \includegraphics[width=0.32\textwidth]{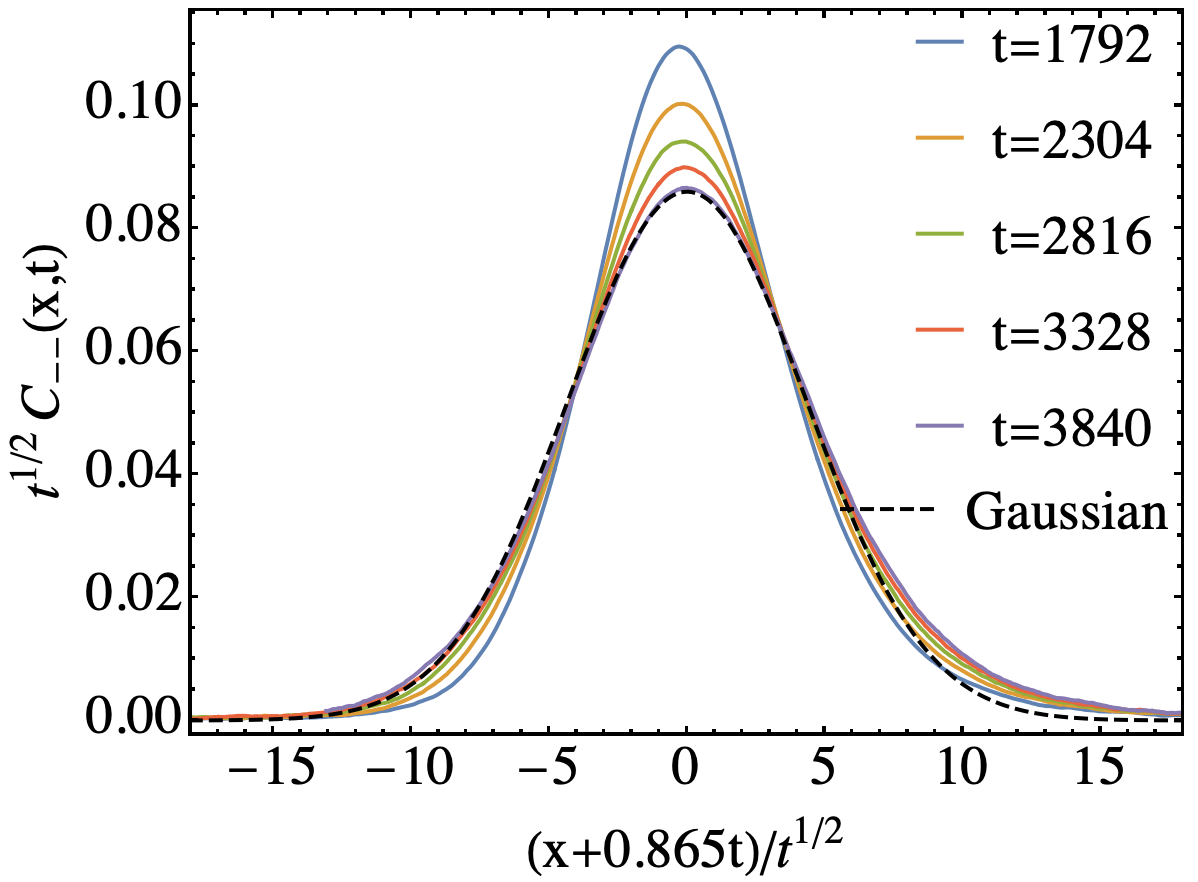}
  \includegraphics[width=0.32\textwidth]{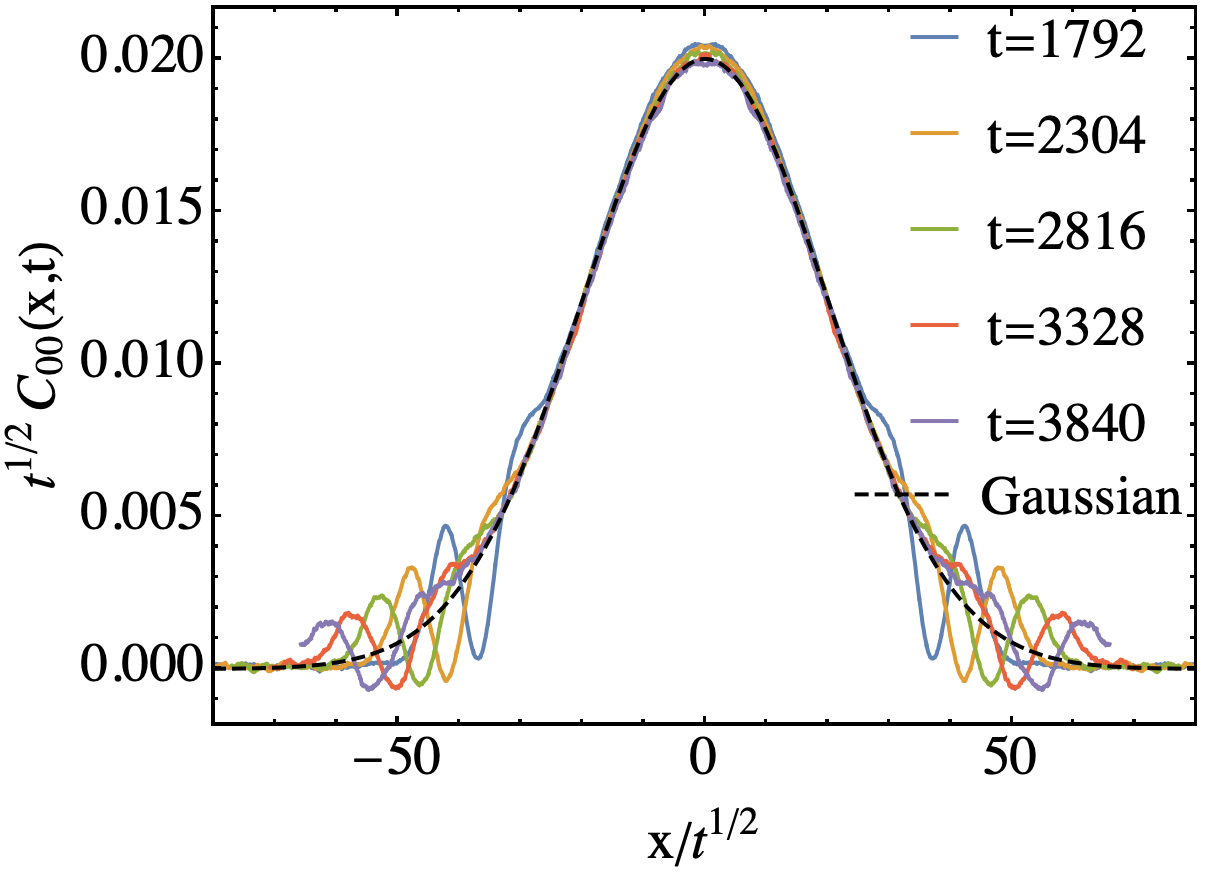}
  \includegraphics[width=0.32\textwidth]{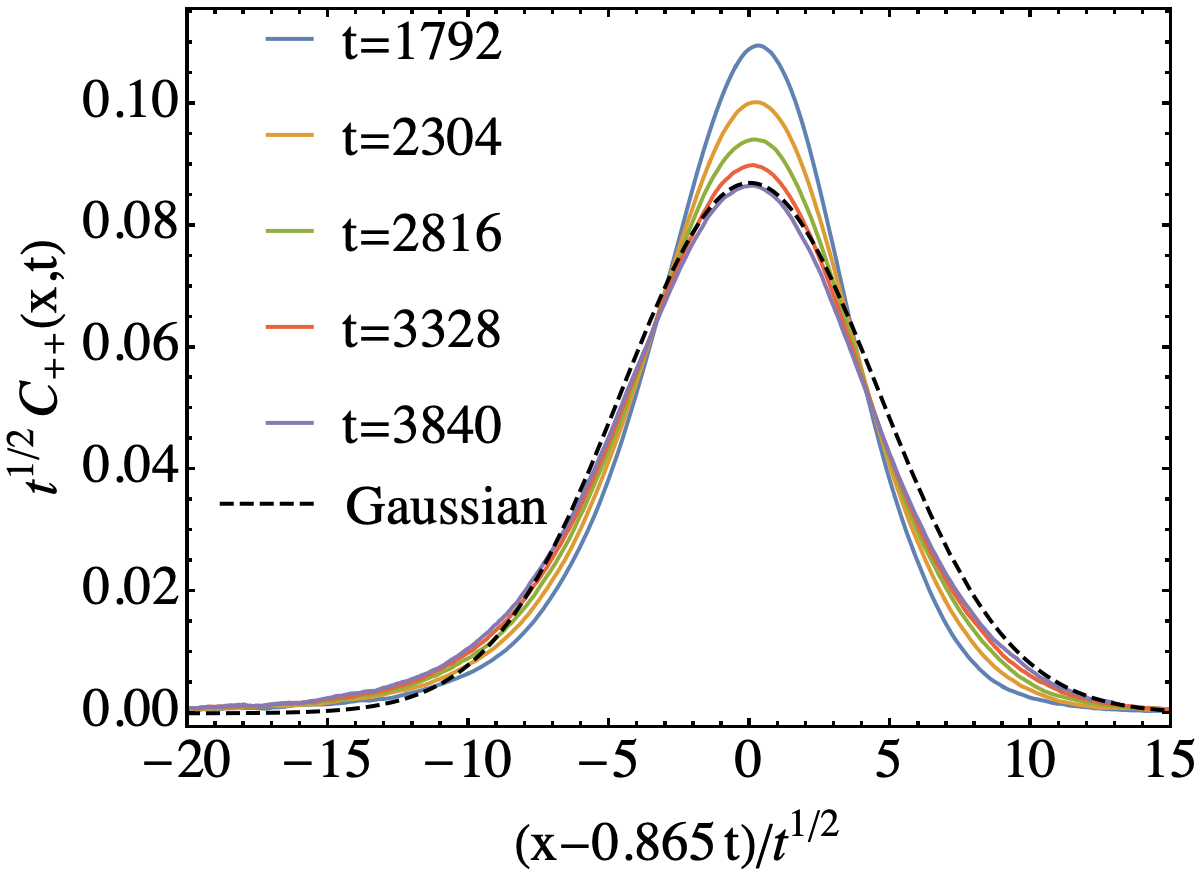}
\caption{Parameters: $\Delta=0.5, \beta=10.0, h=0.3, N=8192$ -- RK4 with $dt=0.005$: Plot of  $C_{++}(x,t),~C_{00}(x,t)$ and $C_{--}(x,t)$, obtained after normal mode transformation. The 2nd row shows the sound modes with KPZ scaling and heat mode with the predicted Levy scaling, while the 3rd row shows diffusive scaling of the same data. Sound speed estimate from theory is $c=0.865$.}
\label{figrk4b0.3SH}
\end{figure} 

\begin{figure}[b]
\includegraphics[width=0.32\textwidth]{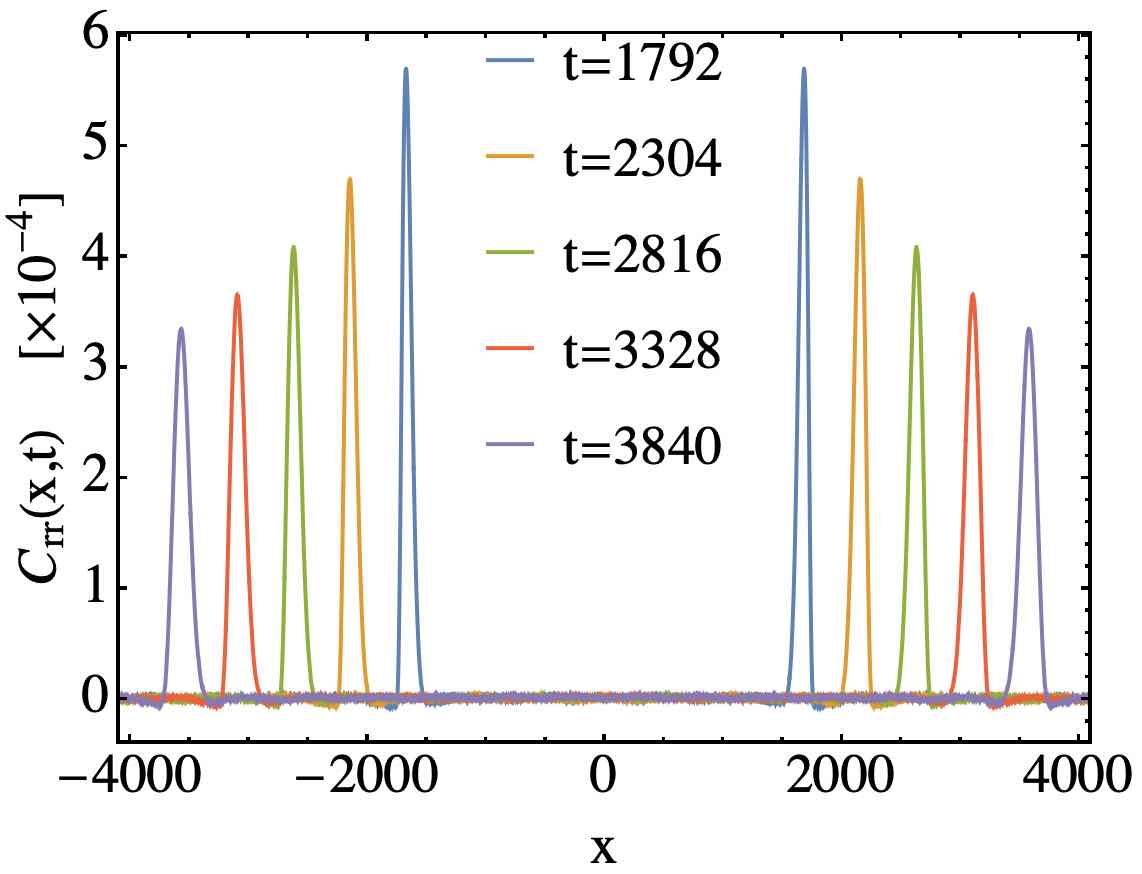}
  \includegraphics[width=0.32\textwidth]{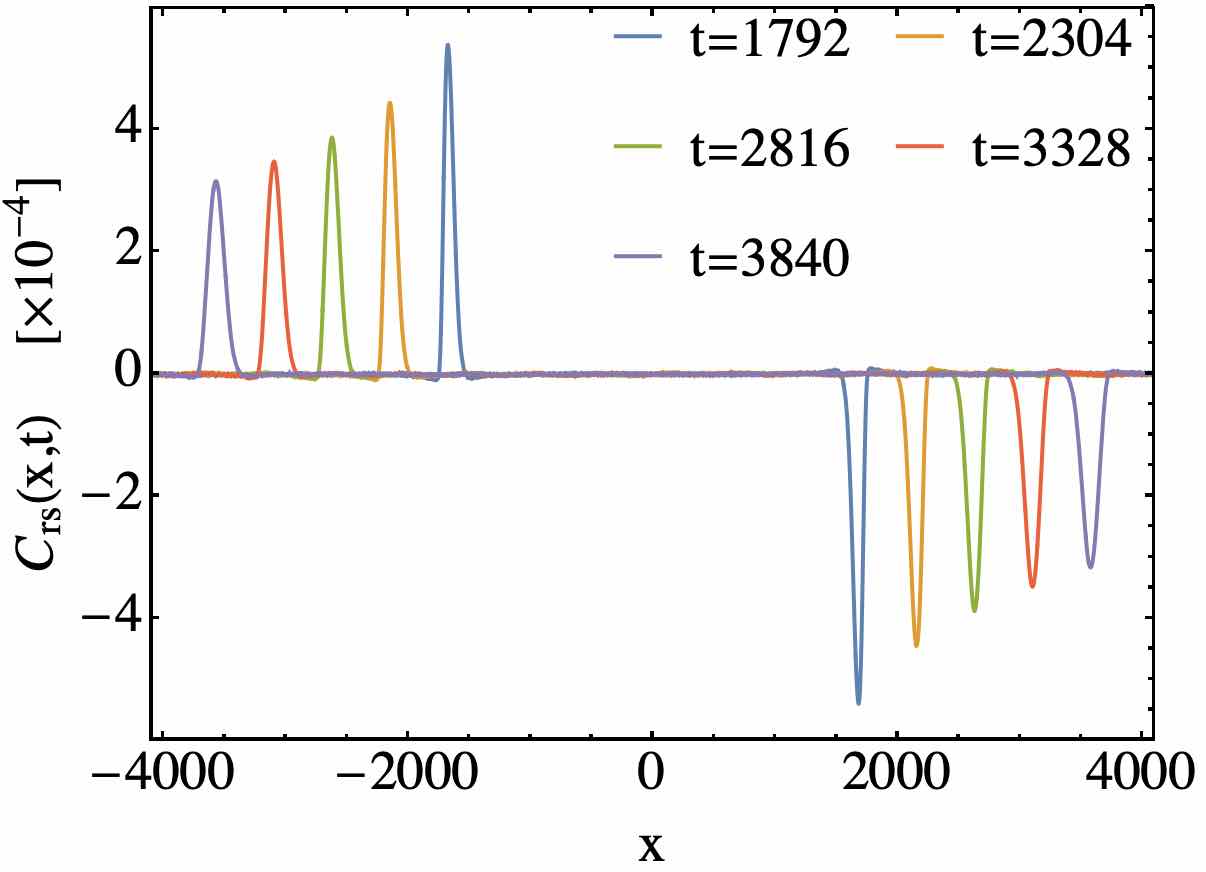}
  \includegraphics[width=0.32\textwidth]{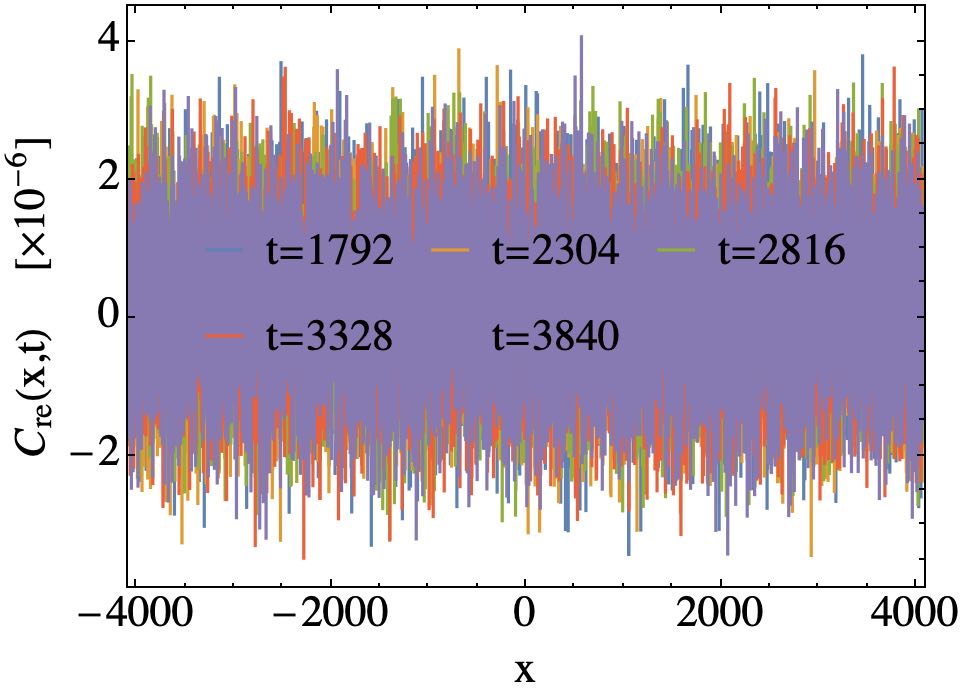}
  
  \includegraphics[width=0.32\textwidth]{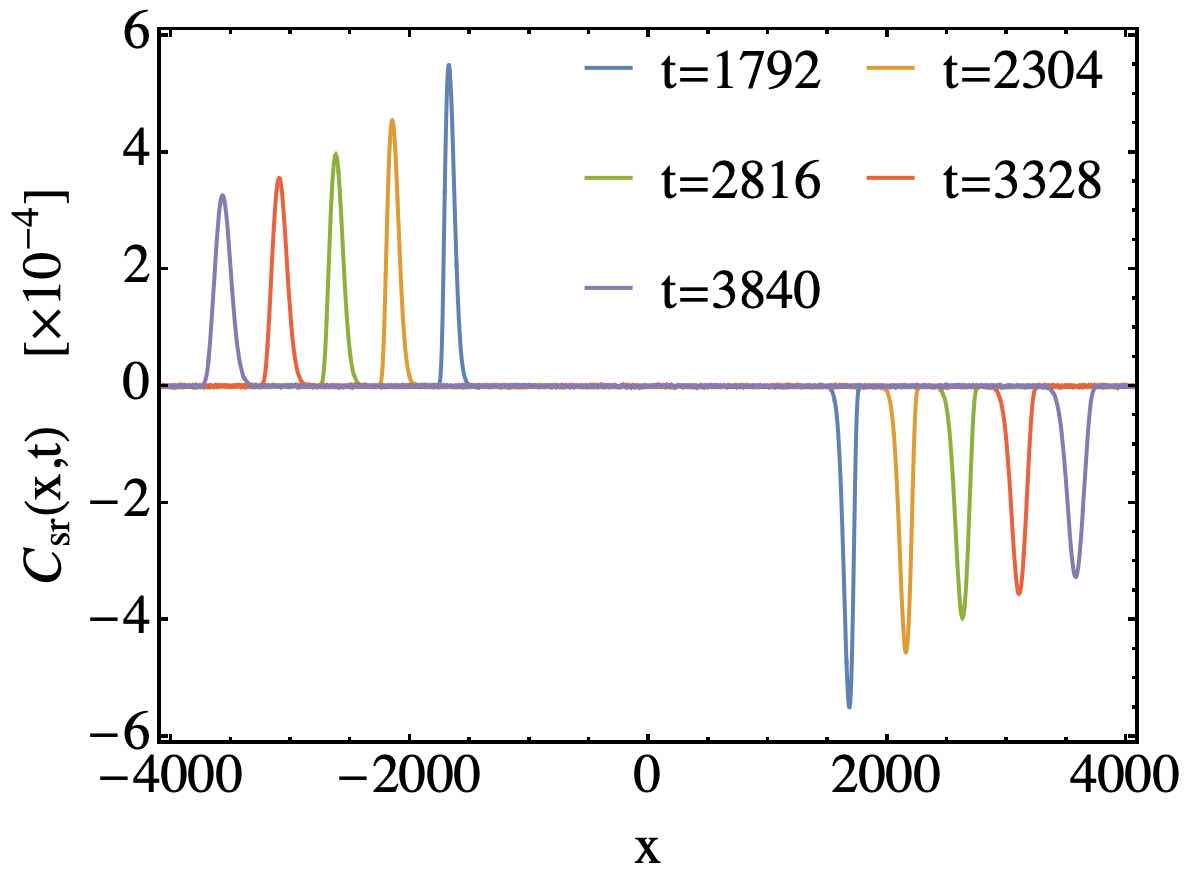}
  \includegraphics[width=0.32\textwidth]{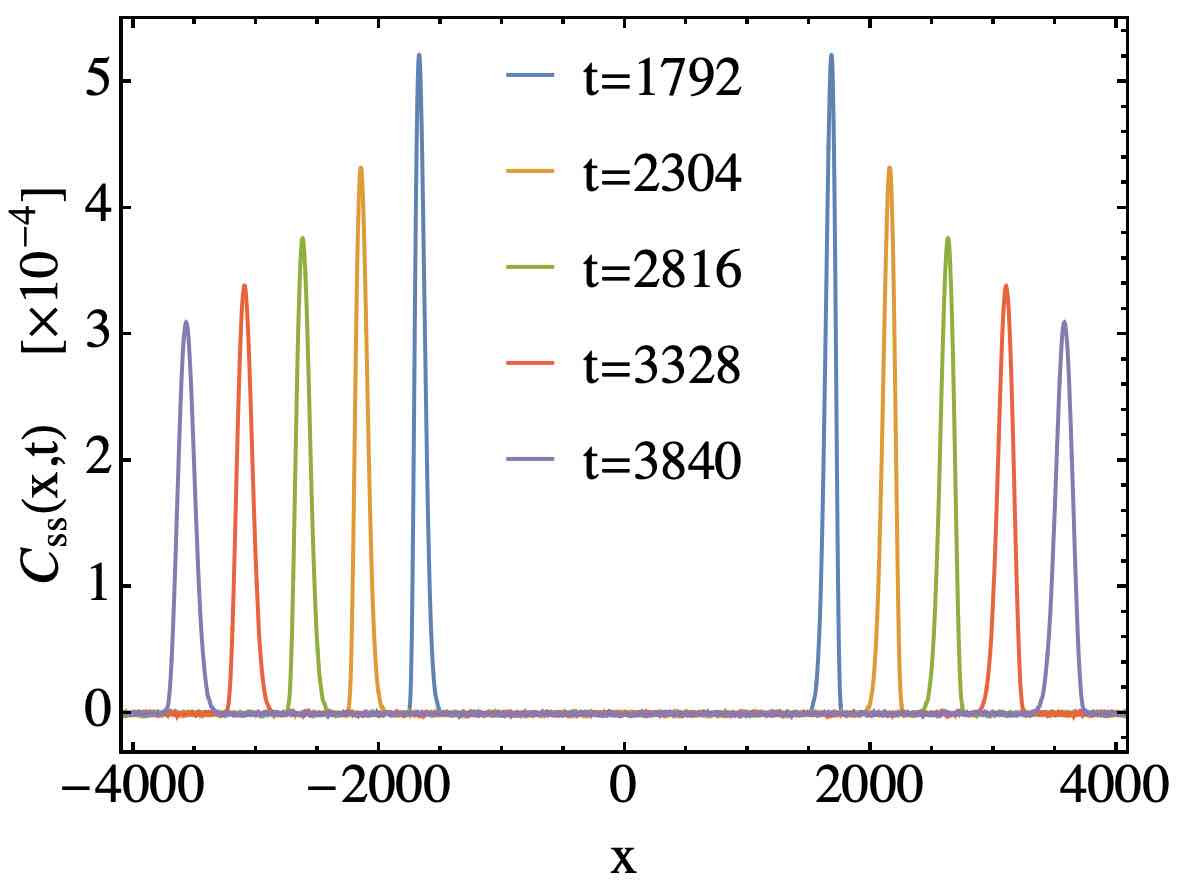}
  \includegraphics[width=0.32\textwidth]{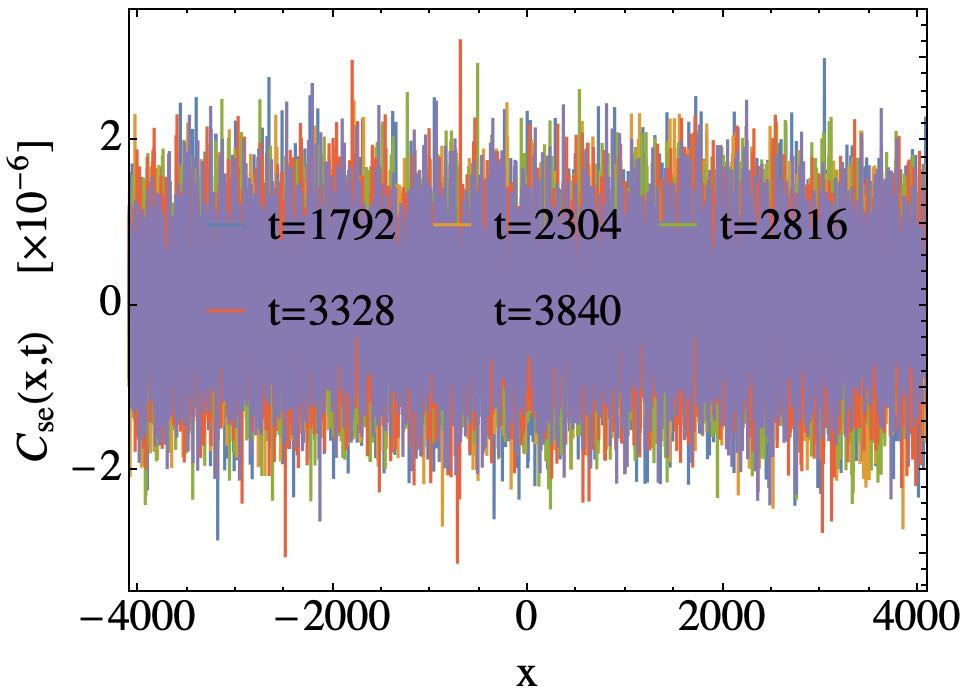}

  \includegraphics[width=0.32\textwidth]{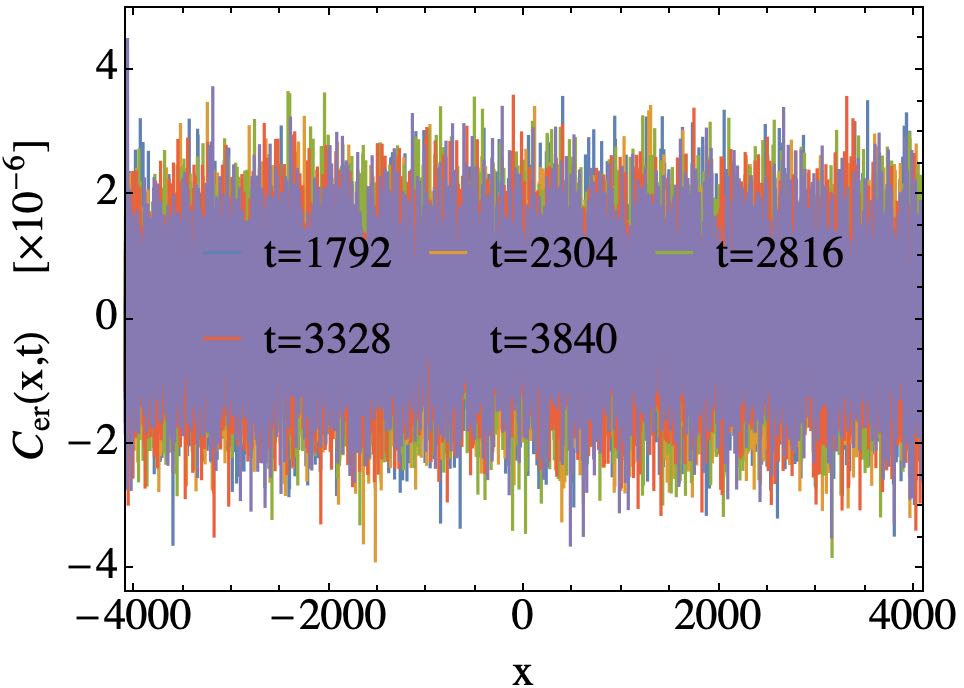}
  \includegraphics[width=0.32\textwidth]{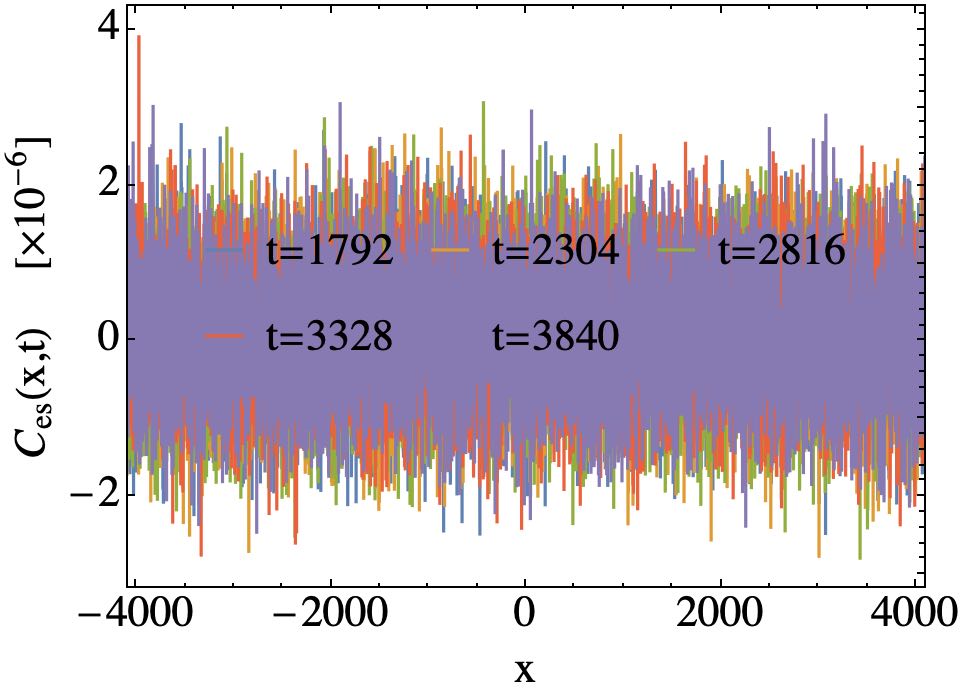}
  \includegraphics[width=0.32\textwidth]{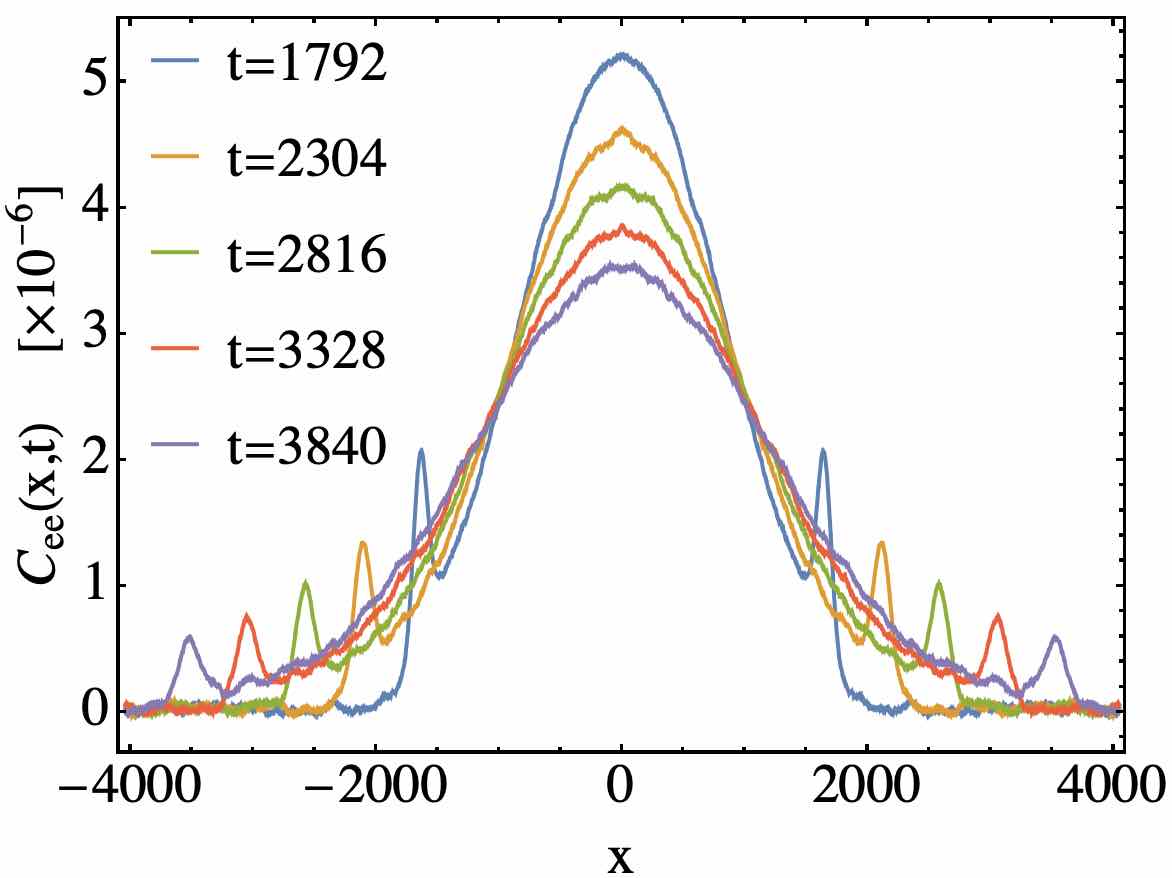}
  
\caption{Parameters: $\Delta=0.5, \beta=10.0, h=0, N=8192$ -- RK-$4$ with $dt=0.005$: Plot of $C_{ab}(x,t)$ at different times. $C_{es}(x,t)$, $C_{se}(x,t)$, $C_{re}(x,t)$ and $C_{er}(x,t)$ are essentially zero, which is expected from symmetry.}
\label{figrk4b0.0}
\end{figure}

We conclude that the $H_\mathrm{lt}$-currents satisfy \eqref{3.18} independently of $U$, provided $U$ diverges sufficiently fast at the two border points so as to have the boundary terms vanish. 

When applying the magic identity, the potential $U$ is specified as the cosine with infinitely high barriers added at $x= \pm \pi$. This does not change the free energy. Thus at the end 
the coupling matrix $\vec{G}$ is given in terms of the LLL free energy. We refer to Appendix A for more details. The chemical potential $\nu 
$ is not a physical control parameter that we will explore. It is needed in the second order expansion, but will be set to $\nu = 0$ afterwards.

\begin{figure}[b]
   \includegraphics[width=0.33\textwidth]{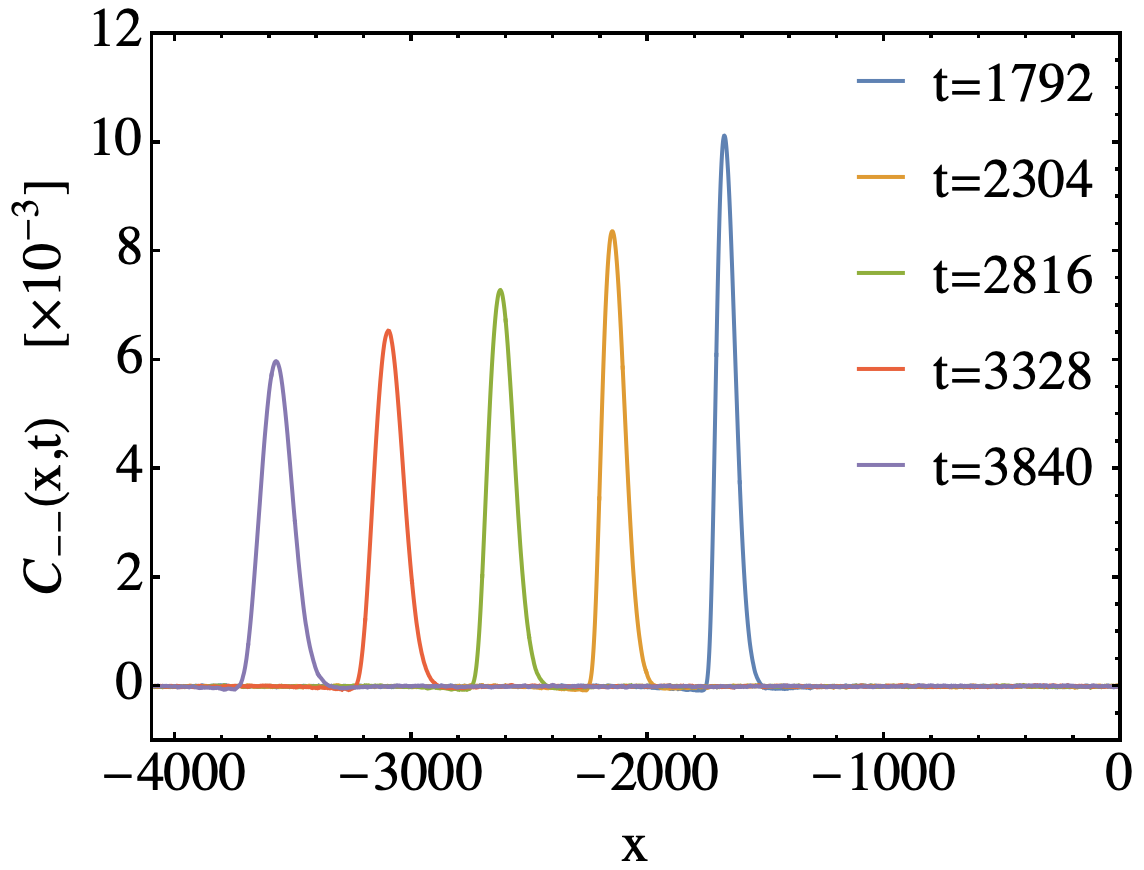}
  \includegraphics[width=0.32\textwidth]{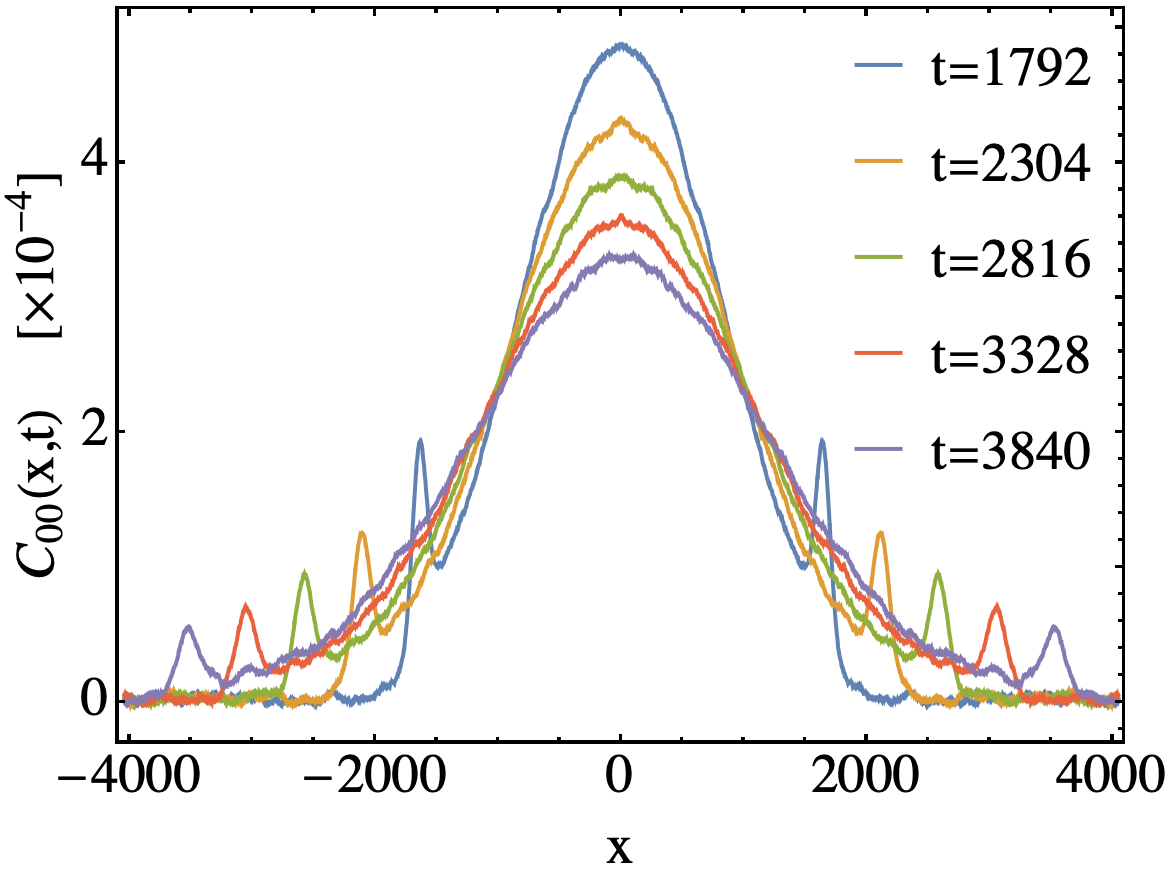}
  \includegraphics[width=0.33\textwidth]{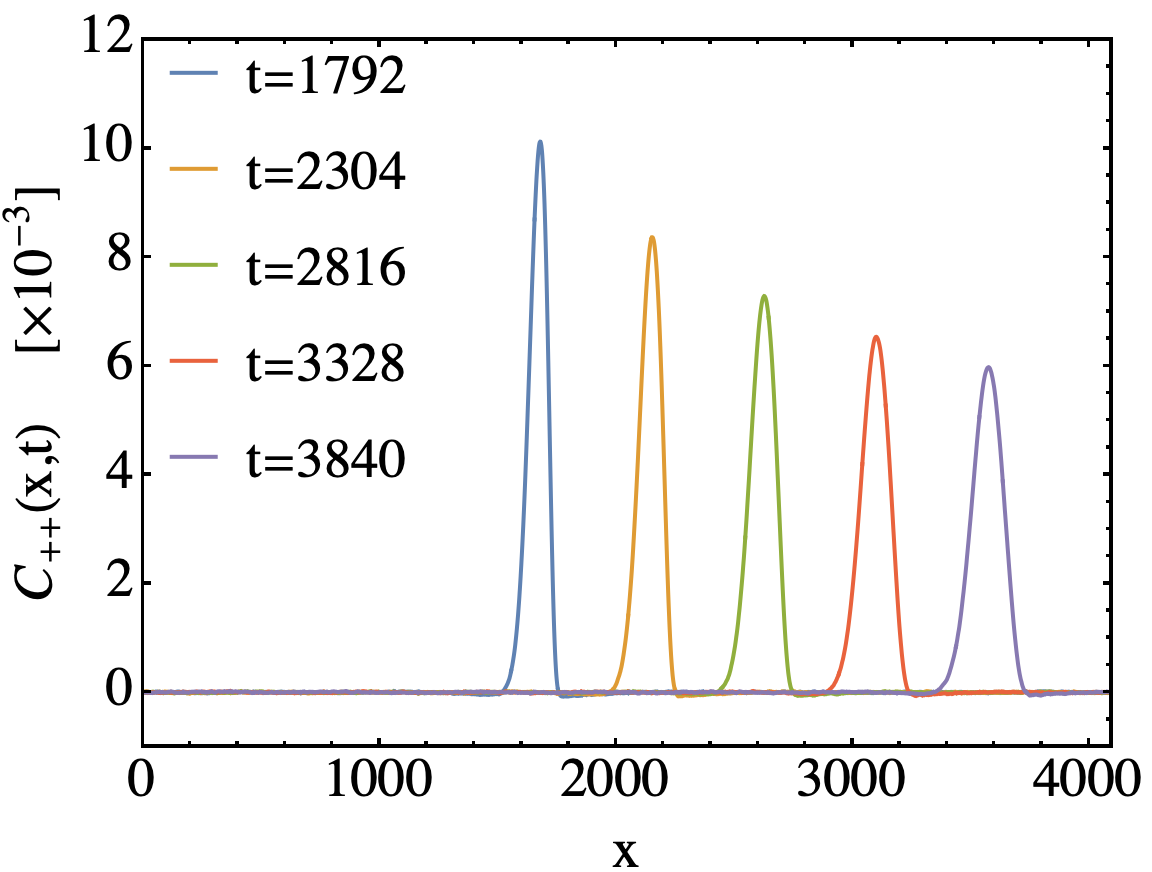}

  \includegraphics[width=0.33\textwidth]{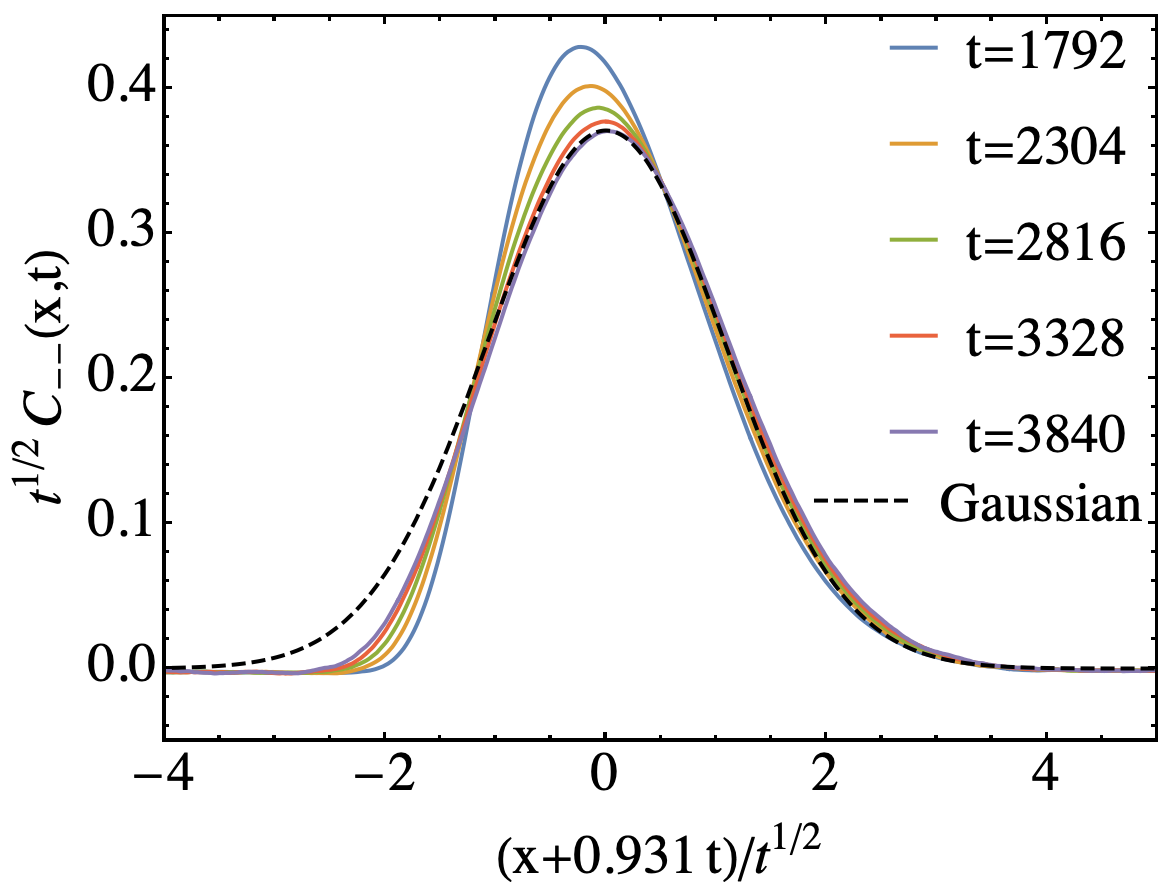}
  \includegraphics[width=0.32\textwidth]{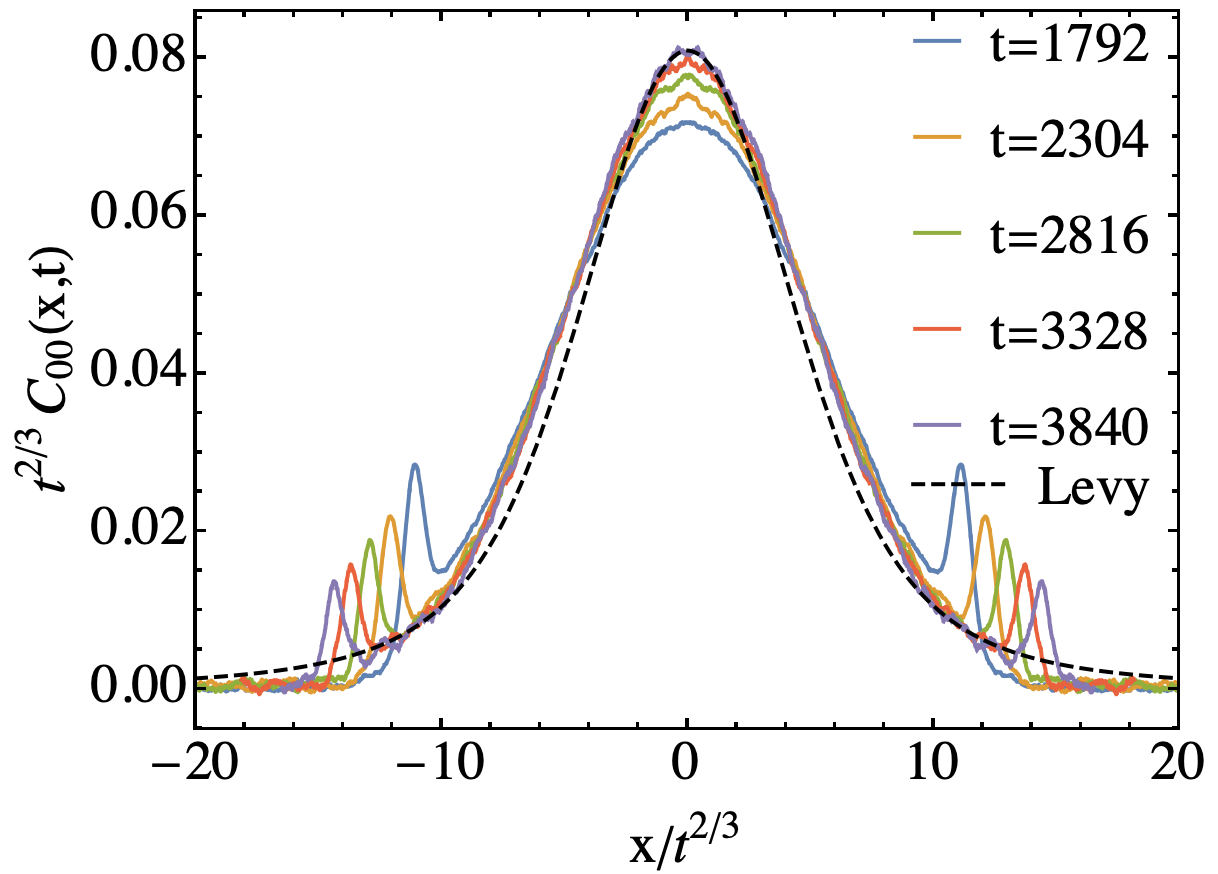}
  \includegraphics[width=0.33\textwidth]{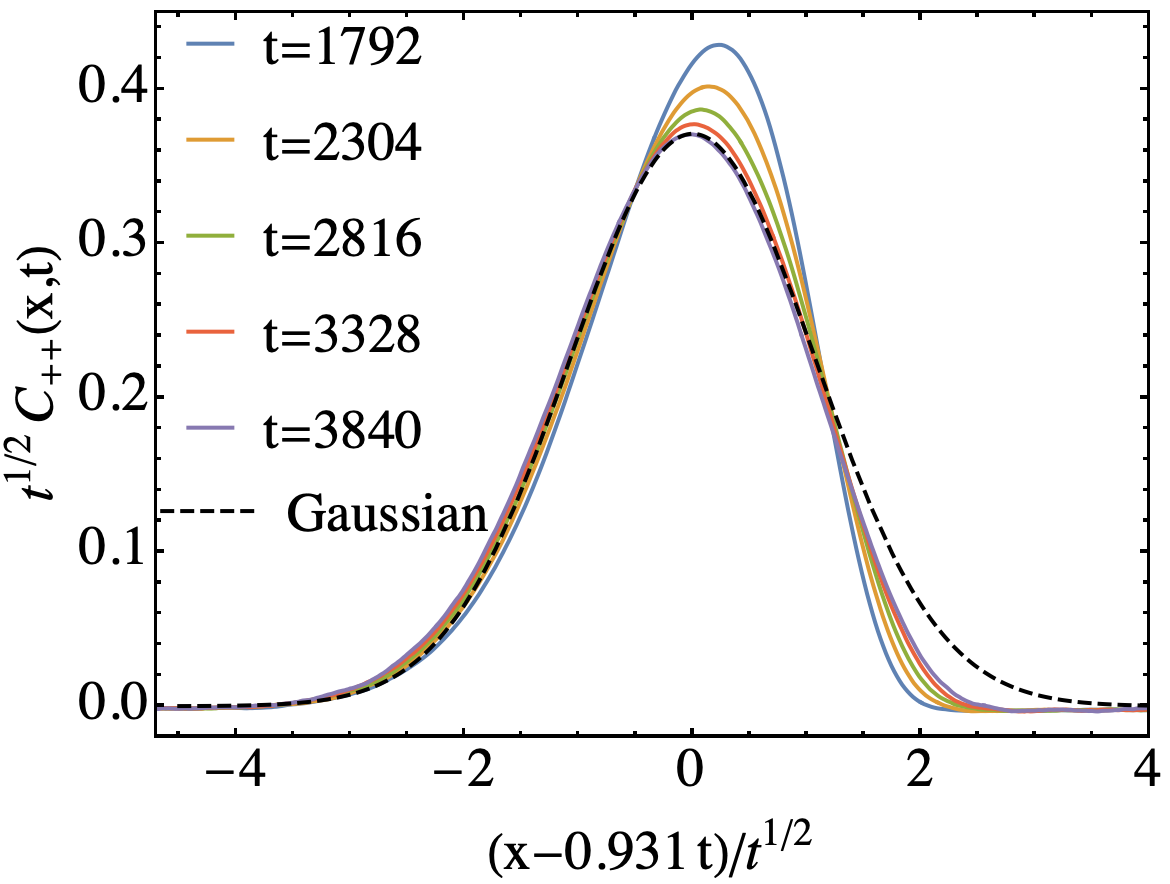}
  
    \includegraphics[width=0.33\textwidth]{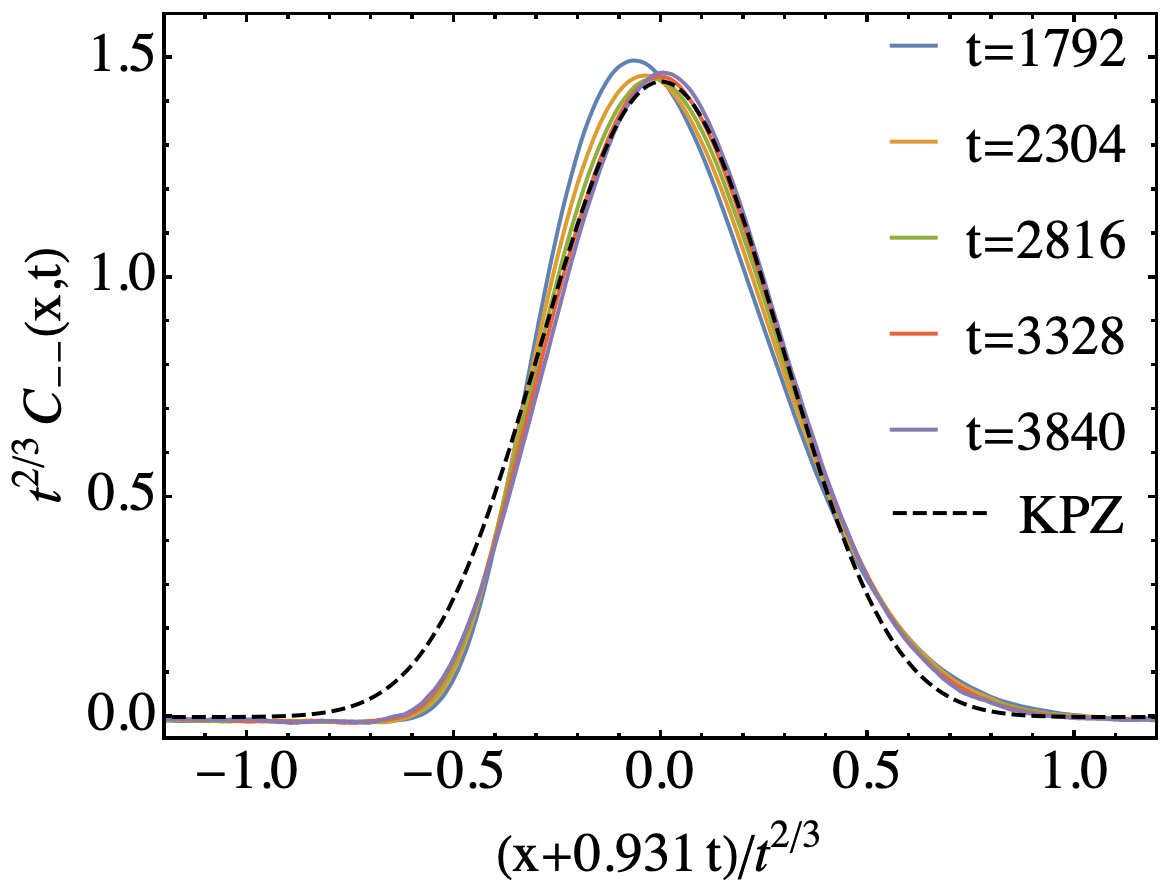}
  \includegraphics[width=0.32\textwidth]{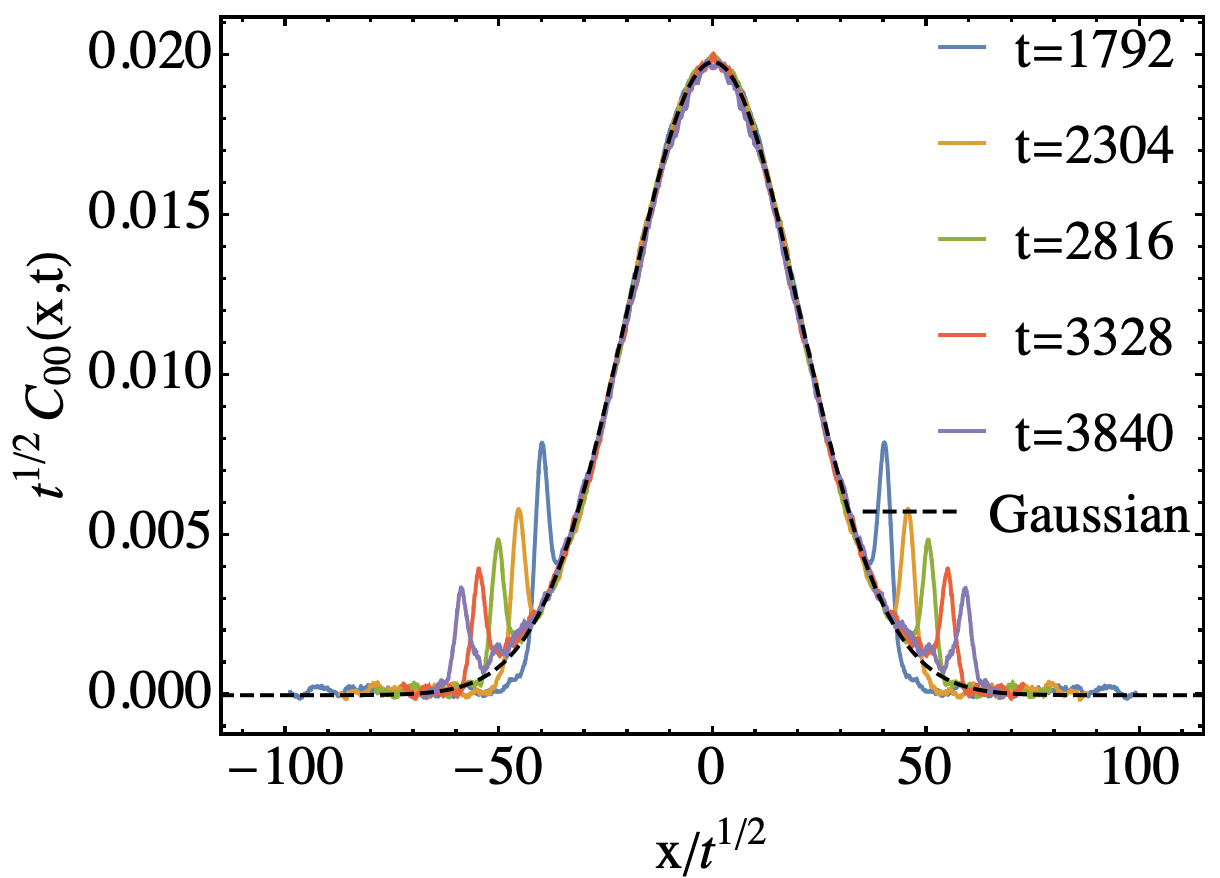}
  \includegraphics[width=0.33\textwidth]{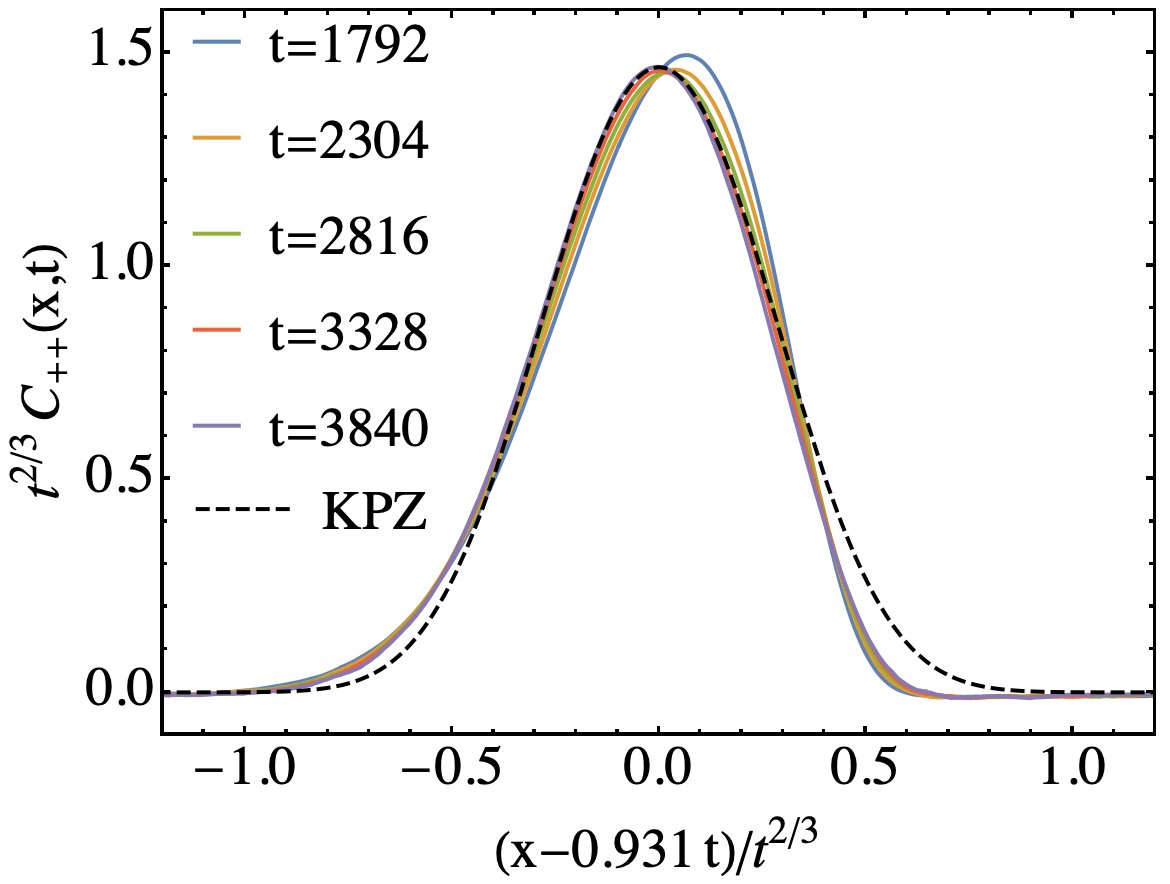}
  
\caption{Parameters: $\Delta=0.5, \beta=10.0, h=0, N=8192$ -- RK-$4$ with $dt=0.005$: Plot of  $C_{--}(x,t),~C_{00}(x,t)$ and $C_{++}(x,t)$, obtained after normal mode transformation. The 2nd row shows the diffusive scaling of both the sound modes and the heat mode with Levy scaling, while the 3rd row shows the same data with KPZ scaling of sound modes and diffusive scaling of the heat mode. Sound speed estimate from theory is $c=0.931$.}
\label{figrk4b0.0SH}
\end{figure}


\section{Numerical results at low temperature}\label{sec4} 
\label{results}\noindent 
We now present numerical results for the dynamical correlations for two cases with (I) finite magnetic field and (II) zero field,  at inverse temperature $\beta=10$. Apart from the spin and energy correlations defined in Eq.~\eqref{corrse} we  now  measure 
\begin{eqnarray}\label{corrse}
C_\mathrm{rr}(j,t) =\langle r_j(t) r_{0}(0) \rangle_{\beta,h}^\mathrm{c},
\end{eqnarray}
corresponding to the extra conserved variable. 
We also compute the normal mode correlations $C_{\sigma,\sigma'}$, where $\sigma,\sigma'=+1,0,-1$ correspond to the modes $\phi_{+1},\phi_0,\phi_{-1}$ respectively. \\ 
{\bf Simulation details}: In the simulations we first generated equilibrium configurations corresponding to the canonical  distribution, specified by $\beta,h$, using a Metropolis Monte Carlo algorithm. The equilibrated configuration was then evolved with our Hamiltonian dynamics using either a fixed step size and sometimes an adaptive step size 4th order Runge-Kutta (RK4) integrator.  This maintains the global conservations (total energy and magnetization) up to an absolute deviation of $\sim 10^{-6}$ and $\sim 10^{-13}$ respectively until the final time of evolution. Individual spin lengths are also conserved up to $\sim 10^{-13}$. Averages were compute over around $10^6$ initial conditions.

In all our simulations we fixed $\beta=10$. At much smaller temperatures, phase slips would be even rarer but the dynamics of small fluctuations about the ordered state is expected to be closer to integrable, and the time required to see the asymptotic scaling becomes inaccessible.

{\bf Case-I} \bm{$(\nu=0,~h= 0.3)$}: In this regime we expect to confirm the  KPZ sound modes and Levy heat modes as predicted by NFH. We have chosen the following parameters: $\Delta=0.5,~ \beta=10.0,~N=8192$. The $z$-component of the total spin, energy, and individual spin length are conserved up to $10^{-15}$ and $10^{-5}$ and $10^{-10}$ respectively. Phase Slip processes are rare but not completely absent.
In Fig.~(\ref{figrk4b0.3}), the correlations $C_{\alpha\beta}(x,t)$ are plotted at different times. 

In Fig.~(\ref{figrk4b0.3SH}), we plot the sound and heat modes obtained after normal mode transformations and with different scalings. Values of the $G$-matrices are given in Table-\ref{table:Gmatrics_h_0.3}.

\begin{table}
\caption{$G$ matrices ($\beta = 10,~h=0.3,~\nu=0$)}
\centering
\begin{tabular}{c c c}
\hline
\hline
$G$ matrix & MD simulation  & NFH \\ [0.5ex]
\hline
\hline
$G^-$ & $\left( \begin{array}{ccc}
0.0355 & -0.411 & 0.0117 \\
-0.411 & -3.961\text{E-}5 & 5.238\text{E-}7 \\
0.0118 & -7.658\text{E-}5 & -0.0118 \end{array} \right)$ & $\left( \begin{array}{ccc}
 0.03536 & -0.4079 & 0.01179 \\
-0.4079 & 0 & 0 \\
0.01179 & 0 & -0.01179 \end{array} \right)$ \\
$G^0$ & $\left( \begin{array}{ccc}
-0.411 & -2.989\text{E-}16 & -3.828\text{E-}5 \\
-3.961\text{E-}5 & 2.057\text{E-}20 & 2.104\text{E-}5 \\
-3.829\text{E-}5 & 3.118\text{E-}16 & 0.411 \end{array} \right)$ & $\left( \begin{array}{ccc}
-0.4079 & 0 & 0 \\
0 & 0 & 0 \\
0 & 0 & 0.4079 \end{array} \right)$ \\
$G^+$ & $\left( \begin{array}{ccc}
0.0118 & -7.657\text{E-}5 & -0.0118 \\
5.280\text{E-}7 & 2.104\text{E-}5 & 0.411 \\
-0.0118 & 0.411 & -0.0352 \end{array} \right)$ & $\left( \begin{array}{ccc}
 0.01179 & 0 & -0.01179 \\
0 & 0 & 0.4079 \\
-0.01179 & 0.4079 & -0.03536 \end{array} \right)$ \\
[1ex]
\hline
\makecell{speed of\\
sound $c$} & \makecell{
0.865 \text{ (maxima of the sound peaks)} \\
0.859 \text{ (normal mode transformation)} } & 0.85217\\
[1ex]
\hline
\hline
\end{tabular}
\label{table:Gmatrics_h_0.3}
\end{table}

As for anharmonic chains, $G^0_{00} = 0$  which implies that the self-coupling term is absent for the heat mode. Also $G^\sigma_{\sigma\sigma}~(\sigma=\pm 1)$ are non-zero which in the context of NFH  is the crucial property for  the KPZ scaling of sound modes and Levy-5/3 scaling of heat mode:
\begin{align}
C_{\sigma\sigma}(x,t) &\sim \dfrac{1}{(\lambda_s t)^{2/3}}f_{\text{KPZ}}\left[ \dfrac{x-c\sigma t}{(\lambda_s t)^{2/3}} \right],\\
C_{00}(x,t) &\sim \dfrac{1}{(\lambda_h t)^{3/5}}f_{\text{Levy}}^{5/3}\left[ \dfrac{x}{(\lambda_h t)^{3/5}} \right],
\end{align}
where $f^\alpha_\text{Levy}(x) = \text{InverseFourier}\left[e^{-|k|^\alpha}\right]\sim \dfrac{1}{\pi}\dfrac{1}{|x|^{\alpha+1}}$ and the universal $f_{\text{KPZ}}$ function is tabulated in \cite{prahofer2004exact}.

From the results of our simulations, shown in Fig.~(\ref{figrk4b0.3SH}), we find that the sound mode is better described by a KPZ-type scaling while the heat mode appears to be closer to diffusive than the expected Levy-$5/3$. The sound speed from the simulations is $c\approx 0.865$ which is close to the theoretical estimate $c=0.85217$.

\textbf{Case-II} \bm{$(\nu=0,~h=0)$}: In this regime, the $r$- interaction potential is symmetric under reflection. This leads to a distinct universality class. On the basis of NFH
the sound modes are expected to be diffusive and the heat mode to be Levy-3/2.  We have chosen the following parameters: $\Delta=0.5$,~$\beta=10.0,~N=8192$. Phase Slip events are absent in this regime. The  $z$-component of the spin, energy, and individual spin length are conserved up to $10^{-12}$ and $10^{-5}$ and $10^{-10}$ respectively. In Fig.~(\ref{figrk4b0.0}), the correlations $C_{\alpha\beta}(x,t)$ are plotted at different times. In Figs.~(\ref{figrk4b0.0SH}), we plot the sound and heat modes obtained after normal mode transformations and with different scalings. The $G$-matrices are given in Table-\ref{table:Gmatrics_h_0}.
\begin{table}[h]
\caption{G matrices ($\beta = 10,~h=0,~\nu=0$)}
\centering
\begin{tabular}{c c c}
\hline
\hline

$G$ matrix& MD simulation & NFH \\ [0.5ex]
\hline
\hline
$G^-$ & $\left( \begin{array}{ccc}
6.655\text{E-}6 & -0.4494 & 1.520\text{E-}4 \\
-0.4494 & -3.415\text{E-}5 & 1.080\text{E-}8 \\
1.768\text{E-}4 & -6.453\text{E-}5 & -1.358\text{E-}5 \end{array} \right)$ & $\left( \begin{array}{ccc}
0 & -0.4488 & 0 \\
-0.4488 & 0 & 0 \\
0 & 0 & 0 \end{array} \right)$ \\
$G^0$ & $\left( \begin{array}{ccc}
-0.4494 & 3.275\text{E-}20 & -3.227\text{E-}5 \\
-3.415\text{E-}5 & 0. & 2.480\text{E-}5 \\
-3.226\text{E-}5 & 6.551\text{E-}20 & 0.4494 \end{array} \right)$ & $\left( \begin{array}{ccc}
-0.4488 & 0 & 0 \\
0 & 0 & 0 \\
0 & 0 & 0.4488 \end{array} \right)$ \\
$G^+$ & $\left( \begin{array}{ccc}
1.768\text{E-}4 & -6.453\text{E-}5 & -1.355\text{E-}5 \\
1.457\text{E-}8 & 2.480\text{E-}5 & 0.4494 \\
2.055\text{E-}5 & 0.4494 & -5.058\text{E-}4 \end{array} \right)$ & $\left( \begin{array}{ccc}
 0 & 0 & 0 \\
0 & 0 & 0.4488 \\
0 & 0.4488 & 0 \end{array} \right)$ \\
[1ex]
\hline
speed of sound $c$ & $\begin{array}{c}
0.931 \text{ (maxima of the sound peaks)} \\
0.930 \text{ (normal mode transformation)} \end{array}$ & 0.92837\\
[1ex]
\hline
\hline
\end{tabular}
\label{table:Gmatrics_h_0}
\end{table}

We have $G^0_{00} = 0$ here as well. Unlike the previous case $G^\sigma_{\sigma\sigma}~(\sigma=\pm1)$ are \textit{zero} which gives rise to diffusive sound mode and Levy-3/2 heat mode in NFH:
\begin{align}
C_{\sigma\sigma}(x,t) &\sim \dfrac{1}{(\lambda_s t)^{1/2}}\exp\left[ \dfrac{(x-c\sigma t)^2}{\lambda_s t} \right]\\
C_{00}(x,t) &\sim \dfrac{1}{(\lambda_h t)^{2/3}}f_{\text{Levy}}^{3/2}\left[ \dfrac{x}{(\lambda_h t)^{2/3}} \right],
\end{align}

As mentioned earlier the expectation from theory is that the sound modes are diffusive while the heat mode is Levy-$3/2$. However, as can be seen from our  simulations, i.e.  Fig.~(\ref{figrk4b0.0SH}), these scalings are not seen conclusively. The sound speed from the simulations is $c\approx 0.931$ which agrees well with the theoretical estimate $c=0.92837$. 

\section{Discrete time LLL dynamics}
\label{discreteLLL}
In this section we discuss a discrete version of the lattice Landau-Lifshitz equations, using an integration
algorithm that explicitly preserves the total angular momentum and the total energy of
the system to machine precision independent of the time step used in the numerical integration, while allowing a bounded error in the fixed-length
constraint on each classical spin (in the anisotropic case with $\Delta \neq 1$, it is of course only the $z$ component of the angular momentum which is thus conserved). As will be apparent from the description below, the smallness of the violation of the fixed-length constraint for each spin is controlled by the size of the time-step used in the numerical integration. Additionally, this error does not grow with the total time over which the system is evolved. Since all our
general arguments and theoretical analysis rely heavily on the existence of a conserved
energy and angular momentum density, and the fixed-length constraint on each
individual classical spin does not play a similar central role in the theoretical analysis, our
procedure allows us to use relatively large time-steps while preserving the universality
class of the dynamics to machine precision.

Our  procedure may be viewed as a modification of the so-called ``odd-even dynamics''
that has been employed previously in the literature \cite{frank1997geometric}. In the odd-even decomposition,
one splits the time-evolution into two steps, one of which sets in motion all the odd spins,
allowing them to precess in the exchange field supplied by the (temporarily) static
even spins, while the other step reverses the role of even and odd spins
to evolve the even spins. Below we refer to these individual steps as odd and even
precessions.

Our modification is suggested by a particular approximation (the Cayley approximation) to the evolution operators
that implement these even or odd precessions over a small time $\delta$ in a way
that preserves the orthogonal nature of the evolution operator. Consider Eqn.~\eqref{2.2} with $j$ odd
and all $B_j$ held fixed (by keeping all even spins fixed). With all $B_j$ fixed,
the right hand side is a $B_j$ dependent rotation of all the odd spins. Let us
denote the corresponding linear operator by $R_{o}(B_{e})$ (where $o$ stands for odd and the subscript $e$ on $B$ reminds us that $B$ depends on the current
configuration of all even spins, held fixed for the duration of this step). Here $R_{o}$
and $R_{e}$ are antisymmetric matrices that serve as the generators of the corresponding
rotations.
The odd spins after a time $\delta$ can be obtained by applying the operator
$\exp(\delta R_o)$ to the initial configuration of odd spins.

We begin with the Cayley approximation to individual precessions,
\begin{equation}\label{ksd1}
  \exp(\delta R_o) \approx (1-\frac{\delta}{2} R_o)^{-1} (1+\frac{\delta}{2}R_o),
\end{equation}
and similarly for $R_e$. With this in hand, we define the operator
\begin{equation}\label{ksd2}
O_1 =  \left[1+\frac{\delta}{2} R_e(B_o)\right] \left[1-\frac{\delta}{2}R_o(B_e)\right]^{-1}.
\end{equation}
Here, we have explicitly displayed the dependence of
$R_o$ ($R_e$) on $B_e$ ($B_o$) in the configuration on which the operator acts. Let us schematically denote the configuration of odd
spins after the action of the first term by $S^{1/2}_o$. The second term is a first
order approximation to the precession of even spins in the exchange field provided
by $S^{1/2}_o$, starting with the initial configuration of even spins, denoted schematically
by $S^{0}_e$. Now, the first term amounts to precessing odd spins backwards
in time (again using the same first order approximation), using the exchange field provided by the configuration $S^{0}_e$, but starting with the configuration $S^{1/2}_o$ in order to end up finally with the initial configuration $S^{0}_o$ of the odd spins.

\begin{figure}[b]
  \includegraphics[width=0.49\textwidth]{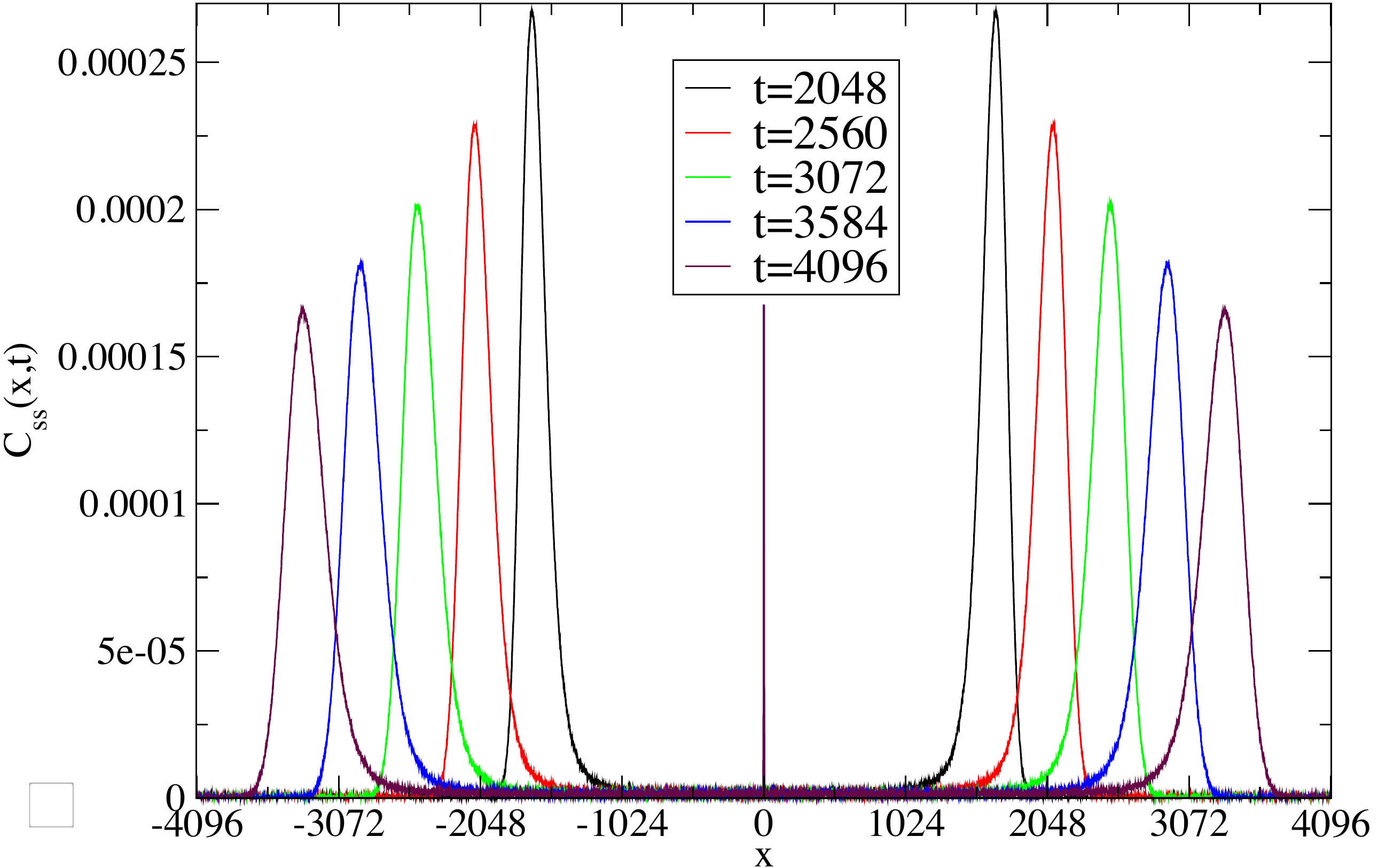}
  \includegraphics[width=0.49\textwidth]{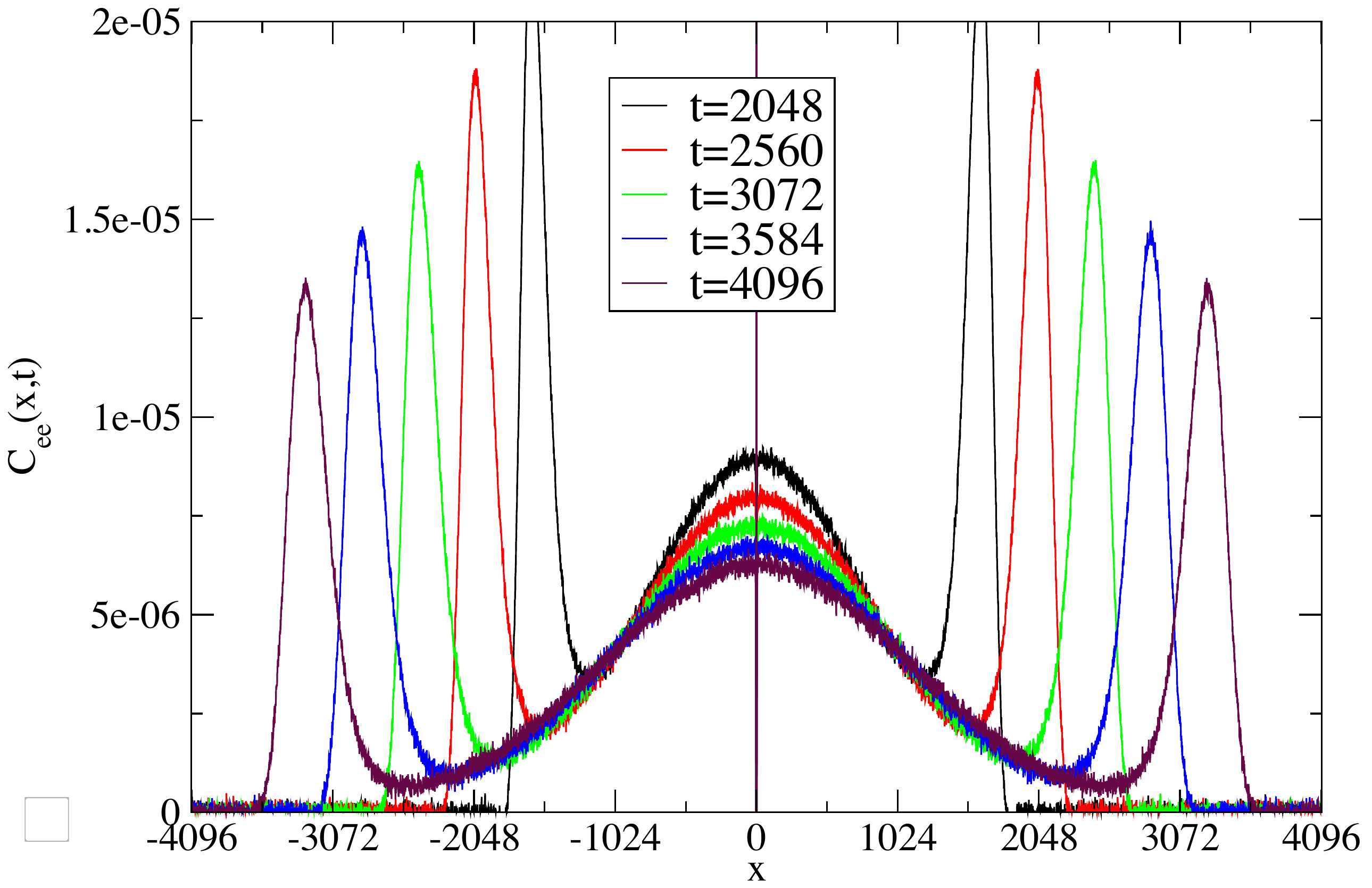}
\caption{Results from discrete dynamics with $dt=1.0$: Plot of $C_{ss}(x,t)$ and $C_{ee}(x,t)$ for parameters $\beta=8.0,h=0.3, N=8192$ at five different times. Note that there is a spurious peak, constant in time, which is an artifact of the discrete dynamics and goes away in the limit $dt \to 0$, however this does not affect the properties of the relevant remaining part of the correlation functions.}
\label{figV}
\end{figure} 

\begin{figure}[b]
  \includegraphics[width=0.49\textwidth]{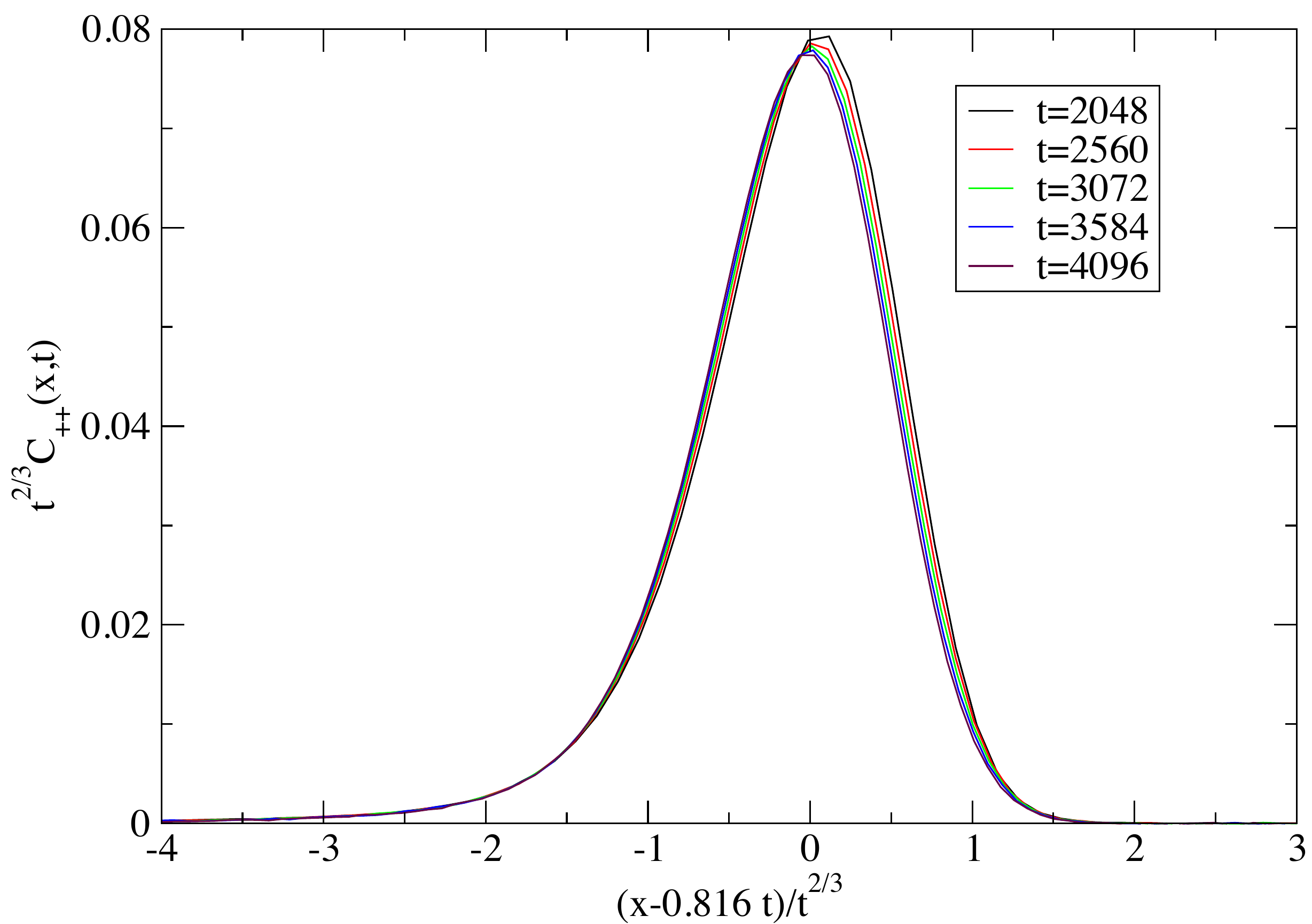}
  \includegraphics[width=0.49\textwidth]{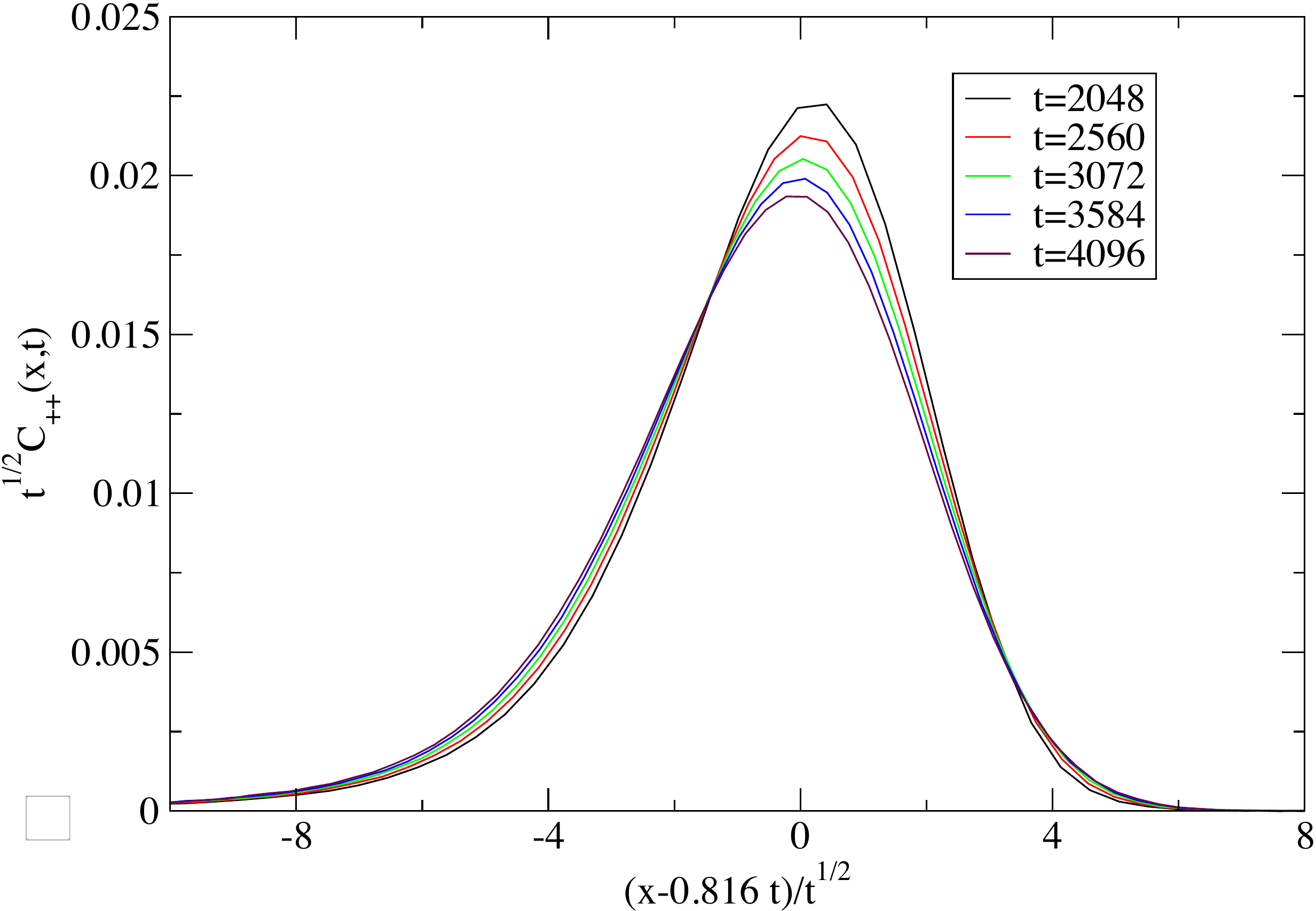}
\caption{ Results from discrete dynamics with $dt=1.0$: Plot of  $C_{++}(x,t)$, obtained after normal mode transformation, with KPZ and diffusive scaling respectively, for parameters $\beta=8.0,h=0.3, N=8192$, at five different times.  Sound estimate from theory is $c=0.809$.}
\label{figVsound}
\end{figure}  

\begin{figure}[t]
  \includegraphics[width=0.49\textwidth]{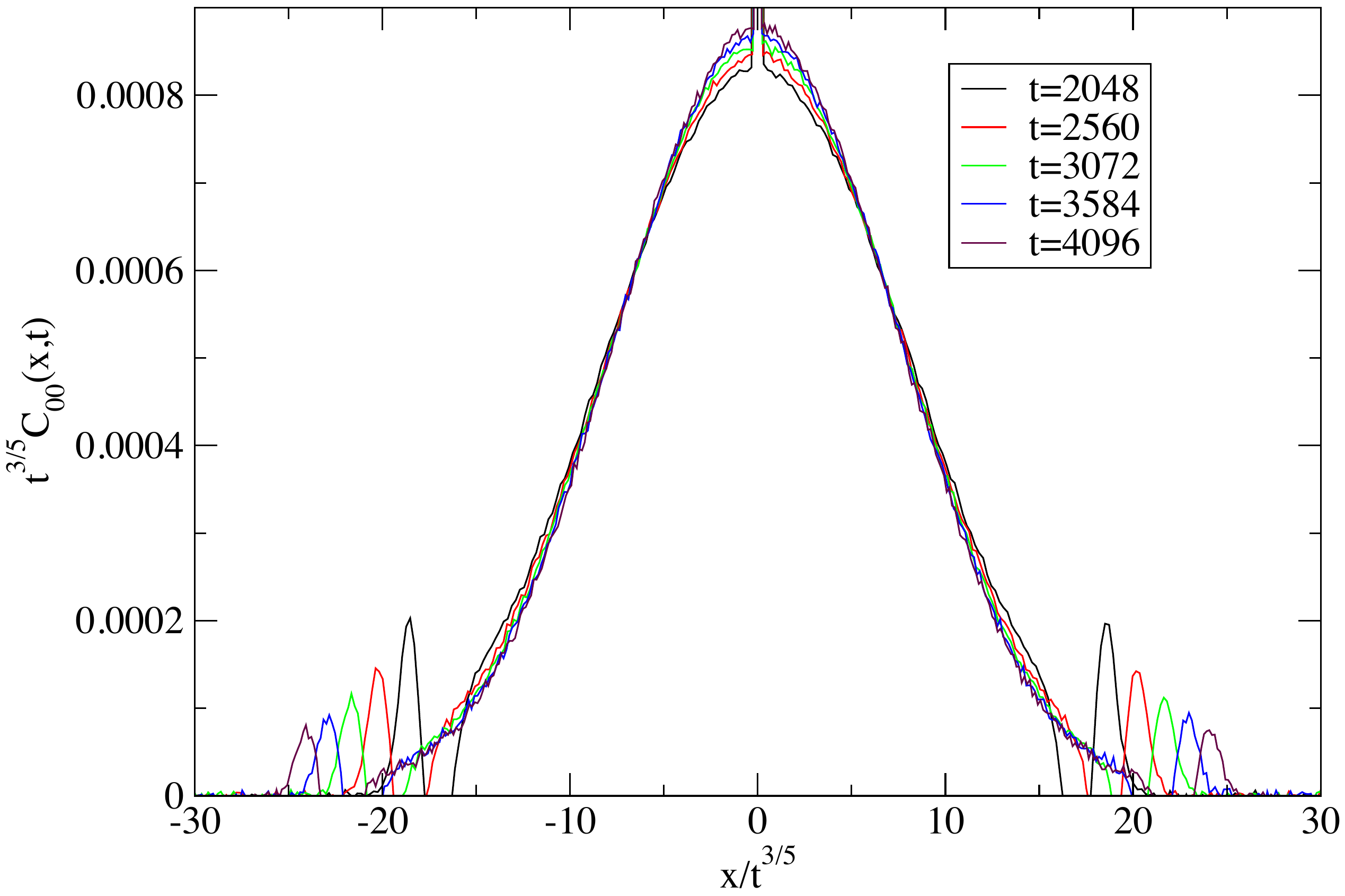}
  \includegraphics[width=0.49\textwidth]{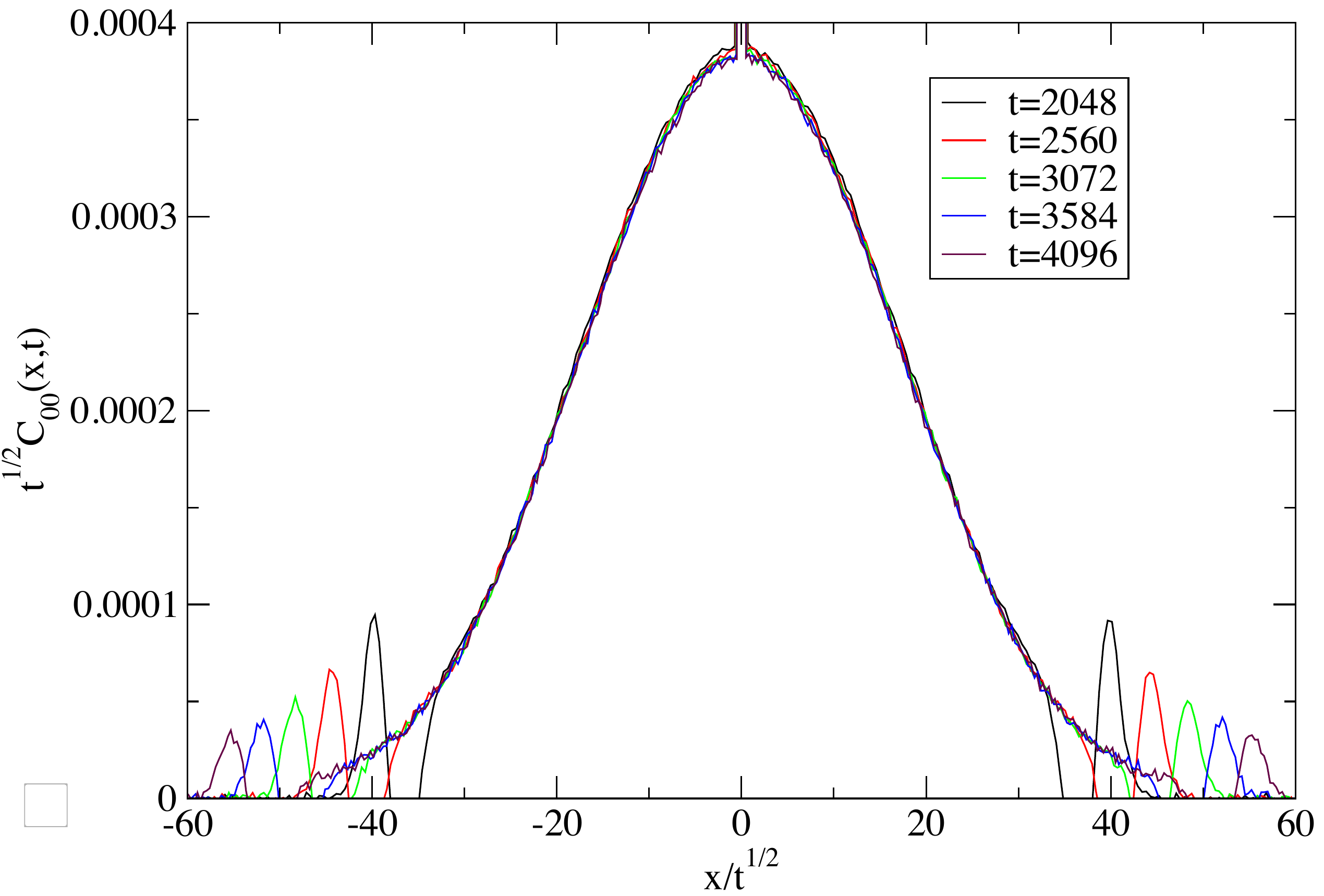}
\caption{Results from discrete dynamics with $dt=1.0$: Plot of  $C_{00}(x,t)$, obtained after normal mode transformation, with Levy-$5/3$ and diffusive scaling respectively for parameters $\beta=8.0,h=0.3, N=8192$, at five different times.}
\label{figVheat}
\end{figure}

As a result, it is easy to see
that $O_1$ explicitly preserves both the total angular momentum (only the $z$ component
of the angular momentum if $\Delta \neq 1$) and the total energy of the system,
independently of the value of $\delta$. However, $O_1$ is not orthogonal.
$O_2$ defined as
\begin{equation}\label{ksd3}
O_2 =  \left[1+\frac{\delta}{2} R_o(B_e)\right]\left[1-\frac{\delta}{2}R_e(B_o)\right]^{-1}
\end{equation}
has the same properties.

Next we note that the product $O_2 O_1$ provides a second order accurate
(in $\delta$) approximation to the actual spin dynamics. This approximation preserves
the total angular momentum and total energy independently of $\delta$, while
preserving the fixed-length constraint on each spin only to second order.
Additionally, our numerical tests reveal that the violation of the fixed length
constraint does not grow with total time of integration if each time step is
implemented by the operator $O_2 O_1$, even for relatively large values of
$\delta$.

Therefore, we integrate the lattice Landau Lifshitz equations using $O_2O_1$ to
evolve the system over one time step. As noted earlier, this can be viewed
as a modification of the standard odd-even dynamics used earliear.

We display our results from discrete time dynamics with $dt=1$.\\
\textbf{Case-III} \bm{$(\beta=8.0, h=0.3, N=8192):$} In Fig.~(\ref{figV}),
we show plots for evolution of the correlations $C_{ss}(x,t)$ and $C_{ee}(x,t)$, while in Figs.~(\ref{figVsound},\ref{figVheat}) we plot the sound mode and heat modes after appropriate translation and with  different scalings.  As with the continuous time dynamics, in Fig.~(\ref{figrk4b0.3SH}), we find again that the sound mode is better described by a KPZ-type scaling while the heat mode appears to be closer to diffusive than the expected Levy-$5/3$. The sound speed from the simulations  
is $c\approx 0.816$ which is close to the theoretical estimate $c=0.809$.

\begin{figure}[t]
  \includegraphics[width=0.49\textwidth]{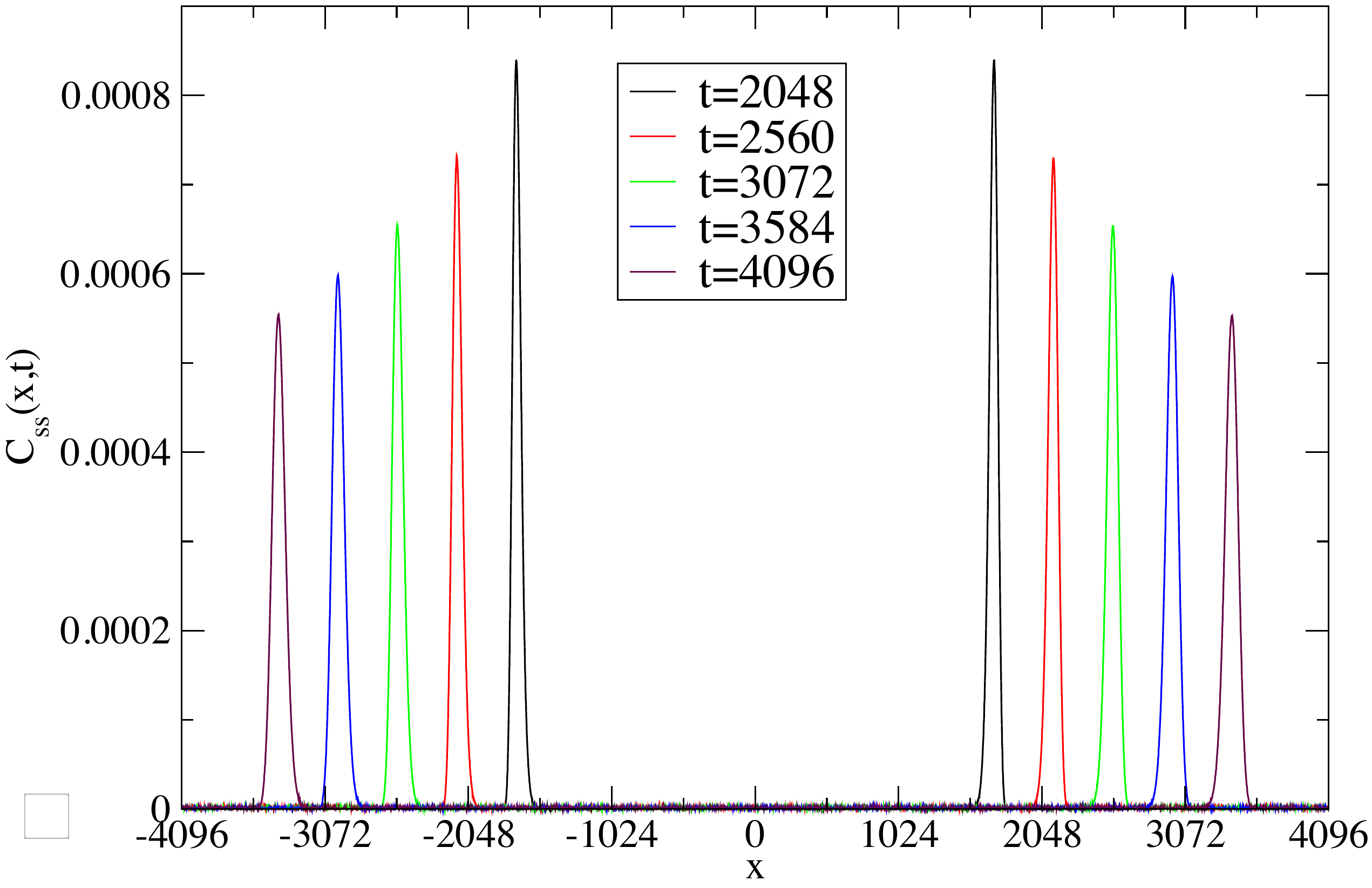}
  \includegraphics[width=0.49\textwidth]{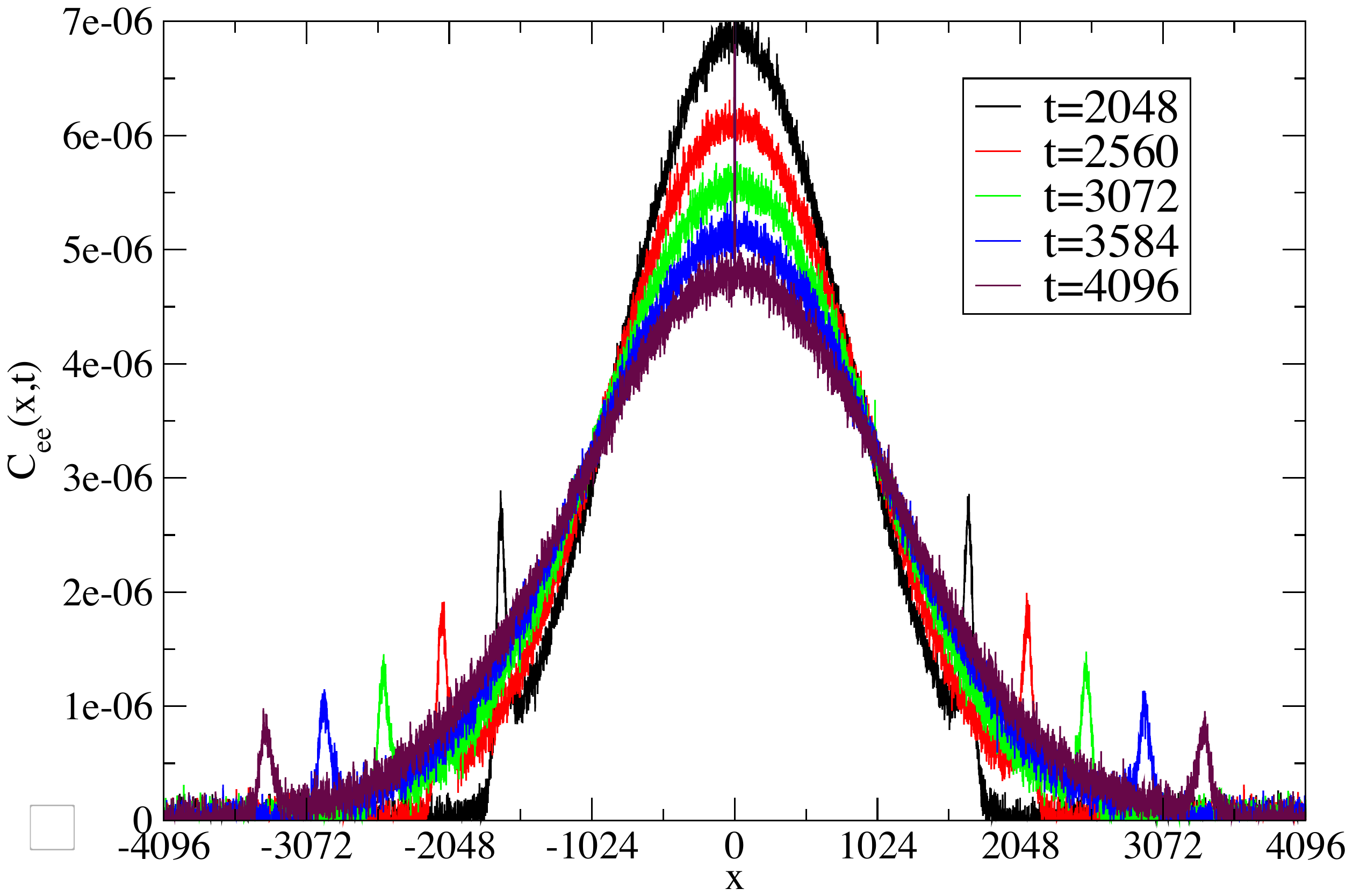}
\caption{Results from discrete dynamics with $dt=1.0$: Plot of $C_{ss}(x,t)$ and $C_{ee}(x,t)$ for parameters $\beta=8.0,h=0.0,N=8192$ at five different times.}
\label{figVI}
\end{figure}

\begin{figure}[t]
  \includegraphics[width=0.49\textwidth]{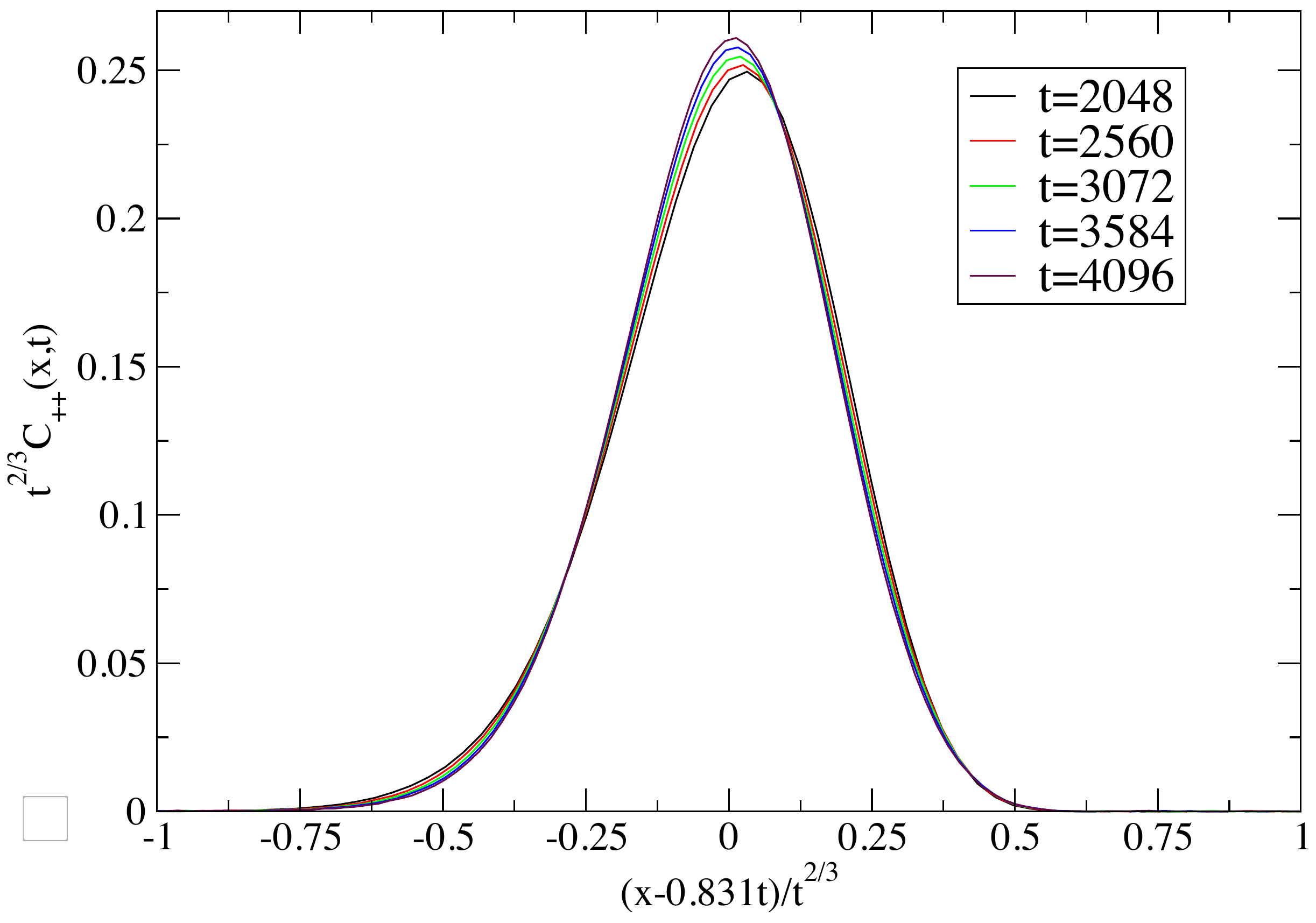}
  \includegraphics[width=0.49\textwidth]{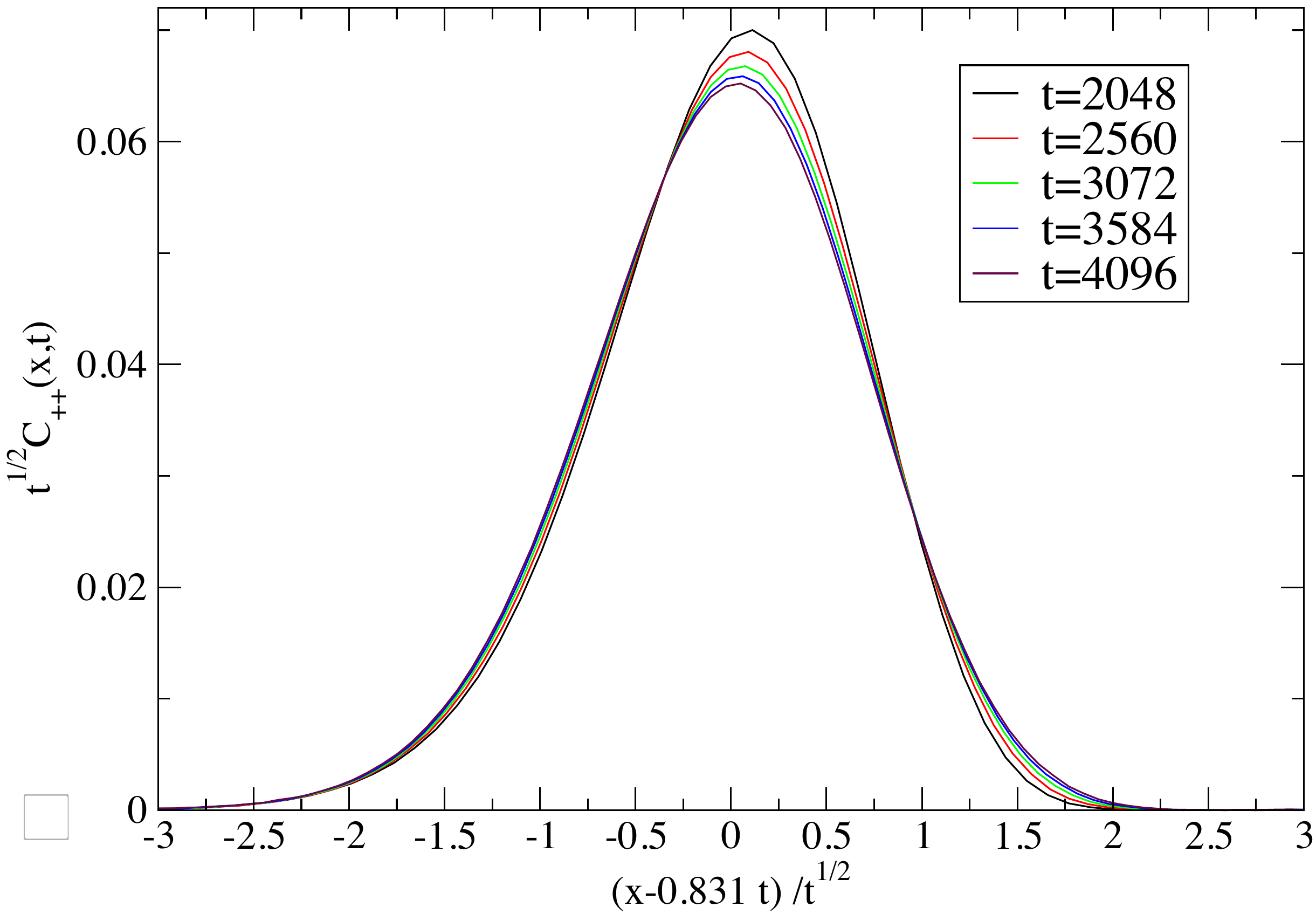}
\caption{Results from discrete dynamics with $dt=1.0$: Plot of  $C_{++}(x,t)$, obtained after normal mode transformation, with KPZ and diffusive scaling respectively, for parameters $\beta=8.0,h=0.0,N=8192$, at five different times.  Sound estimate from theory is $c=0.906$.}
\label{figVIsound}
\end{figure}

\begin{figure}[t]
  \includegraphics[width=0.49\textwidth]{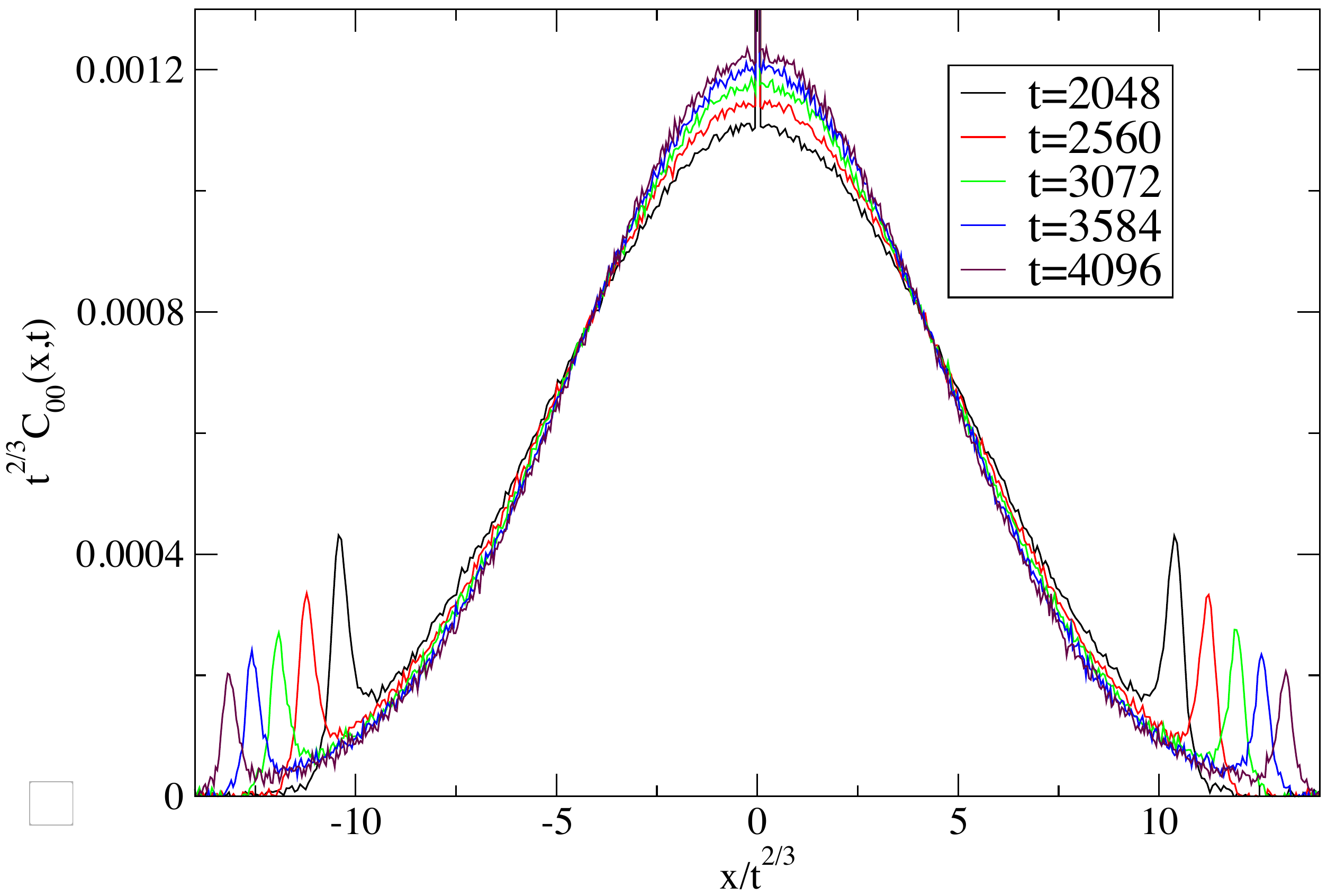}
  \includegraphics[width=0.49\textwidth]{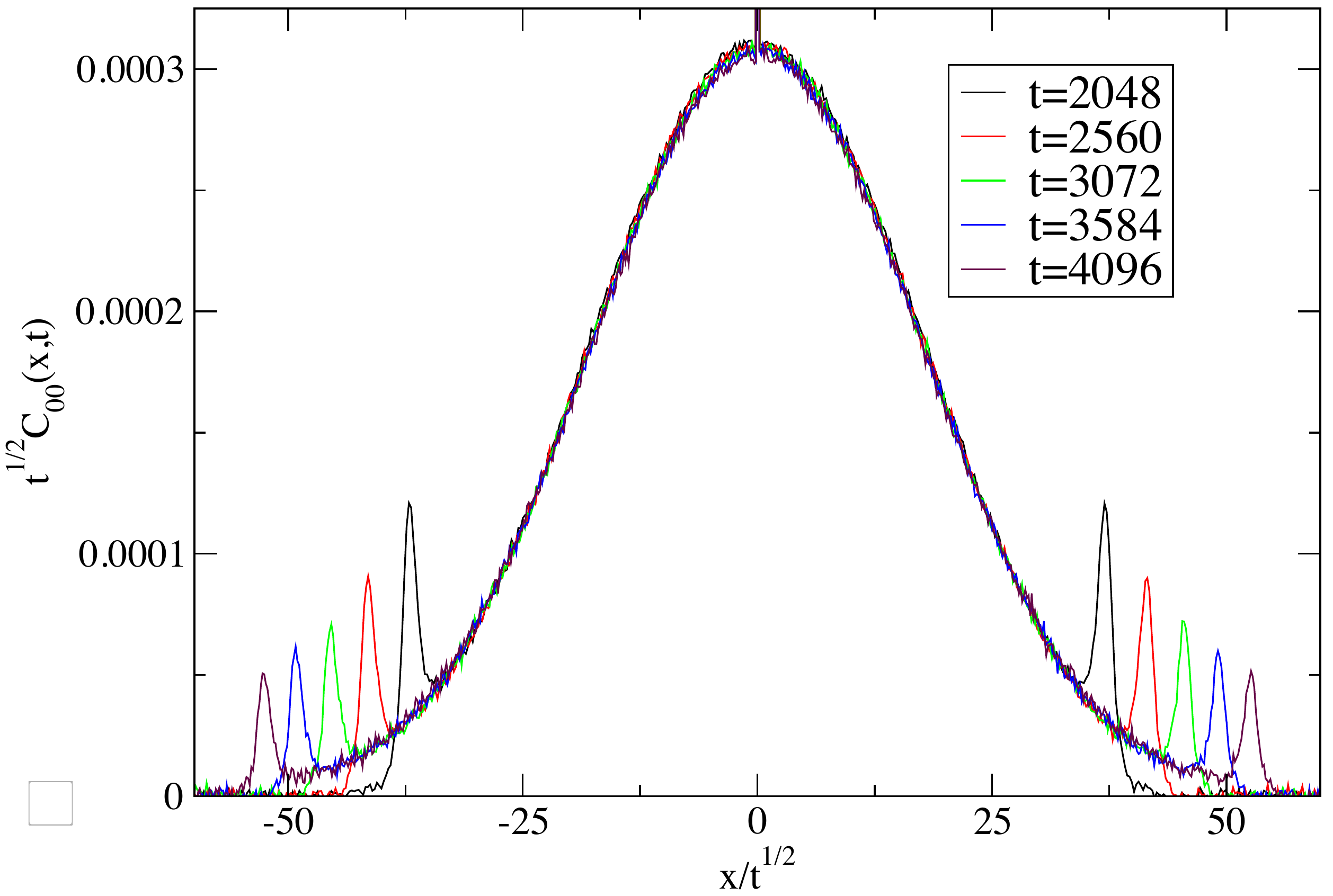}
\caption{Results from discrete dynamics with $dt=1.0$: Plot of  $C_{00}(x,t)$, obtained after normal mode transformation, with Levy-$3/2$ and diffusive scaling respectively for parameters $\beta=8.0,h=0.0,N=8192$, at five different times.}
\label{figVIheat}
\end{figure}

\textbf{Case-IV} \bm{$(\beta=8.0, h=0.0, N=8192): $} 
In Fig.~(\ref{figVI}),
we show plots for evolution of the correlations $C_{ss}(x,t)$ and $C_{ee}(x,t)$, while in Figs.~(\ref{figVIsound},\ref{figVIheat}) we plot the sound mode and heat modes after appropriate translation and with  different scalings.   In this case the expectation from theory is that the sound modes are diffusive while the heat mode is Levy-$3/2$. However, as can be seen  in  Figs.~(\ref{figVIsound},\ref{figVIheat}),  these scalings are not seen conclusively in the simulations. The sound speed from the simulations  
is $c\approx 0.831$. Note that, in contrast to the continuous time model, we do not have a simple construction of the stationary equilibrium state.
Hence even for the sound speed we do not have a reliable prediction.

\section{Conclusions} 
In this work we have mainly studied the classical XXZ or LLL model and an integrable counterpart of it in easy-plane regimes. We are mostly concerned with the basic conserved quantities --- energy and magnetization in both models (integrable and non-integrable) at different regimes of model parameters and external parameters like temperature and magnetic field. For the LLL model, being non-integrable, these two are the only known conserved quantities but at low temperature a third almost conserved quantity emerges, producing sound modes. We have put this into the framework of NFH to get some predictions about the dynamical scaling. We have shown that the non-integrable XXZ chain in the easy-plane regime displays diffusive behaviour at high temperatures, and sound modes with KPZ broadening at low temperatures. The integrable spin chain however demonstrates ballistic behaviour in the easy plane regime. We also find that at high temperature the easy-axis nonintegrable case gives us perfectly diffusive behavior. Our various results are summarized in Table \ref{tab-summary}. For the heat mode, while NFH predicts anomalous Levy broadening, we have not been able to clearly observe this yet in the simulations. It is possible that one requires larger system sizes and longer times to observe this and further studies are necessary. Similar problems related to non-observation of the expected asymptotic behaviour, due to possible finite size or finite time effects, have been reported in other recent transport studies\cite{Zhong12,Das2014,Wang13,Dhar2018,Cintio18}, and understanding their origins is an open problem.

\begin{table}[h]
\caption{Summary of the transport properties observed in this work.}
\begin{center}
\begin{tabular}{c c}
\hline
\hline
\\
\smallskip \textbf{Model and the parameter regime} & \textbf{Observation} \\ [1ex]
\hline
\hline
\\
\multicolumn{2}{c}{\smallskip (Non-integrable LLL chain)}\\[1ex]
\hline
\\
\smallskip Easy plane at high T & Diffusive spin and energy peaks, no sound modes\\
\smallskip Easy plane at low T and $h=0$ & Diffusive sound modes$^*$ and Levy-3/2 heat mode$^*$ \\
\smallskip Easy plane at low T and small h & KPZ sound modes and Levy-5/3 heat mode$^*$ \\
\smallskip Easy axis at high T & Diffusive spin and energy peaks, no sound modes\\
[1ex]
\hline
\\
\multicolumn{2}{c}{\smallskip (Integrable LLL chain)}\\[1ex]
\hline
\\
\smallskip Easy plane at high T & Ballistic spin and energy peaks\\
[1ex]
\hline
\hline
\end{tabular}
\end{center}
* Our results for these cases are not entirely conclusive.
\label{tab-summary}
\end{table}

\section{Acknowledgements}
AD would like to thank the support from the grant EDNHS ANR-14-CE25-0011 of the French National Research Agency (ANR) and from Indo-French Centre for the Promotion of Advanced Research (IFCPAR) under project 5604-2.
MK gratefully acknowledges the Ramanujan Fellowship
SB/S2/RJN-114/2016 from the Science and Engineering Research Board
(SERB), Department of Science and Technology, Government of India. MK also acknowledges support the Early Career Research Award, ECR/2018/002085  from the Science and Engineering Research Board (SERB), Department of Science and Technology, Government of India. MK would like to acknowledge support from the project 6004-1 of the Indo-French Centre for the Promotion of Advanced Research (IFCPAR). DAH was supported in part by (USA) DOE grant DE-SC0016244. The numerical simulations were done on \textit{Mowgli, Mario} and \textit{Tetris} clusters of ICTS-TIFR and \textit{Gaggle} and \textit{Pride} clusters of DTP-TIFR.
 
\appendix
\numberwithin{equation}{section}

\section{Coupling coefficients for nonlinear fluctuating hydrodynamics}\label{appA}

Given the magic identity \eqref{3.18}, one can follow \cite{MendlSpohnNLS2015} to obtain the couplings of the quadratic nonlinearities of NFH.
Our simulations are for $\nu = 0$. Restricting to this case,  the averages below, denoted by $\langle \cdot \rangle$,  $\langle \cdot; \cdot \rangle$,
$\llangle \cdot; \cdot \rrangle$, refer to fixed $h,\beta$ at $\nu = 0$. One obtains 
\begin{equation}\label{A.1}
G^0 = \frac{c_\mathrm{s}}{2 \beta}\, \llangle e_0 - h\,s_0; e_0 - h\,s_0 \rrangle^{-1/2} \begin{pmatrix} -1 & 0 & 0 \\ 0 & 0 & 0 \\ 0 & 0 & 1 \end{pmatrix}
\end{equation}
and
\begin{equation}\label{A.2}
G^{+} = \frac{c_\mathrm{s}}{2 \beta}\, \llangle e_0 - h\,s_0; e_0 - h\,s_0 \rrangle^{-1/2} \left(\Upsilon \begin{pmatrix} -1 & 0 & 1 \\ 0 & 0 & 0 \\ 1 & 0 & 3 \end{pmatrix} + \begin{pmatrix} 0 & 0 & 0 \\ 0 & 0 & 1 \\ 0 & 1 & 0 \end{pmatrix} \right),\quad G^{-} = -(G^{+})^{\mathcal{T}}.
\end{equation}
Here $c_\mathrm{s}$ denotes the speed of sound determined through
\begin{equation}
\label{eq:sound_speed_sq}
c_\mathrm{s} = \frac{1}{\beta}\, (\Gamma \llangle r_0; r_0 \rrangle)^{-1/2}\, \llangle e_0 - h\,s_0; e_0 - h\,s_0 \rrangle^{1/2}, 
\end{equation}
where $\Gamma = \llangle s_0; s_0 \rrangle \llangle e_0; e_0 \rrangle - \llangle s_0; e_0 \rrangle^2$ and we have introduced the shorthand notation
\begin{equation}\label{A.3}
\llangle f_0; g_0 \rrangle = \sum_{j=1}^N \langle f_j; g_0 \rangle
\end{equation}
with $\langle f; g \rangle = \langle f g \rangle - \langle f \rangle \langle g \rangle$ denoting the second cumulant.  Also 
${}^\mathcal{T}$ denotes the transpose relative to the anti-diagonal and 
\begin{equation}
\label{eq:upsilon_def}
\Upsilon = - \llangle s_0; e_0 - h\,s_0 \rrangle (2 \Gamma)^{-1/2}.
\end{equation}

The thermodynamic averages and cumulants can be obtained as appropriate derivatives of the free energy $F(h,\nu,\beta)$ defined in \eqref{3.15} 
with $\nu$-derivative at $\nu = 0$,
\begin{eqnarray}
\label{A.4}
&&\llangle r_0; r_0 \rrangle = -\beta^{-1}\, \partial_{\nu}^2 F,\hspace{76pt}
\llangle s_0; s_0 \rrangle = -\beta^{-1}\, \partial_h^2 F, \nonumber\\
&&\llangle e_0 - h\,s_0; e_0 - h\,s_0 \rrangle = - \partial_{\beta}^2\, (\beta F),\qquad
\llangle s_0; e_0 - h\,s_0\rrangle = \partial_{\beta} \partial_{h} F\,.
\end{eqnarray}
As second order Taylor coefficients, the $G$-matrices are symmetric.

The free energy has to be numerically evaluated. An efficient method is to use   transfer operator techniques \cite{MendlSpohnNLS2015}.    Inserting our simulation parameters yields (rounded to 4 digits):
\paragraph{$\Delta = 0.5$, $\beta = 10$, $h = 0.3$}, speed of sound $c_\mathrm{s} = 0.85217$,
\begin{equation}\label{A.5}
G^{+} = \begin{pmatrix}
 0.01179 & 0      & -0.01179 \\
 0       & 0      &  0.4079  \\
-0.01179 & 0.4079 & -0.03536 \\
\end{pmatrix},
\qquad
G^0 = \begin{pmatrix}
-0.4079 & 0 & 0      \\
 0      & 0 & 0      \\
 0      & 0 & 0.4079 \\
\end{pmatrix}.
\end{equation}

\paragraph{$\Delta = 0.5$, $\beta = 10$, $h = 0$}, speed of sound $c_\mathrm{s} = 0.92837$,
\begin{equation}\label{A.6}
G^{+} = \begin{pmatrix}
 0 & 0      & 0      \\
 0 & 0      & 0.4488 \\
 0 & 0.4488 & 0      \\
\end{pmatrix},
\qquad
G^0 = \begin{pmatrix}
-0.4488 & 0 & 0      \\
 0      & 0 & 0      \\
 0      & 0 & 0.4488 \\
\end{pmatrix}.
\end{equation}
In TABLE I and II these results are compared with data coming from molecular dynamics.

\section{Low temperature approximation}
\label{sec3B}
It is instructive to work out the prefactors of $G$-matrices \eqref{A.1},  \eqref{A.2} in the harmonic approximation.
In principle, also the next order correction could be computed, compare with \cite{MendlSpohnNLS2015}. 
We start from the hamiltonian
\begin{equation} 
\label{B.1}
H = \sum_{j=-N/2}^{N/2-1} \left[ -\sqrt{1 - s_j^2} \sqrt{1 - s_{j+1}^2} \cos(r_j) - \Delta s_j s_{j+1} - h s_j \right]
\end{equation}
with $s_j \in [-1,1]$, phase difference $r_j = \phi_{j+1} - \phi_j$, and external field $h$.

The minimum of $H$ is assumed for $r_j = 0$ and $s_j = \frac{h}{2 (1 - \Delta)}$ for all $j$. We expand $H$ in a Taylor series around this minimum 
and set $t_j = s_j - \frac{h}{2 (1 - \Delta)}$.  Neglecting boundary terms one obtains
\begin{equation}
\label{B.2}
H = N e_\mathrm{g} + H_0 + V + \dots 
\end{equation}
with the ground state energy  $e_\mathrm{g} = -1 - \frac{h^2}{4 (1 - \Delta)}$, the quadratic contribution
\begin{equation}
\label{B.3}
H_0 = \frac{1}{2} \sum_j \left[ a \, t_j^2 - \tfrac{1}{2} b \, t_j (t_{j-1} + t_{j+1}) + c \, r_j^2 \right],
\end{equation}
and the cubic correction
\begin{equation}
\label{B.4}
V = \frac{h}{4 (1 - \Delta)}\sum_j \Big[ \frac{1}{c^2} t_j^2 \, ( - t_{j-1} + 2 t_j - t_{j+1}) - r_j^2\, ( t_j + t_{j+1} ) \Big].
\end{equation}
Here we introduced the shorthands
\begin{equation}
\label{B.5}
a = \frac{2}{c}, \quad b = a - 2 (1 - \Delta), \quad c = 1 - \left( \frac{h}{2 (1 - \Delta)} \right)^2.
\end{equation}

We write $H_0 = \frac{1}{2} \langle x, A x \rangle$ with $x = (t_{-N/2}, \dots, t_{N/2-1}, r_{-N/2}, \dots, r_{N/2-1}) \in \mathbb{R}^{2N}$ and a block-diagonal matrix $A$ consisting of two blocks $A_t$ and $A_r = c\mathbb{I}_N$, with tridiagonal
\begin{equation}
\label{B.6}
A_t = \begin{pmatrix}
 a            & -\frac{b}{2} &              &              &              \\
 -\frac{b}{2} & a            & -\frac{b}{2} &              &              \\
              & \ddots       & \ddots       & \ddots       &              \\
              &              & -\frac{b}{2} & a            & -\frac{b}{2} \\
              &              &              & -\frac{b}{2} & a            \\
\end{pmatrix}.
\end{equation}

The partition function of the quadratic part plus $N e_\mathrm{g}$ is defined as
\begin{equation}
\label{B.7}
Z_0^{(N)}(h, \beta) = \mathrm{e}^{-\beta N e_\mathrm{g}} \int_{([-1,1] \times [-\pi,\pi])^N} \mathrm{e}^{-\beta H_0} \prod_j \mathrm{d} s_j \mathrm{d} r_j.
\end{equation}
For large $\beta$,  the integration domain of $s_j$ and $r_j$ can be approximately extended to $\mathbb{R}$, which yields
\begin{equation}
\label{B.8}
\int_{\mathbb{R}^{2N}} \mathrm{e}^{-\beta H_0} \mathrm{d}^{2N} x = \int_{\mathbb{R}^{2N}} \mathrm{e}^{-\frac{1}{2} \langle x, \beta A x \rangle} \mathrm{d}^{2N} x = \left( \det \frac{\beta A}{2 \pi}\right)^{-1/2}.
\end{equation}
The determinant of the tridiagonal matrix $A_t$ is available in closed form \cite{HuOConnell1996}, such that
\begin{equation}
\label{B.9}
\det A = (\det A_t) (\det A_r) =\left(\frac{b c}{2}\right)^N \frac{\sinh((N+1)\lambda)}{\sinh(\lambda)} \quad \text{with } \cosh(\lambda) = \frac{a}{b}.
\end{equation}
Inserting these relations results in the free energy 
\begin{equation}
\label{B.10}
F_0^{(N)}(h, \beta) \equiv -\frac{1}{\beta N} \log\!\big(Z_0^{(N)}(h, \beta)\big) = e_\mathrm{g} + \frac{1}{\beta} \log\!\Big(\frac{\beta}{2\pi}\sqrt{b c/2}\Big) + \frac{1}{2 \beta N} \log\!\Big(\frac{\sinh((N+1)\lambda)}{\sinh(\lambda)}\Big).
\end{equation}
In the thermodynamic limit $N \to \infty$,
\begin{equation}
\label{B.11}
\begin{split}
F_0(h, \beta) &= \lim_{N \to \infty} F_0^{(N)}(h, \beta) = e_\mathrm{g} + \frac{1}{\beta} \log\!\Big(\frac{\beta}{2\pi}\sqrt{b c/2}\Big) + \frac{\lambda}{2 \beta} \\
&= e_\mathrm{g} + \frac{1}{\beta} \left[ \frac{1}{2}\mathrm{arccosh}\!\left(\frac{a}{b}\right) + \log(\beta \sqrt{b c /2}) - \log(2 \pi) \right].
\end{split}
\end{equation}


With this input we work out the free energy derivatives \eqref{A.4}  to leading order in $1/\beta$ and obtain
\begin{eqnarray}
\label{B.12}
&&-\beta^{-1}\, \partial_{\nu}^2 F = \frac{1}{\beta c}, \qquad -\beta^{-1}\, \partial_h^2 F = 0,\qquad - \partial_{\beta}^2\, (\beta F) = 0, \nonumber\\
&& \partial_{\beta} (\beta\,F_0(h,\beta)) = e_\mathrm{g} + \frac{1}{\beta},\qquad \partial_\beta \partial_{\nu} F = 0,  \qquad\partial_\beta \partial_{h} F = 0.
\end{eqnarray}
Therefore the speed of sound is given by
\begin{equation}
\label{B.13}
c_\mathrm{s} = \sqrt{(a - b) c}
\end{equation}
For $h=0$ and $\Delta=0.5$, we get $c_s = 1$.\\
For the $G$ matrices (at $h=0$) we arrive at
\begin{equation}
\label{B.14}
G^{+} = \frac{c_\mathrm{s}}{2} \begin{pmatrix}
 0 & 0 & 0 \\
 0 & 0 & 1 \\
 0 & 1 & 0 \\
\end{pmatrix},
\qquad
G^0 = \frac{c_\mathrm{s}}{2} \begin{pmatrix}
-1 & 0 & 0 \\
 0 & 0 & 0 \\
 0 & 0 & 1 \\
\end{pmatrix}.
\end{equation}
which for our values of parameter ($\Delta=0.5$) gives, 
\begin{equation}
\label{B.14val}
G^{+} =\begin{pmatrix}
 0 & 0 & 0 \\
 0 & 0 & 0.5 \\
 0 & 0.5 & 0 \\
\end{pmatrix},
\qquad
G^0 = \begin{pmatrix}
-0.5 & 0 & 0 \\
 0 & 0 & 0 \\
 0 & 0 & 0.5 \\
\end{pmatrix}.
\end{equation}



\begin{thebibliography}{10}
\providecommand{\url}[1]{#1}
\csname url@samestyle\endcsname
\providecommand{\newblock}{\relax}
\providecommand{\bibinfo}[2]{#2}
\providecommand{\BIBentrySTDinterwordspacing}{\spaceskip=0pt\relax}
\providecommand{\BIBentryALTinterwordstretchfactor}{4}
\providecommand{\BIBentryALTinterwordspacing}{\spaceskip=\fontdimen2\font plus
\BIBentryALTinterwordstretchfactor\fontdimen3\font minus
  \fontdimen4\font\relax}
\providecommand{\BIBforeignlanguage}[2]{{%
\expandafter\ifx\csname l@#1\endcsname\relax
\typeout{** WARNING: IEEEtran.bst: No hyphenation pattern has been}%
\typeout{** loaded for the language `#1'. Using the pattern for}%
\typeout{** the default language instead.}%
\else
\language=\csname l@#1\endcsname
\fi
#2}}
\providecommand{\BIBdecl}{\relax}
\BIBdecl

\bibitem{aoki2000bulk}
\BIBentryALTinterwordspacing
K.~Aoki and D.~Kusnezov, ``Bulk properties of anharmonic chains in strong
  thermal gradients: non-equilibrium $\varphi$4 theory,'' \emph{Phys. Lett. A},
  vol. \textbf{265}, pp. 250--256, 2000. [Online]. Available:
  \url{https://doi.org/10.1016/S0375-9601(99)00899-3}
\BIBentrySTDinterwordspacing

\bibitem{sirker2011conservation}
\BIBentryALTinterwordspacing
J.~Sirker, R.~G. Pereira, and I.~Affleck, ``Conservation laws, integrability,
  and transport in one-dimensional quantum systems,'' \emph{Phys. Rev. B},
  vol.~\textbf{83}, p. 035115, 2011. [Online]. Available:
  \url{https://doi.org/10.1103/PhysRevB.83.035115}
\BIBentrySTDinterwordspacing

\bibitem{MendlSpohn2013}
\BIBentryALTinterwordspacing
C.~B. Mendl and H.~Spohn, ``Dynamic correlators of {Fermi-Pasta-Ulam} chains
  and nonlinear fluctuating hydrodynamics,'' \emph{Phys. Rev. Lett.}, vol. \textbf{111},
  p. 230601, 2013. [Online]. Available:
  \url{https://doi.org/10.1103/PhysRevLett.111.230601}
\BIBentrySTDinterwordspacing

\bibitem{das2014numerical}
\BIBentryALTinterwordspacing
S.~G. Das, A.~Dhar, K.~Saito, C.~B. Mendl, and H.~Spohn, ``Numerical test of
  hydrodynamic fluctuation theory in the {Fermi-Pasta-Ulam} chain,''
  \emph{Phys. Rev. E}, vol.~\textbf{90}, p. 012124, 2014. [Online]. Available:
  \url{https://doi.org/10.1103/PhysRevE.90.012124}
\BIBentrySTDinterwordspacing

\bibitem{PhysRevLett.84.2144}
\BIBentryALTinterwordspacing
C.~Giardin\`a, R.~Livi, A.~Politi, and M.~Vassalli, ``Finite thermal
  conductivity in 1d lattices,'' \emph{Phys. Rev. Lett.}, vol.~\textbf{84}, pp.
  2144--2147, 2000. [Online]. Available:
  \url{https://doi.org/10.1103/PhysRevLett.84.2144}
\BIBentrySTDinterwordspacing

\bibitem{PhysRevLett.84.2381}
\BIBentryALTinterwordspacing
O.~V. Gendelman and A.~V. Savin, ``Normal heat conductivity of the
  one-dimensional lattice with periodic potential of nearest-neighbor
  interaction,'' \emph{Phys. Rev. Lett.}, vol.~\textbf{84}, pp. 2381--2384, 2000.
  [Online]. Available: \url{https://doi.org/10.1103/PhysRevLett.84.2381}
\BIBentrySTDinterwordspacing

\bibitem{Yang2003}
\BIBentryALTinterwordspacing
L.~{Yang} and P.~{Grassberger}, ``{Are there really phase transitions in 1-d
  heat conduction models?}'' \emph{arXiv: cond-mat/0306173}, 2003. [Online].
  Available: \url{https://arxiv.org/abs/cond-mat/0306173}
\BIBentrySTDinterwordspacing

\bibitem{das2014role}
\BIBentryALTinterwordspacing
S.~G. Das and A.~Dhar, ``Role of conserved quantities in normal heat transport
  in one dimenison,'' \emph{arXiv:1411.5247}, 2014. [Online]. Available:
  \url{https://arxiv.org/abs/1411.5247}
\BIBentrySTDinterwordspacing

\bibitem{spohn2014fluctuating}
\BIBentryALTinterwordspacing
H.~Spohn, ``Fluctuating hydrodynamics for a chain of nonlinearly coupled
  rotators,'' \emph{arXiv:1411.3907}, 2014. [Online]. Available:
  \url{https://arxiv.org/abs/1411.3907}
\BIBentrySTDinterwordspacing

\bibitem{SpohnAHC2014}
\BIBentryALTinterwordspacing
H.~Spohn, ``Nonlinear fluctuating hydrodynamics for anharmonic chains,'' \emph{J.
  Stat. Phys.}, vol. \textbf{154}, pp. 1191--1227, 2014. [Online]. Available:
  \url{https://doi.org/10.1007/s10955-014-0933-y}
\BIBentrySTDinterwordspacing

\bibitem{Spohn2016book}
H. Spohn, ``Fluctuating Hydrodynamics Approach to Equilibrium Time Correlations for Anharmonic Chains", in \emph{Thermal Transport in Low Dimensions: From Statistical Physics to Nanoscale Heat Transfer} (Lecture Notes in Physics), S. Lepri, Ed.~, Springer International Publishing, 2016, pp.
  107--158. [Online].
Available:
  \url{https://doi.org/10.1007/978-3-319-29261-8}
\BIBentrySTDinterwordspacing

\bibitem{faddeev2007hamiltonian}
\BIBentryALTinterwordspacing
L.~Faddeev and L.~Takhtajan, \emph{Hamiltonian methods in the theory of
  solitons}.\hskip 1em plus 0.5em minus 0.4em\relax Springer Science \&
  Business Media, 2007. [Online]. Available:
  \url{https://doi.org/10.1007/978-3-540-69969-9}
\BIBentrySTDinterwordspacing

\bibitem{PhysRevA.88.021603}
\BIBentryALTinterwordspacing
M.~Kulkarni and A.~Lamacraft, ``Finite-temperature dynamical structure factor
  of the one-dimensional bose gas: From the {Gross-Pitaevskii} equation to the
  {Kardar-Parisi-Zhang} universality class of dynamical critical phenomena,''
  \emph{Phys. Rev. A}, vol.~\textbf{88}, p. 021603, 2013. [Online]. Available:
  \url{https://doi.org/10.1103/PhysRevA.88.021603}
\BIBentrySTDinterwordspacing

\bibitem{PhysRevA.92.043612}
\BIBentryALTinterwordspacing
M.~Kulkarni, D.~A. Huse, and H.~Spohn, ``Fluctuating hydrodynamics for a
  discrete {Gross-Pitaevskii} equation: Mapping onto the {Kardar-Parisi-Zhang}
  universality class,'' \emph{Phys. Rev. A}, vol.~\textbf{92}, p. 043612, 2015.
  [Online]. Available: \url{https://doi.org/10.1103/PhysRevA.92.043612}
\BIBentrySTDinterwordspacing

\bibitem{MendlSpohnNLS2015}
\BIBentryALTinterwordspacing
C.~B. Mendl and H.~Spohn, ``{Low temperature dynamics of the one-dimensional
  discrete nonlinear Schr\"odinger equation},'' \emph{J. Stat. Mech.}, vol.
  \textbf{2015}, p. P08028, 2015. [Online]. Available:
  \url{https://doi.org/10.1088/1742-5468/2015/08/P08028}
\BIBentrySTDinterwordspacing

\bibitem{PhysRevE.90.012147}
\BIBentryALTinterwordspacing
C.~B. Mendl and H.~Spohn, ``Equilibrium time-correlation functions for one-dimensional hard-point
  systems,'' \emph{Phys. Rev. E}, vol.~\textbf{90}, p. 012147, 2014. [Online].
  Available: \url{https://doi.org/10.1103/PhysRevE.90.012147}
\BIBentrySTDinterwordspacing

\bibitem{Zagorodny2004}
\BIBentryALTinterwordspacing
J.~Zagorodny, ``Dynamics of vortices in the two-dimensional anisotropic
  heisenberg model with magnetic fields,'' \emph{Ph.D. thesis, Universit\"{a}t
  Bayreuth, Bayreuth, Germany}, 2004. [Online]. Available:
  \url{https://epub.uni-bayreuth.de/id/eprint/955}
\BIBentrySTDinterwordspacing

\bibitem{MendlSpohnCurrent}
\BIBentryALTinterwordspacing
C.~B. Mendl and H.~Spohn, ``Current fluctuations for anharmonic chains in
  thermal equilibrium,'' \emph{J. Stat. Mech.}, vol. \textbf{2015}, p. P03007, 2015.
  [Online]. Available: \url{https://doi.org/10.1088/1742-5468/2015/03/P03007}
\BIBentrySTDinterwordspacing

\bibitem{zhao06}
\BIBentryALTinterwordspacing
H.~Zhao, ``Identifying diffusion processes in one-dimensional lattices in
  thermal equilibrium,'' \emph{Phys. Rev. Lett.}, vol.~\textbf{96}, p. 140602, Apr 2006.
  [Online]. Available:
  \url{https://link.aps.org/doi/10.1103/PhysRevLett.96.140602}
\BIBentrySTDinterwordspacing

\bibitem{PhysRevE.94.062130}
\BIBentryALTinterwordspacing
A.~Kundu and A.~Dhar, ``Equilibrium dynamical correlations in the toda chain
  and other integrable models,'' \emph{Phys. Rev. E}, vol.~\textbf{94}, p. 062130, 2016.
  [Online]. Available: \url{https://doi.org/10.1103/PhysRevE.94.062130}
\BIBentrySTDinterwordspacing

\bibitem{prosen2013macroscopic}
\BIBentryALTinterwordspacing
T.~Prosen and B.~{\v{Z}}unkovi{\v{c}}, ``Macroscopic diffusive transport in a
  microscopically integrable hamiltonian system,'' \emph{Phys. Rev. Lett.},
  vol. \textbf{111}, p. 040602, 2013. [Online]. Available:
  \url{https://doi.org/10.1103/PhysRevLett.111.040602}
\BIBentrySTDinterwordspacing

\bibitem{AvijitFT}
\BIBentryALTinterwordspacing
A.~Das, M.~Kulkarni, H.~Spohn, and A.~Dhar, ``Kardar-parisi-zhang scaling for
  the faddeev-takhtajan classical integrable spin chain,''
  \emph{arXiv:1906.02760}, 2019. [Online]. Available:
  \url{https://arxiv.org/abs/1906.02760}
\BIBentrySTDinterwordspacing

\bibitem{AvijitChaos}
\BIBentryALTinterwordspacing
A.~Das, S.~Chakrabarty, A.~Dhar, A.~Kundu, D.~A. Huse, R.~Moessner, S.~S. Ray,
  and S.~Bhattacharjee, ``Light-cone spreading of perturbations and the
  butterfly effect in a classical spin chain,'' \emph{Phys. Rev. Lett.}, vol.
  \textbf{121}, p. 024101, 2018. [Online]. Available:
  \url{https://doi.org/10.1103/PhysRevLett.121.024101}
\BIBentrySTDinterwordspacing

\bibitem{ongoing}
A.~Das, M.~Kulkarni, A.~Dhar, and D.~A. Huse, \emph{In preparation}, 2019.

\bibitem{prahofer2004exact}
\BIBentryALTinterwordspacing
M.~Pr{\"a}hofer and H.~Spohn, ``Exact scaling functions for one-dimensional
  stationary {Kardar-Parisi-Zhang} growth,'' \emph{J. Stat. Phys.}, vol. \textbf{115},
  pp. 255--279, 2004. [Online]. Available:
  \url{https://doi.org/10.1023/B:JOSS.0000019810.21828.fc}
\BIBentrySTDinterwordspacing

\bibitem{frank1997geometric}
\BIBentryALTinterwordspacing
J.~Frank, W.~Huang, and B.~Leimkuhler, ``Geometric integrators for classical
  spin systems,'' \emph{J. Comput. Phys.}, vol. \textbf{133}, pp. 160--172, 1997.
  [Online]. Available: \url{https://doi.org/10.1006/jcph.1997.5672}
\BIBentrySTDinterwordspacing

\bibitem{Zhong12}
\BIBentryALTinterwordspacing
Y.~Zhong, Y.~Zhang, J.~Wang, and H.~Zhao, ``Normal heat conduction in
  one-dimensional momentum conserving lattices with asymmetric interactions,''
  \emph{Phys. Rev. E}, vol.~\textbf{85}, p. 060102, 2012. [Online]. Available:
  \url{https://doi.org/10.1103/PhysRevE.85.060102}
\BIBentrySTDinterwordspacing

\bibitem{Das2014}
\BIBentryALTinterwordspacing
S.~G. Das, A.~Dhar, and O.~Narayan, ``Heat conduction in the $\alpha$-$\beta$
  fermi--pasta--ulam chain,'' \emph{Journal of Statistical Physics}, vol. \textbf{154},
  pp. 204--213, 2014. [Online]. Available:
  \url{https://doi.org/10.1007/s10955-013-0871-0}
\BIBentrySTDinterwordspacing

\bibitem{Wang13}
\BIBentryALTinterwordspacing
L.~Wang, B.~Hu, and B.~Li, ``Validity of fourier's law in one-dimensional
  momentum-conserving lattices with asymmetric interparticle interactions,''
  \emph{Phys. Rev. E}, vol.~\textbf{88}, p. 052112, 2013. [Online]. Available:
  \url{https://doi.org/10.1103/PhysRevE.88.052112}
\BIBentrySTDinterwordspacing

\bibitem{Dhar2018}
\BIBentryALTinterwordspacing
A.~Dhar, A.~Kundu, J.~L. Lebowitz, and J.~A. Scaramazza, ``Transport properties
  of the classical toda chain: effect of a pinning potential,''
  \emph{arXiv:1812.11770}, 2018. [Online]. Available:
  \url{https://arxiv.org/abs/1812.11770}
\BIBentrySTDinterwordspacing

\bibitem{Cintio18}
\BIBentryALTinterwordspacing
P.~D. Cintio, S.~Iubini, S.~Lepri, and R.~Livi, ``Transport in perturbed
  classical integrable systems: The pinned toda chain,'' \emph{Chaos, Solitons
  \& Fractals}, vol. \textbf{117}, pp. 249 -- 254, 2018. [Online]. Available:
  \url{https://doi.org/10.1016/j.chaos.2018.11.003}
\BIBentrySTDinterwordspacing

\bibitem{HuOConnell1996}
\BIBentryALTinterwordspacing
G.~Y. Hu and R.~F. O'Connell, ``Analytical inversion of symmetric tridiagonal
  matrices,'' \emph{Journal of Physics A: Mathematical and General}, vol.~\textbf{29},
  pp. 1511--1513, 1996. [Online]. Available:
  \url{https://doi.org/10.1088/0305-4470/29/7/020}
\BIBentrySTDinterwordspacing

\end{thebibliography}

\end{document}